\documentclass[aps,prx,reprint]{revtex4-2}
\usepackage{braket}
\usepackage{graphicx}
\usepackage{dcolumn}
\usepackage{bm}
\usepackage{physics}
\usepackage{amssymb}
\usepackage[dvipsnames]{xcolor}
\usepackage[pdftex, pdftitle={Article}, pdfauthor={Author},colorlinks,citecolor=blue,linkcolor=blue,urlcolor=blue]{hyperref}
\usepackage{dsfont}

\begin{document}

\newcommand{\JS}[1]{{\color{Periwinkle} #1}}
\newcommand{\todo}[1]{{\color{RubineRed} #1}}
\newcommand{\nn}{\ensuremath{\vb{n}}}
\newcommand{\kk}{\ensuremath{\vb{k}}}
\newcommand{\kxy}{\ensuremath{\vb{k}}}
\newcommand{\q}{\ensuremath{\vb{q}}}
\newcommand{\rr}{\ensuremath{\vb{r}}}
\newcommand{\RR}{\ensuremath{\vb{R}}}
\newcommand{\GG}{\ensuremath{\vb{G}}}
\newcommand{\g}{\ensuremath{\vb{g}}}
\newcommand{\xx}{\vb{x}}
\newcommand{\yy}{\vb{y}}
\newcommand{\me}{{\rm me}}
\newcommand{\pav}[1]{\textcolor{red}{#1}}
\newcommand{\header}[1]{{\emph{#1}---}}

\preprint{APS/123-QED}


\title{Dimensional localization and topology in 3D supertwisted spirals}

\author{Jeane Siriviboon}
\affiliation{%
 Department of Physics, MIT
}%
\author{Pavel A. Volkov}
\affiliation{%
Department of Physics, University of Connecticut
}%

\date{\today}

\begin{abstract}
We study the electronic structure of multilayer ``spirals" of two-dimensional materials with continuously increasing twist angle. Close to the $\Gamma$ point, the system maps on an in-plane momentum ${\bf k}$-dependent Aubry-Andr\'e  model, allowing to establish its universal electronic properties. The electronic states undergo a 3D-to-2D localization transition on increasing $k$: the state with $k< k_c$ are extended along the spiral axis while the states with $k>k_c$ are localized along the spiral axis due to mismatch between electronic dispersions of the twisted layers. Twist also induces nontrivial topological properties: Fermi arc surface states and bulk Berry phase winding, not requiring spin-orbit coupling or special features of single-layer dispersion. We discuss experimental platforms and the manifestations of the predicted phenomena. The momentum-selective localization leads to a universal suppression of $z$-axis conductance with doping, while the bulk Berry phase induces a topological magnetoconductance correction via mechanism similar to Thouless charge pumping. Our results do not require special band structure features or spin-orbit coupling, establishing twist as a versatile tool to control quantum states of electrons in three-dimensional materials.
\end{abstract}

\maketitle

\header{Introduction} Recent advances in fabrication of van der Waals heterostructures
\cite{Geim2013,Novoselov2016} and two-dimensional moir\'e materials 
\cite{Andrei2021,balents2020review,mak2022semiconductor,pixley2025_rev} have brought about the realization of a number of electronic phases, including superconducting and topological ones \cite{balents2020review,mak2022semiconductor,pixley2025_rev, bernevig2025fractional}. Most results in this field have been obtained in systems with small number of stacked atomic monolayers, or at twisted interfaces between multilayer flakes \cite{waters2023mixed,mullan2023mixing,lucht2025,Zhu_2026}. An intriguing open question is whether twisting can bring about new electronic phases and properties in a three-dimensional setting.

Experimental progress has been achieved in the realization of multilayer twisted spirals of semiconductors via growth \cite{zhao2020supertwisted, wang2024conversion, zhang2023vapour} and advanced assembling techniques \cite{mannix2022robotic,li2024towards}. Moreover, bulk incommensurate superlattices have been recently realized in single crystals (albeit without twist) \cite{nuckolls2026higher}. Most experiments to date \cite{spiral_review} focus on optical properties of supertwisted spirals \cite{Fan2020,ci_2022,Tong2024,Qi_2024_exp,ji2024opto,Liu2025} and scanning tunneling microscopy \cite{sb_spiral,peng_2022}. However, the electronic band structure and transport properties of the bulk of such systems remain relatively unexplored from both experiment and theory. Theoretical work so far is limited to specific models \cite{wu2020three,mele2025,chen2025spinless}. Ref. \cite{mele2025} demonstrated the emergence of edge modes in free electron model with a saddle point potential network potential or networks of them, while \cite{wu2020three,chen2025spinless} has shown the emergence of bulk Weyl points for graphene supertwisted spirals. However, the question of whether 3D twistronics could generically lead to new electronic states in bulk materials remains open.

Here we develop a general theory for the electronic structure of three-dimensional supertwisted spirals (Fig. \ref{fig:Schematics}) around the $\Gamma$ momentum point. The presence of nonzero twist leads to universal emergent wavefunction properties. First, the electronic wavefunctions experience a momentum-selective 3D-to-2D localization along the spiral ($z$) axis (Fig. \ref{fig:Schematics} (b)) : for in-plane momentum less than a threshold, $k<k_c$, the wavefunction is extended along the $z$ axis, while it is localized for $k>k_c$. Furthermore, twist endows the system with nontrivial topology: the surfaces host chiral Fermi arc states (Fig. \ref{fig:Schematics} (b), bottom) and the bulk states have nontrivial Berry phases. The latter leads to delocalization of wavefunction in the presence of magnetic field (Fig. \ref{fig:Schematics} (c)). Our conclusions are enabled by mapping the electronic structure of the twisted spiral for a given in-plane momentum onto the Aubry-Andr\'e (AA) model \cite{aubry1980analyticity,hofstadter,kraus2012topological}, where unavoidable anisotropy of the dispersion in the individual layer plays the role of the (quasi-) periodic potential modulation (Fig. \ref{fig:Schematics} (a)). We develop predictions for transport and spectroscopic experiments allowing to probe the 3D-to-2D localization and topology of supertwisted spirals and discuss the potential candidate platforms and applications beyond electronic properties.

The rest of the paper is organized as follows. In Sec. \ref{sec:AA} we discuss the general physics of electrons in supertwisted spirals near the $\Gamma$ point and mapping to the Aubry-Andr\'e model with momentum-dependent parameters. In Sec. \ref{sec:loc}, we demonstrate the momentum-selective localization of wavefunctions and evaluate its signatures in doping-dependent transport. In Sec. \ref{sec:top} we study the topological properties of the spectrum and wavefunctions: surface Fermi arcs ( Sec. \ref{sec:arcs}) and bulk Berry phase (Sec. \ref{sec:antiloc}) and evaluate their experimental signatures. Finally, in Sec. \ref{sec:disc} we discuss the effects of disorder and moir\'e modulation, as well as candidate materials and platforms.

\begin{figure}
    \centering   \includegraphics[width=1.0\linewidth]{fig/01Schematicsv12.pdf}
    \caption{{\emph{3D supertwisted spirals: localization and topology}}. (a) The schematics of the layer stacking of a crystal with $\theta$ twist between layers. The ellipses depict the electronic Fermi surfaces for each layer. For each in-plane momentum $\kk = (k_x, k_y)$, the system can be modeled as a tight-binding model with varying on-site potential due to Fermi surface mismatch, mapping to the Aubry-Andr\'e model (Sec. \ref{sec:AA}). 
    (b) Wavefunctions of supertwisted spirals: states with  in-plane wavevectors smaller than $k_c$ are extended along $z$ axis, while those with $k>k_c$ are localized (Sec. \ref{sec:loc}). The bottom layer shows a schematic of spectral function, with signatures of disconnected Fermi arcs (Sec. \ref{sec:arcs}).
    (c) In the presence of magnetic field along the spiral axis, wavefunctions delocalize by rotating in momentum space (Sec. \ref{sec:antiloc}), circumventing the mismatch between Fermi surfaces in (a). 
    }
    \label{fig:Schematics}
\end{figure}

\section{From twisted multilayers to the Aubry-Andr\'e model} 

\label{sec:AA}

We begin by constructing a general Hamiltonian to describe electrons in a spiral structure with twist angle $\theta$ between consecutive layers, shown in Fig. \ref{fig:Schematics} (a). We will consider the electronic states with in-plane momenta close to the $\Gamma$ point in each layer and focus on the cases where spin-orbit coupling can be neglected, relegating the discussion of the latter to the Supplementary Materials \cite{supp}. 
The electronic states of individual layers are coupled by tunneling; we assume that only the nearest-neighbor layers are coupled. In the vicinity of the $\Gamma$ point the dominant tunneling process is the one conserving the in-plane crystal momentum \cite{bistritzer2011moire,angeli2021gamma,zhang_fu_2021,hpan_2023}. To highlight that, we split the total Hamiltonian $H$ of the system into two parts: 

\begin{equation}
\begin{gathered}
\label{eq:TB-model}
 \allowdisplaybreaks
    H = H_0  + H_M
    ,
    \\
    H_0=\sum_{\kxy, l} h(\kxy_{l\theta}) c^\dagger_{\kxy l}  c_{\kxy_l} 
    + \sum_{\kxy, l} (t_{\kxy} c^\dagger_{\kxy l+1}c_{\kxy l} + h.c. ),
   \\
   H_{M}= \sum_{\kxy, \g , l}( t_{\g+\kxy} c^\dagger_{\kxy + \g_{l\theta} l+1}c_{\kxy l} +  V_{M, \g+\kxy} c^\dagger_{\kxy + \g_{l\theta} l}c_{\kxy l} + h.c.), 
\end{gathered}
\end{equation}
 where $l$ is the layer index, $h(\kxy)$ is the band dispersion of a single layer, $t$ is the interlayer hopping, $V_{M, \g}$ and $t_{\g}$ are the Fourier components of the moir\'e potential and tunneling corresponding to the reciprocal moir\'e lattice vector $\g$. For the layer $l$, $\kxy_{l\theta} = (k_x \cos l \theta + k_y \sin l \theta, -k_x \sin l\theta + k_y \cos l \theta)$ is the wavevector under $l\theta$ twist.

It is evident that $H_0$ conserves the in-plane momentum $\kxy$, while $H_M$ does not, reflecting the moir\'e superlattice effect. However, at $k\ll g$, the effects of $H_M$ can be treated perturbatively. Since $H_M$ couples the low-energy state $\kxy$ to the state $\kxy+\g$, the perturbative correction to the spectrum will be of the order of ${|V_{M,\g+\kxy}|^2}/{h(\g)}$ and ${|t_{\g+\kxy}|^2}/{h(\g)}$, which is small with respect to $h(\kxy)$ and $t_{\kxy}$ in two important limits: weak interlayer coupling and large twist. First, for interlayer distance $a$ much larger than the in-plane lattice constant $a_\parallel$, $|V_{M,\g+\kxy}|/t_{\kxy},|t_{\g+\kxy}|
/t_{\kxy} \sim e^{-\mathcal{O}(a/a_\parallel)} \ll 1$ \cite{bistritzer2011moire,angeli2021gamma,zhang_fu_2021,hpan_2023}. Moreover, $h(\g)$ grows quickly with twist angle. For a moderate twist angle $\theta \sim 1$, the moir\'e wavevector $\g$ is comparable to the Brillouin zone size, such that $h(\g) \sim W$,  $W$ being the single-layer bandwidth, which is expected to be much larger than $V_g,t_g$. We will thus initially neglect the moir\'e effects and discuss their influence in Sec. \ref{subsec:moire}. There (Sec. \ref{subsec:dis}) we will also address the effects beyond Eq. \eqref{eq:TB-model}, such as disorder.

Expanding the electronic dispersion within individual layers, $h(\kk_{l\theta})$, around the $\Gamma$ point, we can represent the leading terms as:
\begin{align}
    h(\kk_{l\theta}) 
    &= E_{||}(|\kxy|) + \Delta (|\kxy|) \cos(2\pi \beta l + \varphi(\kxy)),
    \label{eq:disp_decomp}
\end{align}
where $\kk = |\kxy| (\cos \phi, \sin \phi)$, $\phi$ being the 2D polar angle in $\kxy$ space; $E_{||}(|\kxy|)$ and $\Delta(\kxy)$ are the isotropic and leading (in $|\kxy|$) anisotropic part of the dispersion, respectively. The form $E_{||}(|\kxy|)$ and $\Delta(\kxy)$ are determined by the point-group symmetry of the layers.
For example, the Fermi surface with $D_{2nh}$ symmetry correspond to $E_{||}(\kxy) \sim |\kxy|^2$, $\Delta \sim |\kxy|^{2n}$, $\beta = n\theta/\pi$,  and $\varphi = 2n\phi$ (for $D_{nh}$ with odd $n$ effects of spin-orbit coupling need to be taken into account \cite{supp}). 
We note that Eq. \eqref{eq:disp_decomp} can also describe the case of band extrema at a finite $k$ away from $\Gamma$ (when $\Delta(|\kxy|)$ is sufficiently large), relevant for T' transition metal dichalcogenides \cite{Varsano2020}, such as WTe$_2$ \cite{Fei2017,Jia2021,pesin2022}.






The eigenstates of $H_0$, Eq. \eqref{eq:TB-model}, can be written as $\sum \psi_{\kxy,l}c^\dagger_{\kxy,l}|0\rangle$. As $\kxy$ is conserved by $H_0$, the problems for different $\kxy$ can be treated independently
, such that $\psi_{\kxy,l}$ satisfies: 
\begin{align}
    \label{eqn:single-channel-TB}
E_z (\kxy) \psi_{\kxy l} &= \epsilon_l \psi_{\kxy l} + t_{\kxy} \psi_{\kxy l+1} + t_{\kxy }^* \psi_{\kxy l-1}
    ,\\
    \epsilon_l &= \Delta(|\kxy|) \cos (2 \pi \beta l + 2\phi)
    ,\\
    E_z(\kxy) &= E - E_{||}(\kxy)
     , 
\end{align}
which coincides with the Aubry-Andr\'e (AA) model \cite{aubry1980analyticity}. Fig. \ref{fig:Schematics} (a) illustrates the mapping of the model onto the AA model: twist and band anisotropy result in a layer-dependent potential with quasi-period $\beta^{-1}$ which competes with the kinetic energy from the hopping term.

The mapping of a generic model of twisted multilayers to the AA model allows to establish a direct connection to two types of emergent electronic physics: localization \cite{aubry1980analyticity,dominguez2019aubry} and topology \cite{hofstadter,kraus2012topological}. In the following sections we elaborate on the consequences of these connections for the physics of supertwisted spirals.

In what follows we will take $t_{\kxy} =t_{\kxy}^*=  t$ for concreteness; generalizing our results for arbitrary $t_{\kxy}$ can be achieved via $t\to t_{\kxy}$ for the respective $\kxy$. The detailed numerical calculations will be presented for two-fold rotational symmetry $D_{2h}$ 
such that
\begin{equation}
    E_{||}(|\kk|) \approx \frac{|\kxy|^2}{2m}, \;
    \Delta(|\kk|) \approx \frac{|\kxy|^2}{2m} \varepsilon, \;
    \beta = \frac{\theta}{\pi}, \;\varphi = 2\phi,
    \label{eq:d2h}
\end{equation}
where $m$ describes the isotropic mass, and $\varepsilon$ defines the ellipticity of the band. The generalization to other symmetry groups is straightforward, as it only changes the definitions \eqref{eq:d2h}, keeping the general mapping to AA model \eqref{eqn:single-channel-TB} intact.

\section{Dimensional localization in supertwisted spirals}
\label{sec:loc}

AA model is well-known for exhibiting localization transition \cite{dominguez2019aubry,aubry1980analyticity} for sufficiently strong quasiperiodic potential. The model \eqref{eqn:single-channel-TB} inherits this physics, up to parameter mapping, when $\theta/\pi$ is irrational. In the AA model, the localization transition occurs at $2t = \Delta$ \cite{aubry1980analyticity,modugno2010anderson}. As $\Delta$ in model \eqref{eqn:single-channel-TB} is a function of in-plane momenta $\kk$, the critical momentum, $|\kxy|=k_c$ is determined by $2t = \Delta(|{\bf k}|)$ , and for the $D_{2h}$ system (Eq. \eqref{eq:d2h}) is explicitly given by:  
\begin{equation}
    k_c = 2 \sqrt{m t/\varepsilon}.
    \label{eq:kc}
\end{equation}
This result is valid as long as $k_c$ is much smaller than the Brillouin zone size, such that expansion in Eq. \eqref{eq:d2h} holds. For $|\kxy| < k_c$, the hopping $t$ dominates over anisotropy, resulting in an extended state along the $z$ axis. On the other hand, for $|\kxy| > k_c$, the single-layer dispersion anisotropy acts as a quasi-periodic on-site potential that localizes the state along the $z$ axis (Fig. \ref{fig:Schematics} (b)). Remarkably, $k_c$ does not depend on the twist angle $\theta$, reflecting the independence of the AA localization transition on the incommensurate modulation period \cite{aubry1980analyticity}.

\begin{figure}\centering\includegraphics[width=0.9\linewidth]{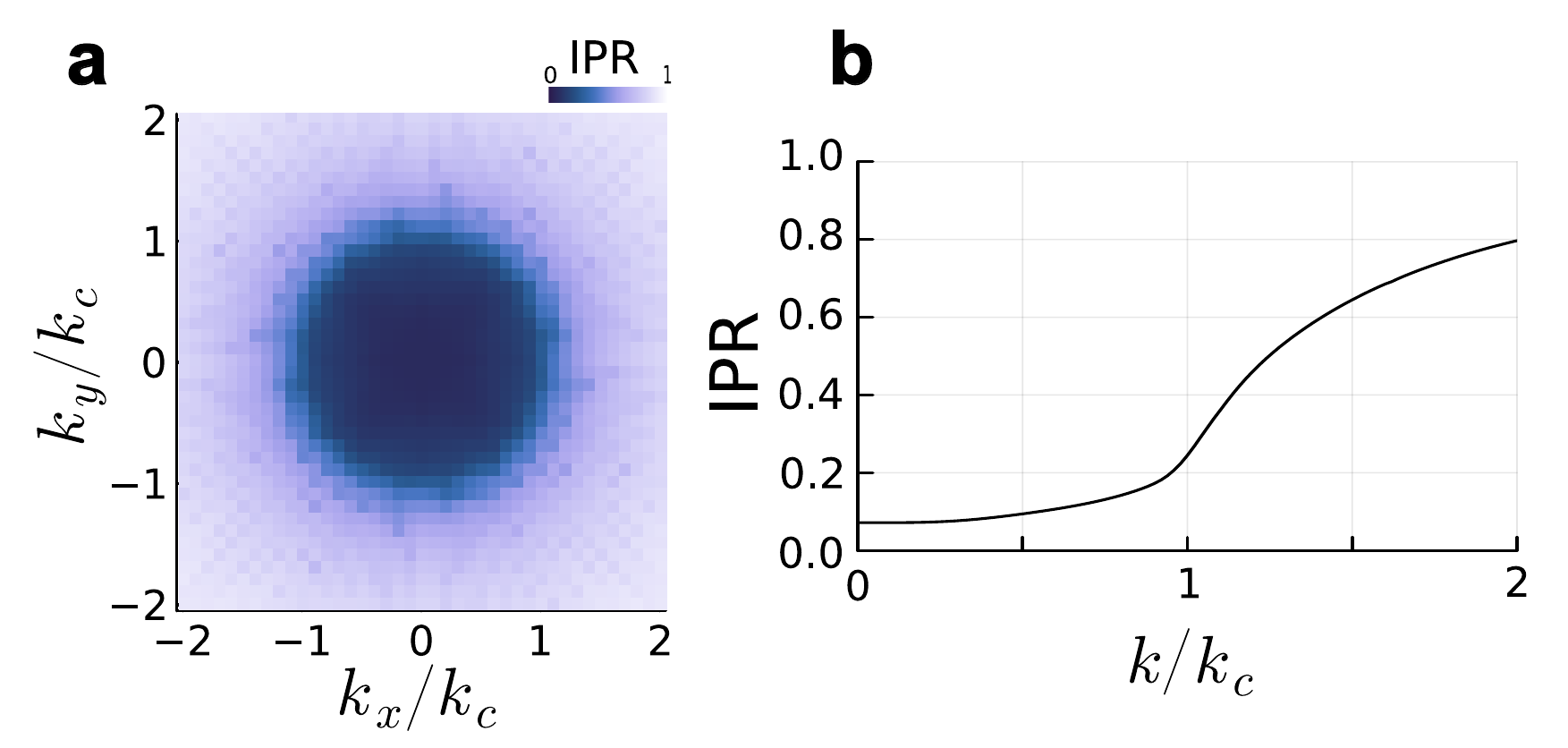}
    \caption{{\emph Localization Transition.} (a) Inverse Participation Ration (IPR) of wavefunctions along the $z$ direction as a function of $k_x$ and $k_y$. (b) average IPR as a function of $k$. (a, b) are calculated at $L = 20$, $\theta = 0.21\pi \approx 39^\circ$, and $\varepsilon = 0.1$, Eq. \eqref{eq:d2h}.}
    \label{fig:IPR}
\end{figure}

We illustrate the above general result by numerically studying the wavefunctions of the model (\ref{eqn:single-channel-TB},\ref{eq:d2h}) with finite layer number $L$. Specifically, we computed the inverse participation ratio (IPR), defined as 
$
    {\rm IPR} = \frac{\sum_{l = 1}^L |\psi_{\kk l}|^4}{\left (\sum_{l = 1}^L |\psi_{\kk l}|^2 \right )^2},
$ as a function of in-plane momentum.
In Fig. \ref{fig:IPR}  we present the IPR averaged over all eigenstates with same $\kxy$ for an $L=20$ system. The averaged IPR does exhibit a notable upturn precisely around $k_c$, while below $k_c$ its value is close to $1/L$, expected for electrons delocalized over all the layers.

The above discussion has been for the aperiodic case, where $\beta$ is irrational, where a localization transition can be rigorously established. For a rational $\beta = p/\lambda$, where $p$ and $\lambda$ are co-prime integers, $\epsilon_l$ acts as a periodic potential with a period of $\lambda$. At finite $\beta$, the anisotropy $\Delta$ opens gaps in the spectrum. Increasing $\Delta$ will suppress the bandwidth , eventually leading to the appearance of almost localized states \cite{zhang2015almost,wilkinson1984,malla2018spinful}. They have large effective mass along the $z$ axis and would be prone to localization due to nonzero long-range hopping, higher-harmonic anisotropy, and disorder \cite{vu2023generic}. For $k>k_c$ all states will become almost localized in this case. 


\begin{figure}
    \centering    \includegraphics[width=0.85\linewidth]{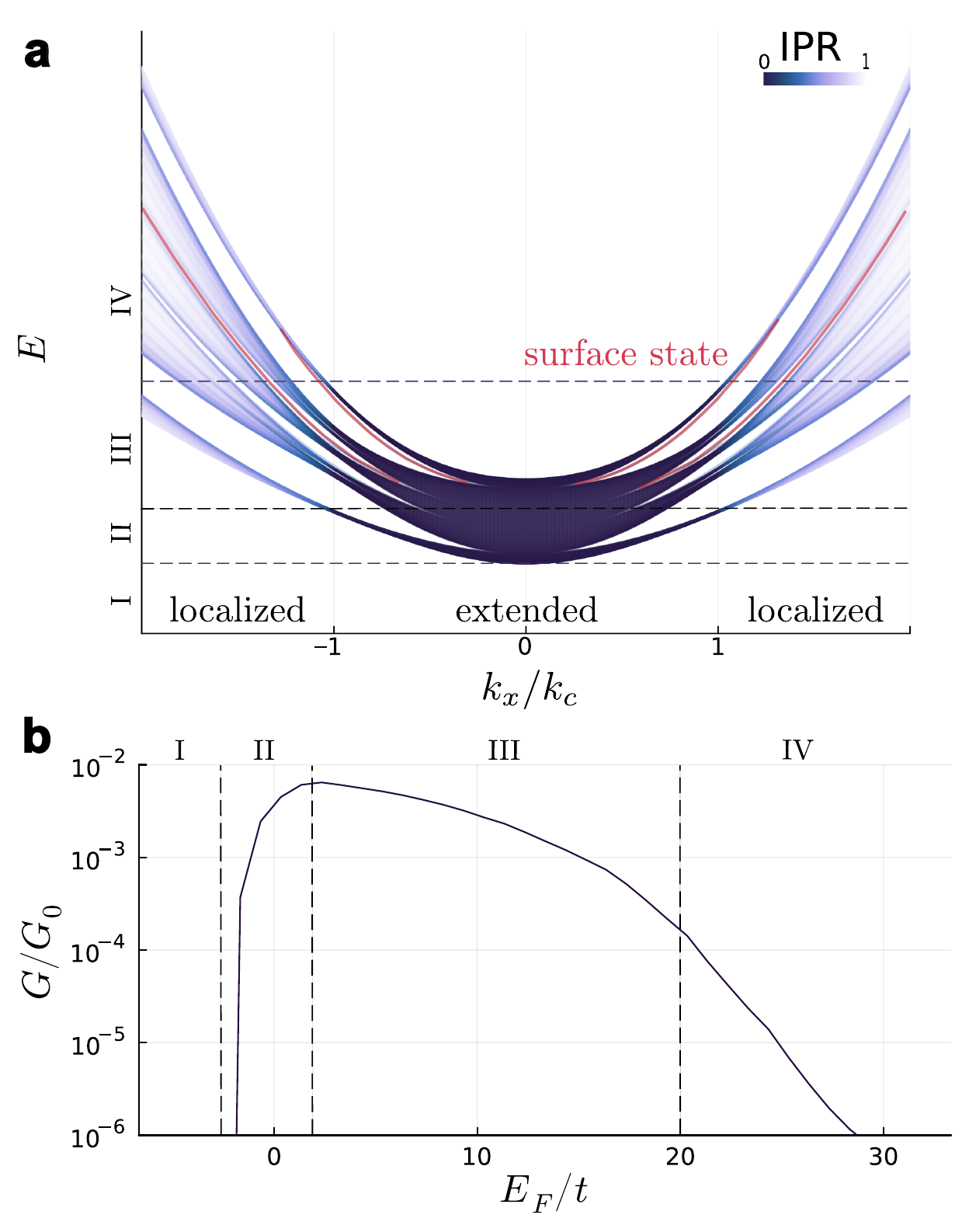}
    \caption{{\emph{Spectra and conductance.}} (a) The energy spectrum of the system at $\theta = (1 + \sqrt{5})\pi/15 \approx 0.21 \pi \approx 39^\circ$. The color indicates IPR of the state. The red line marks the surface states of the system (see Sec. \ref{sec:arcs}). (b) The normalized conductance $G/G_0$ as a function of the Fermi energy $E_F$ at $t =0.03$, $L = 20$, and $\varepsilon = 0.1$. $G_0$ is the ideal ballistic conductance $G_0 = (e^2/h) N_e$, where $N_e = k_{F,lead}^2/(4\pi)$ is the number of states supported by the lead. Roman numerals $I-I_V$ refer to different energy regimes: I: subband energies, II: initial doping regime, where electronic states are extended, III: mixed regime, where localized and extended states coexist and IV: universal decay regime, where all states are localized. }
    \label{fig:spectra}
\end{figure}

\begin{figure}
    \centering
    \includegraphics[width=0.95\linewidth]{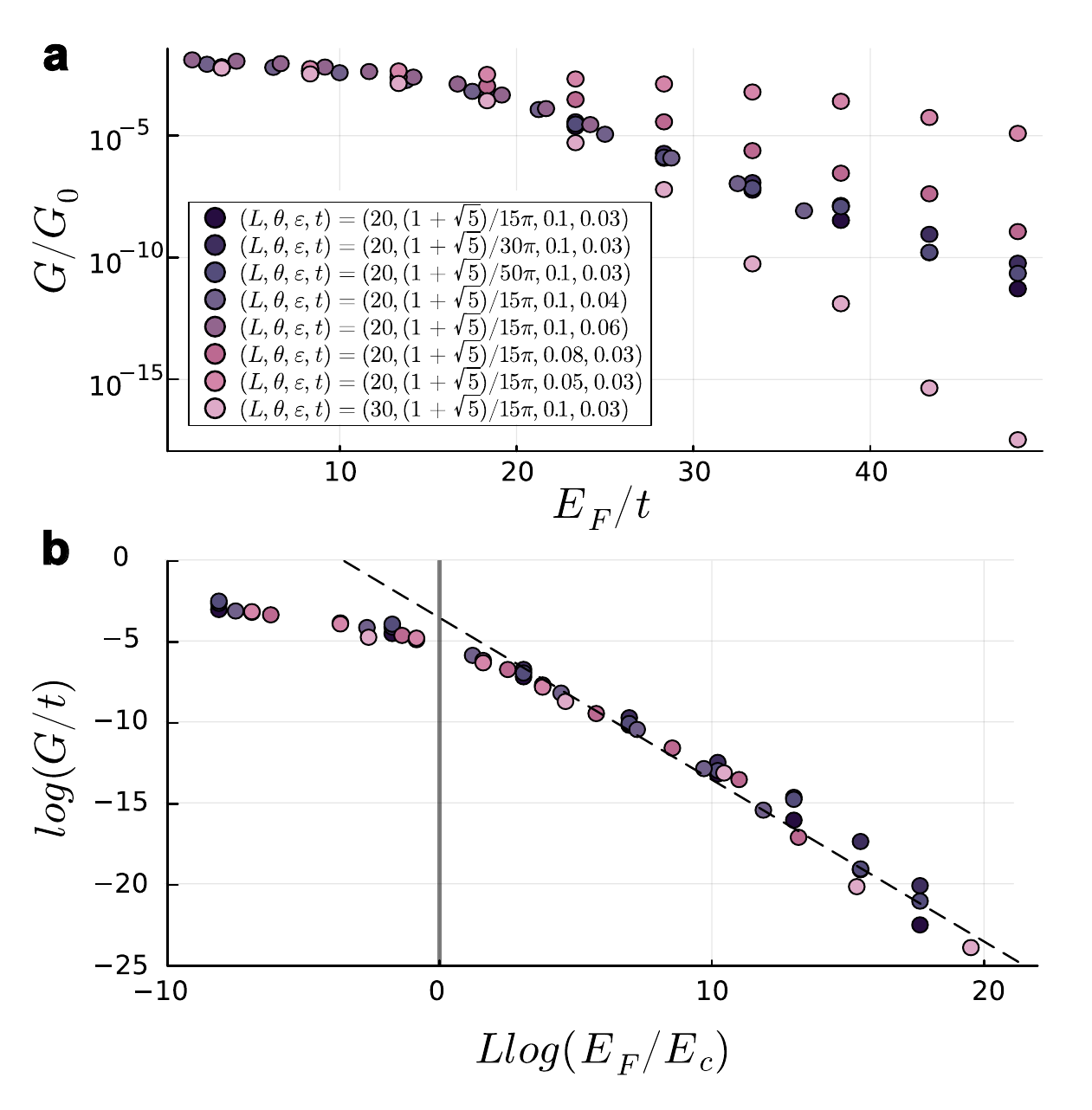}
    \caption{{\emph{Universal conductance suppression}.} (a) Normalized conductance $G/G_0$ as a function of $E_F$ at different $(L, \theta, \varepsilon, t)$ for aperiodic supertwisted spirals with irrational $\theta/\pi$. $G_0$ is the ballistic conductance without twist. (b) logarithm of rescaled conductance $\log G/t$ as a function of $L \log (E_F/E_c)$ at different $(L, \theta, \varepsilon, t)$ (same legend with (a)). The dashed line is the predicted empirical relation $G = 0.8 (e^2/h) \mathcal{N}t (E_c/E_F)^L$, \eqref{eq:univ_g}.} 
    \label{fig:conductance}
\end{figure}

\subsection{Transport signatures of dimensional localization}
We now demonstrate that the 3D-to-2D localization produces distinct signatures in the doping-dependent $z$-axis transport. Specifically, we expect that on increasing the Fermi momentum with doping past $k_c$, the $z$-axis conductance will be suppressed due to the localization effects. We will show below that this mechanism leads to a universal scaling form of conductance.

We consider the spiral multilayer connected to metallic leads at layers $l = 1$ and $l = L$, i.e. $H_{\rm tot} \psi_{\kxy l} = H_l \psi_{\kxy l}$ where $H_l = H_{\rm lead}$ where $l < 1$ and $l > L$ while $H_l = H$ where $1 \leq l \leq  L$. We assume for concreteness the contacts to be made of isotropic metals with a large Fermi surface
$H_{\rm lead}=\frac{|\kxy|^2}{2M} + \frac{k_z^2}{2M} - \frac{k_{F,lead}^2}{2M}$ \cite{supp}. Finally, we assume that $E_F$ within the spiral can be tuned independently from that of the leads by, e.g., doping.

Leveraging the emergent in-plane momentum conservation of $H_0$, Eq. \eqref{eq:TB-model}, we can treat the problems at different $\kxy$ as separate 1D conductance channels along the spiral axis. In this case, the conductance can be calculated using Landauer approach \cite{blanter2000shot}\begin{align}\label{eqn:landauer}
    G(E_F) = \frac{e^2}{h} \sum_{|\kxy|<k_{F,lead}} T(\kxy ,E_F), 
\end{align}
where $T(\kxy, E_F)$ is the transmission eigenvalue of the state $\kxy$ at energy $E_F$ within the spiral. $T(\kxy, E_F)$ is defined by the matrix element $T = |S_{10}(\kxy)|^2$ of the scattering matrix $S(\kxy)$ defined as, 
\begin{align}\label{eqn:transfer}
    \begin{pmatrix}
        \psi_{{\kxy},0-} \\
        \psi_{{\kxy},L+}
    \end{pmatrix}
     = 
    \begin{pmatrix}
        S_{00}(\kxy) & S_{01}(\kxy) \\
        S_{10}(\kxy) & S_{11}(\kxy)
    \end{pmatrix}
      \begin{pmatrix}
        \psi_{{\kxy},0+} \\
        \psi_{{\kxy},L-}
    \end{pmatrix},
\end{align}
where $\psi_{l+}$($\psi_{l-}$) is the right-moving (left-moving) wavefunction at layer $l$. Assuming $k_{F,lead}$ to be much larger than the one in spiral, we take the boundary conditions for $\psi_{l \geq L \& l\leq 0} \propto  e^{i k_F l}$ for all $\kxy$. From Eqs. \eqref{eqn:landauer} - \eqref{eqn:transfer}, we calculate the conductance \cite{supp,blanter2000shot,luo2021transfer} as a function of Fermi level $E_F$ within the spiral structure and system size $L$. For numerical calculations we take $k_F = 5$ in units of $a^{-1}$, $a$ being the spacing between layers of the spiral.  


Fig. \ref{fig:spectra} (b) shows the conductance as a function of the spiral Fermi energy $E_F$, set by the doping. The conductance features can be understood by comparing $E_F$ to the electronic spectrum of the twisted spiral shown in Fig. \ref{fig:spectra} (a). In region I, $E_F$ is below the conduction band minimum $E < -2t$ and the tunneling is classically forbidden, leading to a strongly suppressed $G$. In region II, the conductance increases with doping due to increase in carrier concentration that opens conduction channels at $E_F$. In region III, despite the increase in charge carrier concentration, the conductance decreases. In this regime, some states at the Fermi energy become localized in the $z$ direction (highlighted by the increasing IPR in Fig. \ref{fig:spectra} (a)). Moreover, the conductance of formally extended states also decreases: the wavefunctions with energies $|E|>2|t|-|\Delta(|\kxy|)|$ have been shown to be ``almost localized" in the $\beta \ll 1$ limit \cite{zhang2015almost,wilkinson1984,malla2018spinful} with strongly enhanced effective mass. Finally, in region IV, where $E_F \gtrsim E_c =  k_c^2/(2m
) = 2t/\varepsilon$, all the states at $E_F$ are localized and the non-zero conductance only occurs due to the finite size of system - it is expected to vanish in the bulk $L\to \infty$ limit. 

In Fig. \ref{fig:conductance} (a), we present the details of the dependence of the conductance in region IV on $E_F$ for varying parameters, such as ellipticity $\varepsilon$, (irrational) twist angle $\theta$, tunneling $t$ and system size $L$; additional results are reported in Supplementary Material \cite{supp}.
Remarkably, we find that the numerical results can be well-captured by the following empirical formula:
\begin{align}
    G  \approx 0.8 \frac{e^2}{h} \mathcal{N} t \left( \frac{E_c}{E_F}\right)^{ L},
    \label{eq:univ_g}
\end{align}
where $\mathcal{N} =  A m/(2\pi\hbar^2)$ is the electronic density of states within the spiral, $A$ being the layer area. The prefactor $\mathcal{N} t$ reflects the dependence of conductance on interlayer tunneling, while $\left( {E_c}/{E_F}\right)^{L}$ captures the localization effect. Indeed, when $\beta$ is irrational, the conductance is expected to decay as $G \sim \exp (- L/\xi)$ in the localized phase with the localization length $\xi \sim \log (E_F/E_c)^{-1}$. In Fig. \ref{fig:conductance} (b) we demonstrate that, when the conductance is rescaled as a function of $L\log(E_F/E_c)$, it collapses on a single curve and follows Eq. \eqref{eq:univ_g} for $E_F>E_c$ (dashed line) and all parameter values.

\begin{figure}
    \centering
\includegraphics[width=0.85\linewidth]{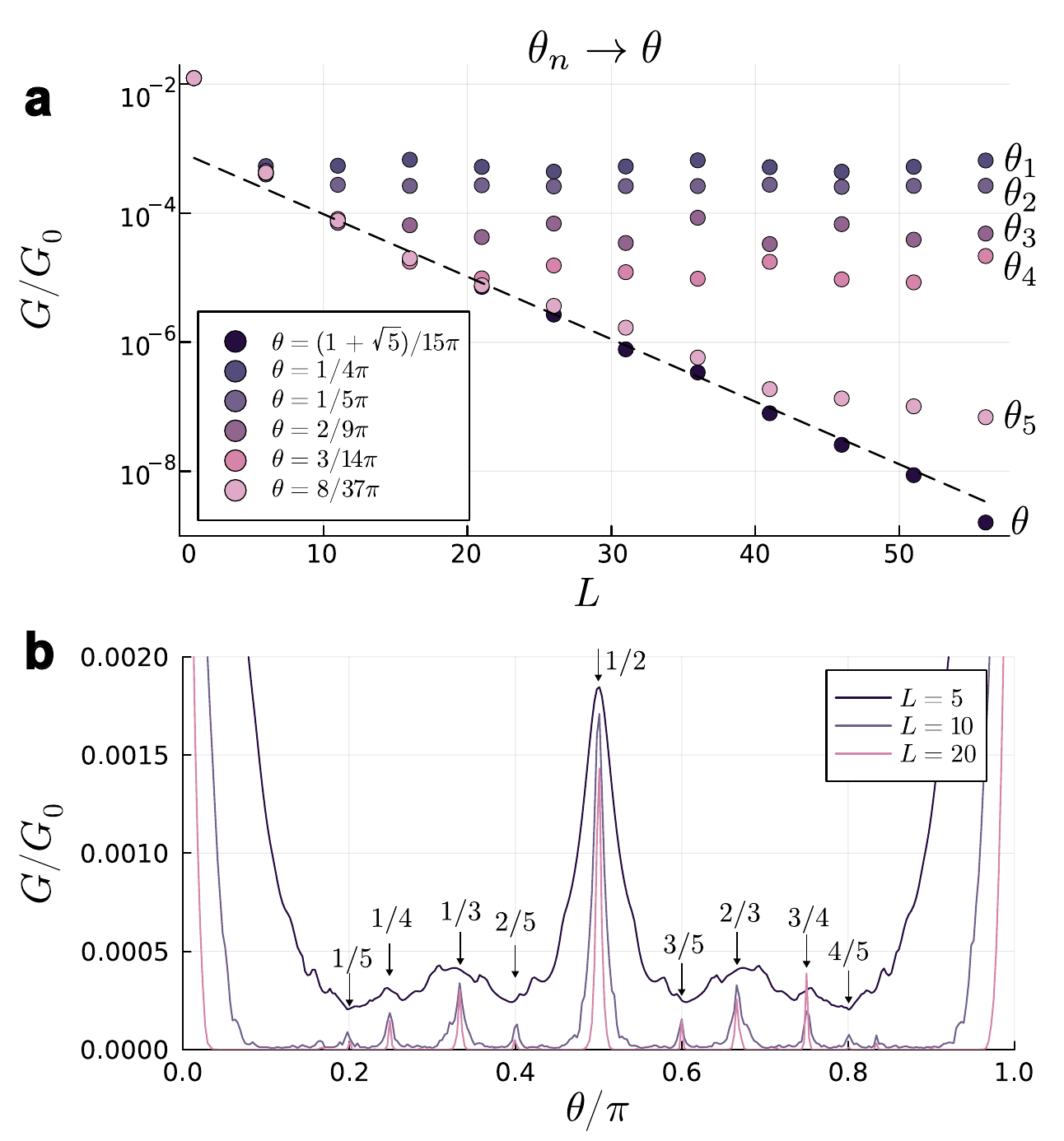}
    \caption{{\emph{Twist commensuration effects in conductance}.} (a) $G/G_0$ as a function of $L$ for a series of $\theta$ which are rational approximants of $\theta = (1+\sqrt{5})/15 \pi$. (b) Normalized conductance $G/G_0$ as a function of $\theta$ at different $L$. The calculation is performed at $t = 0.03$, $ \varepsilon = 0.1$, and $E_F = 1.0$. Notable peaks are labeled by $\beta$ as simple fractions.} 
    \label{fig:commensurate}
\end{figure}

In the above we have kept the angle value irrational, such that the spiral is aperiodic. For rational $\beta$, the spiral is periodic, and the electronic states form bands of formally extended states.
Thus for $L\gtrsim \lambda$ the conductance should remain nonzero as $L \rightarrow \infty$. Increasing $E_F$, however, should still suppress conductance due to decrease of the effective bandwidth. To understand the relationship between the rational and irrational $\beta$, we consider a series of rational approximants  $\beta_n = q_0, q_0 + \frac{1}{q_1}, q_0 + \frac{1}{q_1 + \frac{1}{q_2}}, ...$ where $q_n$ is an integer and $\beta_n \rightarrow \beta$ as $n \rightarrow \infty$. Fig. \ref{fig:commensurate} (a) shows the conductance $G/G_0$ as a function of $L$ at different approximant $\beta_n$. We observed that for each approximant $\beta_n = p_n/\lambda_n$, the conductance resemble the irrational $\beta$ up to $L \lesssim \lambda_n$. Hence, the comparison between Eq. \eqref{eq:univ_g} (dashed line) and numerical calculations in Fig. \ref{fig:commensurate} (a) reveals a good agreement up to $L \lesssim \lambda_n$. 

 
To study the effects of commensurability in more detail, we present the normalized conductance as a function of twist for several system sizes in  Fig. \ref{fig:commensurate} (b). At large $L$, the conductance peaks at the rational $\beta$ and decays rapidly away from the peaks as the system loses its periodicity (and where the scaling form Eq. \eqref{eq:univ_g} holds). At small $L$, the conductance peak broadens as the distinction between periodic and quasi-periodic potential becomes less well-defined \cite{supp}, resulting in recovery of universal behavior \eqref{eq:univ_g} away from the conductance peak.

Thus, the 3D-to-2D localization transition has a direct signature in the z-axis conductance: its scaling dependence on $L$ and $E_F$, defined by Eq. \eqref{eq:univ_g} and plotted in Fig. \ref{fig:conductance} (b), as long as twist angles are not too close to the major commensurate twist peaks in  Fig. \ref{fig:commensurate} (b). The results on conductance near the commensurate peaks are provided in the supplementary material \cite{supp}.  

\section{Topology in supertwisted spirals}
\label{sec:top}

We now demonstrate that twist endows electronic states of supertwisted spirals with topological properties. The AA model, to which we mapped the electronic structure of supertwisted spirals in  Sec. \ref{sec:AA}, exhibits topological properties \cite{lang_2012,kraus2012topological,ganeshan_2013,gottlob2025quasiperiodicity}, rooted in its formal equivalence to the Harper-Hofstadter model for quantum Hall effect on a square lattice \cite{hofstadter,kraus2012topological}.

\subsection{Surface Fermi Arcs}
\label{sec:arcs}
\begin{figure}
\centering\includegraphics[width=1.0\linewidth]{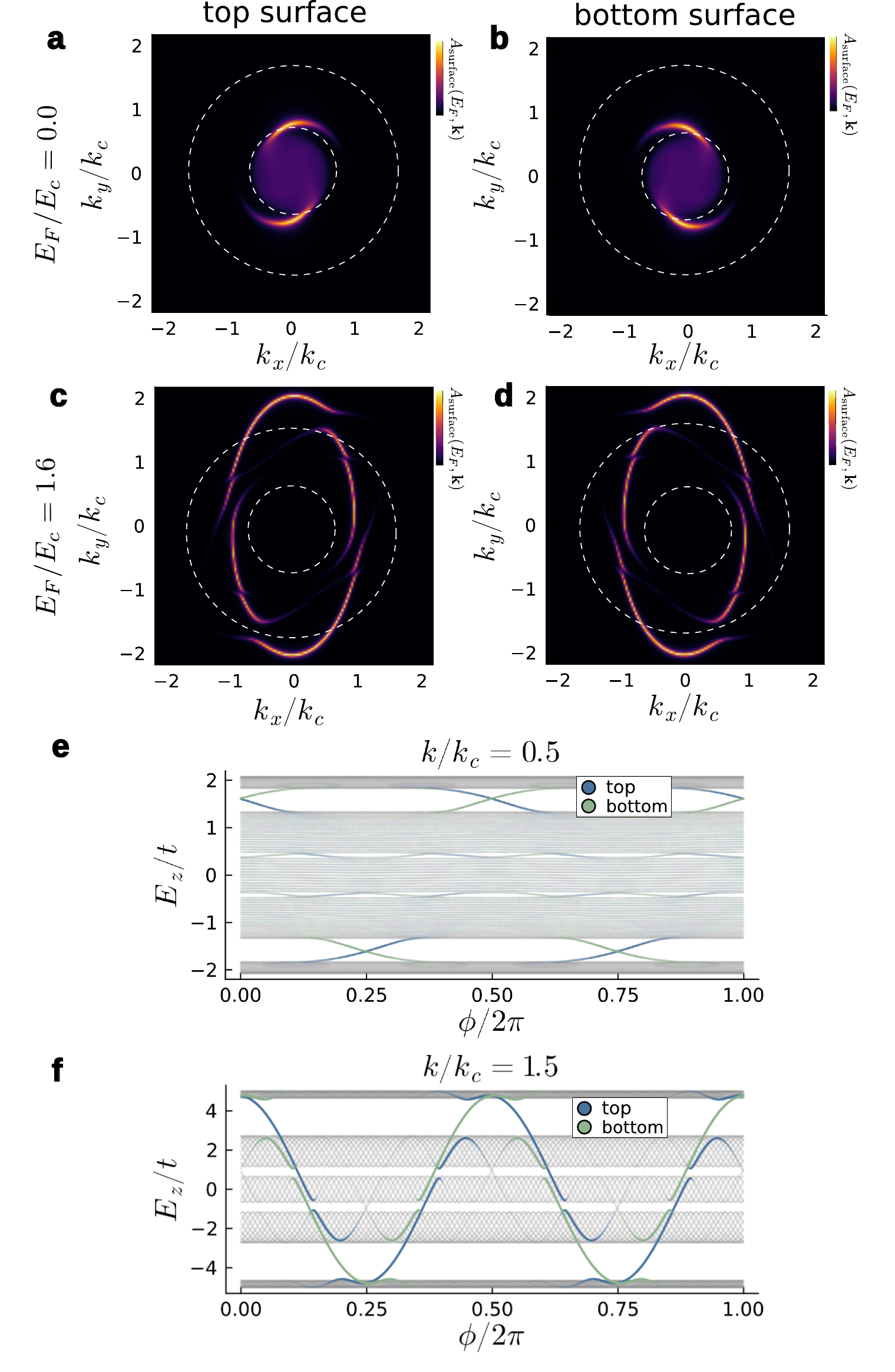}
    \caption{{\emph{Fermi arc surface states.}} (a-b) Surface spectral function $A(\kk, E_F)$ as a function of $\kk$ for (a, b) $E_F/E_c = 0.0$ and (c, d) $E_F/E_c = 1.6$. (a, c) shows the top and (b, d) --- the bottom layer. (e-f) Spectrum of the system on a contour (e) $k/k_c = 0.5$ (f) $k/k_c = 1.5$ (white dashed line) as a function of $\phi = \tan^{-1} (k_y/k_x)$. The bulk states are labeled by gray lines. Parameters used in (a-f) are: $\varepsilon = 0.6, \eta = 0.02, L = 103, \theta = (1+\sqrt{5})/15 \pi \approx 0.21 \pi \approx 39^\circ$.
    }
    \label{fig:surface-state}
\end{figure}

We first demonstrate the presence of the Fermi arc surface states in supertwisted spirals. In Fig. \ref{fig:surface-state} (a-d), we present the spectral function of the system on the top (a,c) and bottom (b,d) layers defined as 
\begin{align}
    A_{\rm surface} (\omega, \kk) &= - \frac{1}{\pi} \sum_j {\rm Im}  \frac{|\psi^j_{\kk l}|^2}{\omega - E_j(\kk) + i \eta} ,
\end{align}
where $l = 1$ ($l = L$) defines the top (bottom) surface,  $E_j(\kk)$ defines the j$^{\rm th}$ eigenvalue of the Hamiltonian at $\kk$, and $\eta = 0.02$ taken for the numerical calculation. The surface states are manifested as sharp disjointed features indicating the presence of surface Fermi arcs in supertwisted spirals. 

Qualitatively, their appearance is expected from the mapping between AA model, Eq. \eqref{eqn:single-channel-TB}, and quantum Hall effect on the square lattice \cite{hofstadter,kraus2012topological}. 
In the latter, the free momentum in the Landau gauge \cite{hofstadter} spanning $\left[-\frac{\pi}{a}, \frac{\pi}{a} \right]$ maps directly on the polar angle in momentum space $\varphi({\bf k}) \in [0,2\pi]$. The role of the spatial direction is played by the layer number $l$. As a result, for certain values $\varphi_{min}^i<\varphi({\bf k})<\varphi_{max}^i$ \cite{kraus2012topological}, surface states (localized on the top $l=1$ or bottom $l=L$ surface) are expected to emerge within the spectral gaps. 

Indeed, surface-localized states are observed in the numerical spectrum of Fig. \ref{fig:spectra} (red lines) within clear spectral gaps. One notes that these modes occur throughout the extended/localized boundary in $|k|$; their existence is independent of the  localization properties, discussed in Sec. \ref{sec:loc}. As the intervals of $\varphi({\bf k})$, where the surface modes exist, are finite, the edge modes would merge with the bulk states at the edges of this interval, explaining the formation of a disconnected Fermi arc at the surface. 

The Fermi arcs persist both for energies where bulk states are extended (Fig. \ref{fig:surface-state} (a,b)) and localized (Fig. \ref{fig:surface-state} (c,d)). In the extended phase, the bulk states also contribute to the surface spectral functions, seen as weaker broad features in the middle of Fig. \ref{fig:surface-state} (a,b). The Fermi arcs, however, are clearly separated in momentum space from the bulk states in this case. In the localized case, Fig. \ref{fig:surface-state} (c,d), one notes that the arcs taken together start to resemble an ellipse. Indeed, as the localization length further decreases with increasing $\Delta(|\kk|)$, the wavefunctions would eventually localize in individual layers, with spectral functions equivalent to that of decoupled layers. The gaps between individual Fermi arcs on approaching this limit can then be attributed to the hybridization between the surface layer and the next layer leading to avoided crossings at the points, where Fermi surfaces of the surface and next layers  intersect.

Each arc in Fig. \ref{fig:surface-state} also shows a preferred chirality: they tend to extend outward in $|{\bf k}|$ as a function of $\varphi({\bf k})$ along the arc, breaking all in-plane mirror symmetries. The chiralities of the Fermi arcs are opposite between top (Fig. \ref{fig:surface-state} (a,c)) and bottom (Fig. \ref{fig:surface-state} (b,d))  surfaces. This observation can be understood from the spectral properties of the AA model. Consider a fixed $|{\bf k}|$ contour (white dashed lines in Fig.  \ref{fig:surface-state} (a-d)). In Fig.  \ref{fig:surface-state} (e-f), we show the spectrum of the system along this contour (parametrized by the polar angle $\varphi({\bf k})$), with surface states highlighted in color. As energy always increases with $|\kk|$, owing to $\varepsilon<1$, Eq. \eqref{eq:d2h}, the edge mode dispersion within the band gaps (Fig. \ref{fig:surface-state} (e-f) ) directly translates into outward or inward extension of the constant-energy contours with $\varphi$. In the localized regime of the AA model (Fig. \ref{fig:surface-state} (f)) we also note that the surface states remain localized beyond the gaps and may change the chirality there. As a result, in the localized state (Fig. \ref{fig:surface-state} (c,d)) the Fermi arcs bend over, rather than uniformly extending out in the extended regime (Fig. \ref{fig:surface-state} (a,b)).

The presence of the Fermi arc states and their chirality can be detected directly via angle-resolved photoemission  \cite{xu2015discovery} or quasiparticle interference \cite{Yuan2019} experiments. Their presence raises the questions of their magnetotransport properties, as the semiclassical trajectories appear to be opened like Fermi arcs in Weyl semimetals \cite{potter2014quantum,zhang2016quantum}. The signatures of the Fermi arcs in quantum oscillations within a semiclassical approach is discussed in  the supplementary material \cite{supp}.

\subsection{Bulk topology and field-induced delocalization}
\label{sec:antiloc}
\begin{figure}
\centering\includegraphics[width=\linewidth]{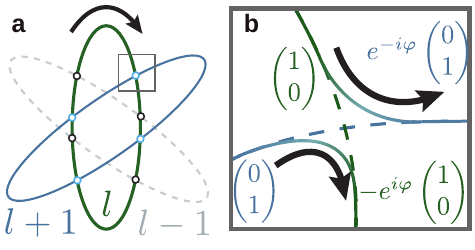}
    \caption{{\emph{Field-induced interlayer tunneling.}} (a) Illustration of the crossing between the Fermi surfaces of neighboring layers in the spiral (marked by gray and blue) in momentum space. Arrows show the direction of momentum evolution in the presence of magnetic field. (b) Zoom-in on the intersection in (a) showing a tunneling-induced avoided crossing along with the phase gain from the wavefunction evolution.}
    \label{fig:fscross}
\end{figure}

We now show that topological properties of the electronic states in supertwisted spirals lead to an emergent physical effect in the bulk: delocalization of electrons in the presence of magnetic field. This phenomenon can be understood as follows. In the absence of magnetic field, the energies of electrons in the adjacent layers do not match except for high symmetry lines, leading to suppressed, non-resonant tunneling. With magnetic field, however, a wavepacket centered at a particular in-plane momentum can evolve in momentum space along an equal-energy contour and approach the degeneracy point, where hybridization between layers will allow its transfer to the other layer (Fig. \ref{fig:fscross} (a)). As a result, tunneling between any pair of layers becomes allowed leading to a screw-like trajectory in the mixed ${\bf k}_\parallel$-$z$ space, illustrated in Fig. \ref{fig:Schematics} (c).

\begin{figure}
\centering\includegraphics[width=1.0\linewidth]{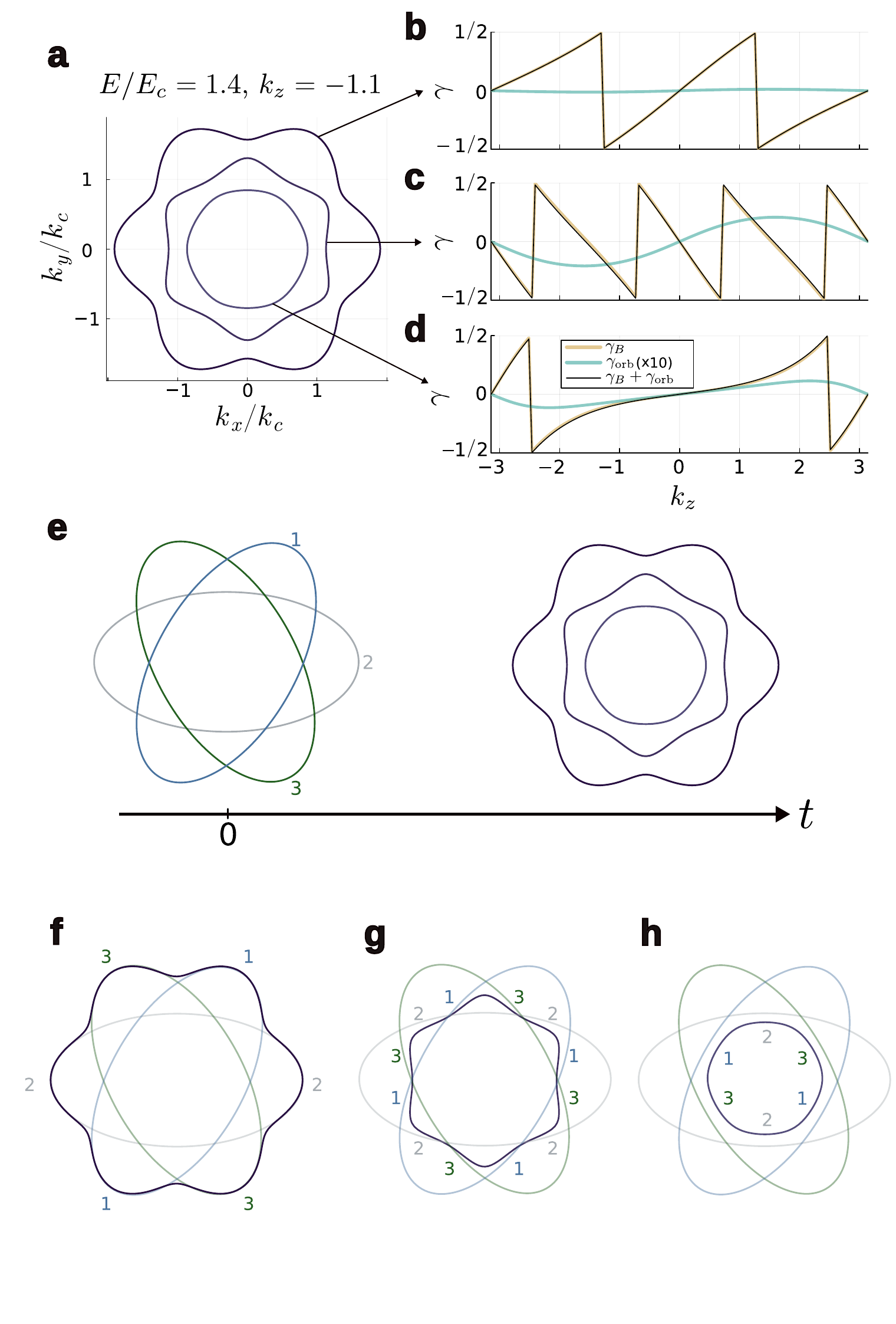}
    \caption{{\emph{Onsager phase offset.}} (a) The Fermi surfaces of supertwisted spiral at $\beta = 1/3$, $E/E_c = 1.4$, and $k_z = -1.1$. (b-d) phase offset $\gamma$ (black) 
    as a function of z-axis momenta $k_z$. The contribution could be decomposed into the Berry phase (orange) and the orbital magnetization (green). The orbital contribution is multiplied by a factor of 10 for clarity. (e) The schematics of Fermi surfaces as a function of $t$ where anti-crossings form at the intersections of layer's $t=0$ Fermi surfaces. The comparison of the (f) outer, (g) middle, and (h) inner Fermi surfaces to the Fermi surfaces of each layer. The number shows the layer within the spiral unit cell, that the wave function resides in at each point in the Fermi surface.} 
    \label{fig:phase}
\end{figure}

We will now  make this intuition precise and demonstrate its connection with the electronic topology of the bulk states. We first consider the commensurate case $\lambda \in \mathbb{N}$. Without magnetic field, $k_z$ is a good quantum number and $\lambda$ closed equal-energy contours form for each $k_z$ (Fig. \ref{fig:phase} (a) for $\lambda=3$). While localization does not occur in this case, one expects the bandwidths of the bands (and the velocity) to scale as $(t/\varepsilon E)^{\lambda-1}$ strongly suppressing the transport along the $z$ axis.

The field-induced delocalization can be understood by analyzing the change in the energy spectrum in presence of an external magnetic field. At low fields, the energies can be obtained using the semiclassical quantization condition \cite{lifshitz1956,shoenberg1984magnetic,gao2017zero,alexandradinata2018revealing,glazman_rev}:
\begin{align}
\label{eqn:semiclassical-quantization}
    S(E_n, k_z) = \frac{2\pi e B}{\hbar c} \left (n +  \gamma(E_n, k_z) +\frac{1}{2} \right) ,
\end{align}
where $S(E_n, k_z)$ is the area defined by the semiclassical orbit in k-space, $n$ is an integer, and $1/2$ is the Maslov phase for contour that is deformable to a circle. $\gamma(E, k_z)$ is the phase offset that arises from wavefunction geometry and orbital magnetization:
 \begin{gather}
    \gamma(E, k_z) = \gamma_B (E, k_z) + \gamma_{\rm orb} (E, k_z)
    \\
    \gamma_B (E, k_z) = \frac{1}{2\pi} \int_{\mathcal{C}} d\kk \cdot \mathcal{A}(\kk, k_z),
    \\
    \gamma_{\rm orb} (E, k_z) =  - \phi_0 \int d^2 \kk \delta(E - E(\kk, k_z)) m(\kk, k_z),
\end{gather}
where $\mathcal{C}$ is the equal-energy contour, $\phi_0 = h/e$ is the magnetic flux quantum, $\mathcal{A}(\kk, k_z)$ is the Berry connection, $m(\kk, k_z)$ is the orbital magnetization operator \cite{Niu1996}. Eq. \eqref{eqn:semiclassical-quantization} results in the quantization of energies: $E\to E_n(k_z)$. The electron velocity along $z$ for levels crossing the Fermi energy $E_n\approx E_F$ is determined by:
\begin{equation}
    \begin{aligned}
     v_z = \frac{\lambda a}{\hbar} \frac{\partial E_{n}}{\partial k_z}  
     =
    \frac{\lambda a}{\hbar}
    \left ( 
    \left \langle 
    \frac{\partial E(\kk, k_z)}{\partial k_z}
    \right \rangle_{FS} + \left ( \frac{\partial S}{\partial E}\right)^{-1}
    \frac{eB}{\hbar c} \frac{\partial (2\pi \gamma)}{\partial k_z}
    \right ) 
\end{aligned}
\label{eq:vz}
\end{equation}
where $k_z\in [-\pi, \pi]$ in the unit of $1/(\lambda a)$ ($a$ being interlayer spacing), and $\langle{...}\rangle_{FS}$ defines an average over the Fermi surface at fixed $k_z$. The first term in the brackets arises from the zero-field $z$-axis dispersion (strongly suppressed by twist) and is independent of magnetic field. The second term, however, indicates that the dependence of $\gamma$ on $k_z$ may enhance $v_z$ at non-zero magnetic field.

As an example, we consider a spiral with $\beta=1/3$. Fig. \ref{fig:phase} (b-d) shows $\gamma(E, k_z) \mod 1$ as a function of $k_z$ for the three Fermi contours shown in Fig. \ref{fig:phase} (a) for $ k_z=-1.1$ in the unit of $(\lambda a)^{-1}$. The main contribution to $\gamma$ arises from the Berry phase, monotonically dependent on $k_z$. The dependence is close to $2\pi \gamma_B \approx 2k_z$ for (b) and $-4 k_z$ for (c), while the orbital contribution appears negligible in all cases. The resulting correction to $v_z$ from the Berry phase, Eq. \eqref{eq:vz} is of the order $\lambda a \frac{e B}{m c}$ (approximating equal-energy contours with a circle). Given that the $z$-axis dispersion is strongly suppressed by twist at zero field, this result directly relates the delocalization (appearance of nonzero $v_z$) of electrons in magnetic field to their bulk Berry phase.

The $k_z$-dependent Berry phase $2\pi \gamma_B \propto  k_z$ can be further demonstrated to have a topological origin. General arguments can be developed in two particular limits. First, consider the equal-energy contour $\mathcal{C}$ to be a $k_z$-independent circle. The problem, Eq. \eqref{eqn:single-channel-TB}, then maps exactly onto the Harper model \cite{hofstadter,kraus2012topological} with the momentum in Landau gauge varying from $0$ to $4 \pi$ (for $D_{2h}$ point group). The winding of the Berry phase  $2\pi (\gamma(2\pi^-) -\gamma(0^+))$ is then equal to the Chern number $C$ of the corresponding band. $C$ can be obtained from Diophantine equation \cite{tknn,kohmoto_1989}. E.g., for $\theta = \frac{\pi}{2r+1}$ we find $C= 1$ for all bands except the middle ($r$-th) one having $C=-2r$ \footnote{This approach leads to an ambiguity for even-denominator twist angles and two innermost trajectories; in\cite{supp} we demonstrate that $C=(1-r)$ for both orbits in that case.}. The specific form $\gamma \propto k_z$ also arises in this limit explicitly.
Introducing a hybrid Wannier representation along the $z$ axis, the Chern number is directly related to the shift of the wavefunction over a closed loop in parameter space
\cite{vanderbilt2018berry}. As a result, the evolution of a wavefunction along the closed circle in $\kxy$ corresponds to the shift $\Psi(l) \to \Psi(l + 2 C \lambda)$ (Fig. \ref{fig:Schematics} (c)). The magnetic field thus leads to the realization of Thouless charge pumping \cite{citro2023thouless} in the superspiral. As a result, the wavefunction acquires a phase factor $e^{i 2 C k_z}$ by Bloch's theorem. The factor of 2 originates from the $D_{2h}$ symmetry of the assumed Fermi surface and can be generalized by substituting two with an appropriate factor corresponding to the Fermi surface rotational symmetry. This argument implies the Berry phase is equal to $2C k_z$:
\begin{align}
\label{eq:ideal-winding}
   2\pi  \gamma_{B,circ} \approx 2 C k_z.
\end{align}
In general, Fermi contours are not circles (Fig. \ref{fig:phase}) and their size may depend on $k_z$, resulting in deviations from this equation. However, the overall winding $[\gamma(k_z+2\pi)-\gamma(k_z)]$ is an integer and thus remains robust as long as no gaps between bands close (equivalent to the equal-energy contours not touching for any $k_z$ value). Indeed, all orbits in Fig. \ref{fig:phase} are consistent with a winding proportional to $C=1,-2,1$. The notable deviations from Eq. \eqref{eq:ideal-winding} in the innermost contour (Fig. \ref{fig:phase} (d)) could be explained by the strong Fermi surface dependence on $k_z$ along with the higher concentration of Berry curvature near the $\Gamma$ point.

A different perspective can be obtained in the $t \ll \Delta({\kk})$ limit, which also provides a path for generalization to incommensurate (irrational $\beta$) systems. In this case, the effects of $t$ are negligible away from momenta where Fermi surfaces of different layers cross, Fig. \ref{fig:fscross} (b). Near the crossing points between layer $l$ and $l'$, the Hamiltonian projected to the respective states with fixed $k_z$ can be written as:
\begin{equation}
    H(l, l',k_z) \approx \begin{pmatrix}
        {\bf v}_{l} \cdot {\bf k} & \tilde{t}_{l,l'}(k_z)\\
          \tilde{t}^*_{l,l'}(k_z) &{\bf v}_{l'} \cdot {\bf k}
    \end{pmatrix},
    \label{eq:2layer}
\end{equation}
where ${\bf v}_{l}$ is the Fermi velocity of the $l$th layer near the crossing and $|\tilde{t}_{l,l'}|$ is the effective Hamiltonian matrix element, connecting the two states. Microscopically, the model, Eq. \eqref{eq:TB-model}, contains only nearest-layer tunneling, the non-nearest neighbor hopping is perturbative and decays as a function of hopping distance $|\tilde {t}_{l,l'}|\propto t^{|l-l'|}$. Furthermore, $\tilde t_{l,l'}$ is complex with the phase factor of $e^{-ik_z}$ ($e^{ik_z}$) when the closest layer is in the next (previous) unit cell of the spiral \footnote{The case when there are two closest layers is relevant for even $\lambda = 2m$ and is treated in \cite{supp}.}. Away from the crossing, the eigenstates of Eq. \eqref{eq:2layer} are the layer eigenstates which are chosen to be $(1,0), (0,1)$ on the left of the crossing (Fig. \ref{fig:fscross}) by gauge fixing. Adiabatically following the eigenstates along each contour, one obtains that after the transition they evolve into $(0,e^{-i\varphi})$ and $-(e^{i\varphi},0)$, where $\varphi=0,\pm k_z$. As a result, they acquire a Berry phase of $-\varphi/2\pi$ and $\varphi/2\pi$, respectively. In what follows we'll omit the $-$ sign in the second case; due to $C_2$ symmetry in our case these always occur an even number of times, canceling out. Moreover, in this case we can find $\gamma_{orb}$ vanishes exactly 
\cite{supp}.

Thus, computing $2\pi \gamma(k_z)$ reduces in this case to counting the number of interlayer crossings and adding $\pm k_z$ when the tunneling occurs between the unit cells. This is equivalent to the number of unit cells that the electron advances over in $z$ axis when going over the closed constant-energy contour, multiplied by $k_z$. The case $\beta=1/3$ is illustrated in Fig. \ref{fig:phase}. Panel (e) highlights the relevant momentum points, where tunneling hybridizes crossing Fermi surfaces of different layers. Panels (f-h) shows the contribution of each layer within unit cell to the resulting hybridized Fermi surface for a fixed $k_z$. To calculate the Berry phase, we need to fix the direction of motion along the equal-energy contours in Fig. \ref{fig:phase} (f-h) and count how many unit cells along $z$ does the state traverse when going through the full contour. For clockwise motion along the orbit in Fig. \ref{fig:phase} (f) the state evolves as $1\to2\to3\to1\to2\to3$, leading to total Berry phase of $2k_z$. Analogously, for clockwise motion along (g) the result is $-4 k_z$, while (h)  has the same pattern as (f), leading to $2 k_z$. These results coincide with the ones calculated from the Chern number. Such counting can be performed systematically for any periodic twist (see \cite{supp}) leading to the result Eq. \eqref{eq:ideal-winding}.

Increasing magnetic fields can lead to magnetic breakdown occurring at certain Fermi surface intersections, such that interlayer tunneling is suppressed. Specifically, the probability of magnetic breakdown for 
Eq. \eqref{eq:2layer} in presence of magnetic field along $z$ is given by \cite{stark_falicov,kaganov1983coherent,haidemenakis2014} $P_{MB}= \exp\left[-\frac{\pi c |\tilde{t}_{l,l'}|^2}{\hbar |e| B |{\bf v}_+| |{\bf v}_-|}\right]$, where ${\bf v}_\pm = ({\bf v}_l\pm {\bf v}_{l'})/2$. Since we expect $\tilde{t}_{l,l'}$ to be strongly suppressed with $|l-l'|$ for significant layer anisotropy and twist $t\ll \varepsilon E_F \theta$, there is a wide range of fields where breakdown dominates for all except nearest-layer crossings where $|\tilde{t}_{l,l'= l\pm 1}| = t$. For small twist angles $\theta\ll 1$, one can define a formal small parameter characterizing energy mismatch between rotated Fermi surfaces: $\alpha \approx \frac{t}{\varepsilon E_F \theta}$. The regime of dominant breakdown for all but nearest-layer tunneling amounts to $ t^2 \gg \hbar \frac{|e| B v_+ v_-}{c} \gg \alpha^2 t^2$.

In this limit, a general analysis for both commensurate and incommensurate cases can be developed. Specifically, for each layer there are only $8$ intersections between neighboring Fermi surfaces, dividing the Fermi surface into $8$ segments (see Fig. \ref{fig:fscross}(a) and \cite{supp}). The semiclassical motion of an electron wavepacket in the presence of a magnetic field along $z$ would consist of motion along such segments. When reaching the endpoint of a segment, the wavepacket transitions to a neighboring layer and propagating along a Fermi surface segment in the new layer (Fig. \ref{fig:fscross}(b)). One can systematically analyze the resulting trajectories \cite{supp}; for the periodic case the resulting closed orbits are also characterized by nonzero Chern numbers, similarly to the case of Fig. \ref{fig:phase} without magnetic breakdown.

For the incommensurate (irrational $\beta$) case, Eq. \eqref{eq:2layer} can still be applied; since the spiral period formally becomes infinite, no tunnelings acquire a phase besides $0$ or $\pi$. The resulting electron motion consists of monotonic propagation of wavepackets along the Fermi surface segments in Fig. \ref{fig:fscross}. At the end of each segment, electron transfers to a neighboring layer and the motion repeats. Importantly, the length of the segments does not change during the process. As a result, the $z$-axis velocity is given by $|v|_z^i =\frac{a}{T_i}$, where $T_i$ is the time required to traverse a particular segment. The total time for traversing all segments is $\sum_{i=1}^8 T_i=\frac{2\pi}{\omega_c}$, where $\omega_c=\frac{e B\sqrt{1-\varepsilon^2}}{m c }$ is the single-layer cyclotron frequency.

We can now estimate the magnitude of field-induced correction to the conductance in supertwisted spirals. {In the ballistic (clean) limit, the chemical potential  of up-moving electrons is increased by $e V$ and the current is given by $I_z = e A \sum_{v_z^i>0} \frac{\nu_i}{a} e V v_z^i$, where $\nu_i$ is the 2D density of states associated with a particular segment. Due to relation between density of states fraction and $T_i$: $ \frac{A \nu_i}{\mathcal{N}'}=\frac{\omega_c T_i}{2 \pi}$, $\mathcal{N}'=\mathcal{N}/\sqrt{1-\varepsilon^2}$ being the single-layer density of states \cite{supp}, summing over all the segments with positive velocity reduces $I_z$ to $\frac{2 \omega_c a}{\pi}\frac{\mathcal{N}'}{a} e ^2V$. The final result is:
\begin{equation}
    \delta G(B) \approx  \frac{2 \omega_c \mathcal{N}' e ^2}{\pi}.
    \label{eq:cond_cor}
\end{equation}
To estimate the magnitude of this correction, recall that Eq. \eqref{eq:cond_cor} holds for $t^2 \gg \hbar \frac{|e| B v_+ v_-}{c}  \sim_{\theta\ll 1} \hbar \omega_c \varepsilon E_F \theta$, so $ \delta G(B) \ll G_0 \left (\frac{t}{\varepsilon E_F \theta} \right )$, where $G_0 = \mathcal{N}' e ^2 \frac{t}{\hbar}$ is of the order of c-axis conductance {\it without twist}. Given that in the absence of field the conductance along the $z$ axis is strongly suppressed by twist $G(B = 0) \sim G_0 \left(\frac{2t}{\varepsilon E_F}\right )^L$ (Sec. \ref{sec:loc}), $\delta G(B)$ can conceivably become larger than the zero-field conductance even in spirals with relatively few twisted layers. We note that on increasing the field beyond $  B\gtrsim B_{MB} =  \frac{t^2 c}{\hbar |e| v_+ v_-}$, magnetic breakdown will become relevant even for nearest-layer crossings, which limits the linear-in-field contribution, Eq. \eqref{eq:cond_cor}.}

\section{Discussion}
\label{sec:disc}

Above we have established the spectral and transport behavior of supertwisted spirals by exploring the connection of the approximate Hamiltonian around the $\Gamma$ point \eqref{eqn:single-channel-TB} with the Aubry-Andr\'e model. We now discuss the realistic effects that have been not treated above.

\subsection{Effects of disorder}
\label{subsec:dis}

Here we discuss the qualitative effects of three physical mechanisms of disorder in supertwisted spirals: interlayer disorder (variation of energy or tunneling between individual layers arising, from e.g. unequal doping), twist-angle disorder, and intralayer disorder due to impurities in individual layers. We note also, that random shifts between the layers (that can appear if the supertwisted spiral is assembled) do not affect $H_0$ in \eqref{eq:TB-model} and are not expected to affect our results, but see Sec. \ref{subsec:moire} for the discussion of interplay with the moir\'e effects.

The first two mechanisms conserve the in-plane momentum ${\bf k}_\parallel$ but lead to random modulations of site energy, tunneling or the quasiperiodic potential phase (for twist disorder) in the language of AA model Eq. \eqref{eqn:single-channel-TB}. All of the above introduces randomness known to lead to localization in one dimension at any strength \cite{mirlin_rev}. However, the localization length would depend on the disorder strength. In contrast, the twist-induced localization length $\xi(k)$ abruptly becomes of the order $a$ at $k>k_c$. As a result, weak disorder of this type with $l_{loc}> \xi(k)$ would not preclude the observation of the dimensional localization effects, such as conductance suppression with doping, Fig. \ref{fig:spectra} (b). Furthermore, for twist disorder the amplitude of the resulting random potential would be $\propto |\kk|^{2n}$ (for $D_{2nh}$ point group), such that the localization length would diverge at low $|\kk|$. Finally, the topological phenomena discussed in Sec. \ref{sec:top}, such as edge states and field-induced delocalization due to nontrivial Berry phase of closed trajectories, are protected by the topological gaps in the spectrum (also for incommensurate case \cite{gottlob2025quasiperiodicity}) as long as the disorder strength is weaker than energy scale of the topological gap.

Let us now discuss the effects of disorder within the planes that introduces scattering between states with different ${\bf k}_\parallel$. Standard treatment of disorder (such as Born approximation \cite{lee_review}) leads to mixing of states within
$\delta k \sim l_{mfp}^{-1}$ ($l_{mfp}$ being the mean free path) of the Fermi energy; as a result, we expect the signatures of twist-induced dimensional localization in wavefunction structure to be qualitatively robust as long as $k_c l_{mfp}\gg 1$ (Eq. \eqref{eq:kc}). In $z$-axis transport, the in-plane disorder will also lead to an additional contribution, because the electrons can be scattered towards momenta where the can tunnel resonantly (marked in Fig. \ref{fig:fscross} (a)). {To estimate this effect, consider a simplified model with additional diffusive tunneling between layers that does not conserve momentum in the form of local impurities with concentration $n_{imp}$ and local tunneling amplitude $t_{imp}$. For two consecutive layers, the additional conductance can be estimated perturbatively \cite{ryndyk2015theory}, as $\delta G_1 \sim \mathcal{N}' \frac{e^2}{h} \frac{t_{incoh}^2(n_{imp} a_\parallel^2)}{W}$, where $W$ - is the bandwidth within a layer. For a $L$-layer supertwisted spiral this leads to $\delta G = \delta G_1/ L$. This value will introduce a lower bound to conductance; however, as long as it is much lower than the peak ballistic conductance value in Fig. \ref{fig:spectra}, it will not hinder the observation of localization onset. Compared to Eq. \eqref{eq:univ_g}, this suggests an estimate $\frac{t_{incoh}^2(n_{imp} a_\parallel^2)}{W} \sim \frac{\hbar}{\tau} \ll t$, where $\tau$ is the characteristic scattering time, as the condition for observability of localization effects in transport. Finally, for localized (along $z$) states at $|{\bf k}_\parallel|>k_c$ two-dimensional weak localization corrections \cite{mirlin_rev,lee_review} may become relevant, leading to localization for the in-plane transport too; we leave such study to future work.}

Let us now discuss the effect of in-plane disorder on topological properties presented in Sec. \ref{sec:top}. While topological gaps at a given ${\bf k}_\parallel$ are robust, scattering between different Fermi surface segments effectively introduces hybridization between edge modes of opposite chirality, Fig. \ref{fig:surface-state} (e,f), as both can coexist at the same energy due to $k$-dispersion of the bands. In particular, the presence of rotational symmetries in the plane implies multiple Fermi arcs generated at each surface ($2$ at least), allowing backscattering. At energies where bulk states have negligible spectral weight (e.g., Fig.  \ref{fig:surface-state}(c,d)), we expect that addition of disorder will not destroy the Fermi arcs by hybridizing them with bulk states. In principle, in presence of a strong spin-orbit coupling one can construct supertwisted spiral models, where $C_2$-related Fermi arcs will be spin-polarized, such that time-reversal symmetry will protect them from backscattering (see \cite{supp}). 

In Sec. \ref{sec:antiloc}, we have established the connection between field-induced delocalization and the Berry phase of the closed orbits during quantum oscillations. This suggests that the magnetic-field induced delocalization effects should be observable at least up for $\omega_c \tau \gtrsim 1$ \cite{shoenberg1984magnetic,glazman_rev}. The specific form of magnetoconductance, Eq. \eqref{eq:cond_cor}, is expected to be more fragile, relying on the ballistic transport assumption. Qualitatively, it requires the electron to be able to traverse $L$ layers before scattering, $\frac{L a}{v_z}\ll\tau$. Using the $v_z$ estimate from above, this condition translates to $\omega_c \tau \gg L$. We highlight that this is only the condition for applicability of Eq. \eqref{eq:cond_cor}, which falls within the range where we expect the magnetoconductance corrections to be observable $\omega_c \tau \gtrsim 1$;  evaluating these specific forms would be considered elsewhere.

\subsection{Effects of moir\'e superlattice}
\label{subsec:moire}

In the analysis above, we have neglected the effects of the moir\'e modulation of tunneling and potential $H_M$ \eqref{eq:TB-model}. The strongest effects of the moir\'e modulation would be for the states satisfying $|\kk| \sim |\g|/2$, where the scattering is resonant. Let us now address two specific limits.


First, consider $g\gg k_c$, such that the wavefunctions are localized within individual layers of the spiral, $\psi_{\kxy l}\to\psi_{\kxy l}^z\propto \delta_{zl}$, where $z$ defines the layer index the electron localized in. Projecting $H_M$ on these eigenstates, only the terms coming from the moir\'e patterns of the closest neighboring layers would contribute. Therefore, each layer of the spiral structure is subject to a moir\'e potential of the same magnitude but different orientation. Note, however, that the moir\'e patterns formed between a layer and its next and previous neighbor don't match exactly. Such setup has been studied in the context of helical trilayer graphene and strongly depends on lattice relaxation: without relaxation such structure would form a quasicrystal
\cite{kaxiras2020,mora2023}, while relaxation effects may lead to the formation of commensurate domains \cite{Devakul2023} or an effective Bloch band structure \cite{bernevig2019,popov2023}. In all of these cases, one expects formation of enhanced density of states due to gap opening from moir\'e umklapp scattering in every layer, promoting correlation effects. As a result, the systems will function as a collection of 2D layers with enhanced density of states, weakly coupled along the $z$ axis; a more detailed analysis of the resulting electronic structure and correlation effects would be pursued elsewhere.

In the other limit, $g\ll k_c$, we can approximate the states affected by $H_M$ as plane waves along the $z$ axis  $\psi_{\kxy l}^{k_z}\propto \frac{1} {\sqrt{L}} e^{i k_z l}$. As we show in \onlinecite{supp}, the moir\'e umklapp scattering is suppressed in this case. Moreover if we allow the rotation axis not to be aligned in all the layers of the spiral, as expected in a layer-by-layer assembly,  $\langle \psi_{\kxy l}^{k_z}|H_M|\psi_{\kxy l}^{k_z}\rangle$ vanishes after averaging over its position fluctuations along the spiral. Such fluctuations, however, do not affect the magnitude of the momentum-conserving interlayer tunneling leading to Eq. \eqref{eqn:single-channel-TB}. 

Overall, we find that the moir\'e effects will produce a stronger impact on the localized states for $g\gg k_c$, which requires larger twist angles for their observation.

\subsection{Material candidate platforms}
\label{sec:cand}

Let us now discuss the candidate platforms. We envision finite-size spirals made of exfoliable materials with band extrema near $\Gamma$ point. For sufficiently small number of layers, gating may be sufficient to obtain signatures of doping-induced localization in transport (Fig. \ref{fig:conductance}). On the other hand, with fixed chemical doping, the dependence of conductance on the layer number (Fig. \ref{fig:commensurate}) can be used. Specific materials include TiS$_3$ \cite{island2016titanium, yi2018band}, black phosphorus \cite{wang2015highly,xu2019recent, wu2021large}, T' TMDs \cite{Varsano2020}, such as WTe$_2$ \cite{Fei2017,Jia2021,pesin2022} (see also \cite{jiang20242dtheoreticallytwistablematerial} for computationally proposed materials).  TiS$_3$ contains a Fermi surface around $\Gamma$, while black phosphorus shows pockets along $\Gamma-Y$ in the hole-doped side but a Fermi surface around $\Gamma$ at the electron-doped side. Alternatively, materials such as GaSe \cite{ben2018valence} can be a candidate for a material with higher rotational symmetry $D_{3h}$ which contains both $\Gamma$-point Fermi surface and Fermi pockets depending on the doping. In addition to materials with band extrema at $\Gamma$ in monolayers, a number of transition metal dichalcogenides show the band edge moving from $K$-point to $\Gamma$-point on increasing the layer number \cite{foutty2023,ma2025relativistic,campbell2024interplay}. Therefore, it appears possible that supertwisted spirals of such materials \cite{zhao2020supertwisted,ji2024opto} may have band extrema at the $\Gamma$ point. While we focused on the vicinity of the $\Gamma$ point for analytical transparency, our results do suggest that moving away from $\Gamma$ leads to localization of the eigenstates. We therefore speculate that signatures of 3D to 2D localization may be observed in spirals of metallic materials with larger pockets centered around $\Gamma$, such as NbSe$_2$ \cite{Xi2015ising} or TaS$_2$ \cite{NavarroMoratalla2016}; while these materials contain pockets also around $K$ points, the $z$-axis transport is likely to be dominated by the $\Gamma$-pocket. 

Finally, our results are straightforward to generalize to the propagation of waves (optical, acoustic or of collective excitations, such as excitons \cite{ji2024opto}) in supertwisted spirals. We, therefore, expect the out-of-plane localization transition with the in-plane propagation wavevector $q_\parallel$ along with the emergence of Fermi arc surface states. We also highlight that in optical or acoustic platforms, where individual layers are homogeneous on the wavelength scale, moir\'e effects can be strongly suppressed compared to the electronic case, leading to a more faithful representation of the physics of $H_0$, \eqref{eq:TB-model}.

\section{Conclusion}

We have demonstrated that continuously increasing twist between layers in a bulk material leads to universal emergent properties of the electronic wavefunctions. Specifically, increasing in-plane momentum induces a 3D-to-2D localization transition for electronic states around $\Gamma$ point, Fig. \ref{fig:Schematics} (Sec. \ref{sec:loc}). The transition  can be controlled experimentally by doping or gating, and is driven by the unavoidable anisotropy of the in-plane electronic dispersion. Signatures of the 3D-to-2D localization can be observed in the $z$-axis electronic conductance, which exhibits a universal suppression with doping and layer number \eqref{eq:univ_g} in the localized regime (Figs. \ref{fig:conductance} (b), \ref{fig:commensurate} (a)). Furthermore, twist also leads to emergent topological properties: chiral Fermi arc states (Fig. \ref{fig:surface-state}) and delocalization under magnetic field (Figs. \ref{fig:Schematics} (c) and \ref{fig:phase}). Our theoretical results are applicable to a wide range of materials and platforms (Sec. \ref{sec:cand}), driven by a universal mapping of the low-energy physics to the Aubry-Andr\'e model (Sec. \ref{sec:AA}). Our results do not rely on any special properties of the monolayers, such as Dirac dispersion or spin-orbit coupling, establishing the 3D-to-2D localization and nontrivial topology as a universal emergent property of wave propagation in supertwisted spirals.

\header{Acknowledgement}
Jeane Siriviboon was supported by funding from the Tushar Shah and Sara Zion Fellowship. We thank M. Li, P. Jarillo-Herrero, A. Vishwanath, J. H. Wilson, Y. Onishi, E. J. Mele and J. H. Pixley for comments and discussions. 

\bibliographystyle{unsrt}
\bibliography{refs}

@article{aubry1980analyticity,
  title={Analyticity breaking and Anderson localization in incommensurate lattices},
  author={Aubry, Serge and Andr{\'e}, Gilles},
  journal={Ann. Israel Phys. Soc},
  volume={3},
  number={133},
  pages={18},
  year={1980}
}

@article{zhao2020supertwisted,
  title={Supertwisted spirals of layered materials enabled by growth on non-Euclidean surfaces},
  author={Zhao, Yuzhou and Zhang, Chenyu and Kohler, Daniel D and Scheeler, Jason M and Wright, John C and Voyles, Paul M and Jin, Song},
  journal={Science},
  volume={370},
  number={6515},
  pages={442--445},
  year={2020},
  publisher={American Association for the Advancement of Science}
}

@article{ji2024opto,
  title={Opto-twistronic Hall effect in a three-dimensional spiral lattice},
  author={Ji, Zhurun and Zhao, Yuzhou and Chen, Yicong and Zhu, Ziyan and Wang, Yuhui and Liu, Wenjing and Modi, Gaurav and Mele, Eugene J and Jin, Song and Agarwal, Ritesh},
  journal={Nature},
  volume={634},
  number={8032},
  pages={69--73},
  year={2024},
  publisher={Nature Publishing Group UK London}
}

@article{wang2024conversion,
  title={Conversion of chirality to twisting via sequential one-dimensional and two-dimensional growth of graphene spirals},
  author={Wang, Zhu-Jun and Kong, Xiao and Huang, Yuan and Li, Jun and Bao, Lihong and Cao, Kecheng and Hu, Yuxiong and Cai, Jun and Wang, Lifen and Chen, Hui and others},
  journal={Nature Materials},
  volume={23},
  number={3},
  pages={331--338},
  year={2024},
  publisher={Nature Publishing Group UK London}
}

@article{modugno2010anderson,
  title={Anderson localization in Bose--Einstein condensates},
  author={Modugno, Giovanni},
  journal={Reports on progress in physics},
  volume={73},
  number={10},
  pages={102401},
  year={2010},
  publisher={IOP Publishing}
}

@article{luo2021transfer,
  title={Transfer matrix study of the Anderson transition in non-Hermitian systems},
  author={Luo, Xunlong and Ohtsuki, Tomi and Shindou, Ryuichi},
  journal={Physical Review B},
  volume={104},
  number={10},
  pages={104203},
  year={2021},
  publisher={APS}
}

@article{blanter2000shot,
  title={Shot noise in mesoscopic conductors},
  author={Blanter, Ya M and B{\"u}ttiker, Markus},
  journal={Physics reports},
  volume={336},
  number={1-2},
  pages={1--166},
  year={2000},
  publisher={Elsevier}
}

@misc{supp,
  note = "See Supplemental Material."
}

@incollection{markos2008wave,
  title={Wave propagation: from electrons to photonic crystals and left-handed materials},
  author={Markos, Peter and Soukoulis, Costas M},
  booktitle={Wave Propagation},
  year={2008},
  publisher={Princeton University Press}
}

@article{zhang2015almost,
  title={The almost mobility edge in the almost Mathieu equation},
  author={Zhang, Yi and Bulmash, Daniel and Maharaj, Akash V and Jian, Chao-Ming and Kivelson, Steven A},
  journal={arXiv preprint arXiv:1504.05205},
  year={2015}
}

@article{vu2023generic,
  title={Generic mobility edges in several classes of duality-breaking one-dimensional quasiperiodic potentials},
  author={Vu, DinhDuy and Das Sarma, Sankar},
  journal={Physical Review B},
  volume={107},
  number={22},
  pages={224206},
  year={2023},
  publisher={APS}
}

@article{island2016titanium,
  title={Titanium trisulfide (TiS3): a 2D semiconductor with quasi-1D optical and electronic properties},
  author={Island, Joshua O and Biele, Robert and Barawi, Mariam and Clamagirand, Jos{\'e} M and Ares, Jos{\'e} R and S{\'a}nchez, Carlos and van der Zant, Herre SJ and Ferrer, Isabel J and D’Agosta, Roberto and Castellanos-Gomez, Andres},
  journal={Scientific reports},
  volume={6},
  number={1},
  pages={22214},
  year={2016},
  publisher={Nature Publishing Group UK London}
}

@article{wang2015highly,
  title={Highly anisotropic and robust excitons in monolayer black phosphorus},
  author={Wang, Xiaomu and Jones, Aaron M and Seyler, Kyle L and Tran, Vy and Jia, Yichen and Zhao, Huan and Wang, Han and Yang, Li and Xu, Xiaodong and Xia, Fengnian},
  journal={Nature nanotechnology},
  volume={10},
  number={6},
  pages={517--521},
  year={2015},
  publisher={Nature Publishing Group UK London}
}

@article{ben2018valence,
  title={Valence band inversion and spin-orbit effects in the electronic structure of monolayer GaSe},
  author={Ben Aziza, Zeineb and Z{\'o}lyomi, Viktor and Henck, Hugo and Pierucci, Debora and Silly, Mathieu G and Avila, Jos{\'e} and Magorrian, Samuel J and Chaste, Julien and Chen, Chaoyu and Yoon, Mina and others},
  journal={Physical Review B},
  volume={98},
  number={11},
  pages={115405},
  year={2018},
  publisher={APS}
}

@article{xu2019recent,
  title={Recent progress in black phosphorus and black-phosphorus-analogue materials: properties, synthesis and applications},
  author={Xu, Yijun and Shi, Zhe and Shi, Xinyao and Zhang, Kai and Zhang, Han},
  journal={Nanoscale},
  volume={11},
  number={31},
  pages={14491--14527},
  year={2019},
  publisher={Royal Society of Chemistry}
}

@article{wu2021large,
  title={Large-scale growth of few-layer two-dimensional black phosphorus},
  author={Wu, Zehan and Lyu, Yongxin and Zhang, Yi and Ding, Ran and Zheng, Beining and Yang, Zhibin and Lau, Shu Ping and Chen, Xian Hui and Hao, Jianhua},
  journal={Nature materials},
  volume={20},
  number={9},
  pages={1203--1209},
  year={2021},
  publisher={Nature Publishing Group UK London}
}

@article{yi2018band,
  title={The band structure of the quasi-one-dimensional layered semiconductor TiS3 (001)},
  author={Yi, Hemian and Komesu, Takashi and Gilbert, Simeon and Hao, Guanhua and Yost, Andrew J and Lipatov, Alexey and Sinitskii, Alexander and Avila, Jose and Chen, Chaoyu and Asensio, Maria C and others},
  journal={Applied Physics Letters},
  volume={112},
  number={5},
  year={2018},
  publisher={AIP Publishing}
}

@article{mele2025,
  title = {Squeezing quantum states in three-dimensional twisted crystals},
  author = {Phong, V\~o Tien and Kunkelmann, Kason and De Beule, Christophe and Al Ezzi, Mohammed M. and Slager, Robert-Jan and Adam, Shaffique and Mele, E. J.},
  journal = {Phys. Rev. B},
  volume = {111},
  issue = {24},
  pages = {245156},
  numpages = {17},
  year = {2025},
  month = {Jun},
  publisher = {American Physical Society},
  doi = {10.1103/cj2q-f9q2},
  url = {https://link.aps.org/doi/10.1103/cj2q-f9q2}
}

@article{wu2020three,
  title={Three-dimensional topological twistronics},
  author={Wu, Fengcheng and Zhang, Rui-Xing and Das Sarma, Sankar},
  journal={Physical Review Research},
  volume={2},
  number={2},
  pages={022010},
  year={2020},
  publisher={APS}
}

@article{spiral_review,
author = {Wang, Qian and Wang, Xinchao and Lou, Qianwen and Jiang, Ying and Fan, Xiaopeng},
title = {Two-Dimensional Spiral: A Promising Moiré Superlattice},
journal = {Laser \& Photonics Reviews},
volume = {19},
number = {6},
pages = {2401368},
doi = {https://doi.org/10.1002/lpor.202401368},
url = {https://onlinelibrary.wiley.com/doi/abs/10.1002/lpor.202401368},
eprint = {https://onlinelibrary.wiley.com/doi/pdf/10.1002/lpor.202401368},
year = {2025}
}

@article{Tong2024,
  title = {Giant Second Harmonic Generation in Supertwisted WS2 Spirals Grown in Step-Edge Particle-Induced Non-Euclidean Surfaces},
  volume = {18},
  ISSN = {1936-086X},
  url = {http://dx.doi.org/10.1021/acsnano.4c02807},
  DOI = {10.1021/acsnano.4c02807},
  number = {33},
  journal = {ACS Nano},
  publisher = {American Chemical Society (ACS)},
  author = {Tong,  Tong and Chen,  Ruijie and Ke,  Yuxuan and Wang,  Qian and Wang,  Xinchao and Sun,  Qinjun and Chen,  Jie and Gu,  Zhiyuan and Yu,  Ying and Wei,  Hongyan and Hao,  Yuying and Fan,  Xiaopeng and Zhang,  Qing},
  year = {2024},
  month = aug,
  pages = {21939–21947}
}

@article{Qi_2024_exp,
author = {Minru Qi and Tong Tong and Xiaopeng Fan and Xiangdong Li and Shen Wang and Guofeng Zhang and Ruiyun Chen and Jianyong Hu and Zhichun Yang and Ganying Zeng and Chengbing Qin and Liantuan Xiao and Suotang Jia},
journal = {Opt. Express},
number = {6},
pages = {10419--10428},
publisher = {Optica Publishing Group},
title = {Anomalous layer-dependent photoluminescence spectra of supertwisted spiral WS2},
volume = {32},
month = {Mar},
year = {2024},
url = {https://opg.optica.org/oe/abstract.cfm?URI=oe-32-6-10419},
doi = {10.1364/OE.516177},
}

@article{Fan2020,
  title = {Mechanism of Extreme Optical Nonlinearities in Spiral WS2 above the Bandgap},
  volume = {20},
  ISSN = {1530-6992},
  url = {http://dx.doi.org/10.1021/acs.nanolett.0c00305},
  DOI = {10.1021/acs.nanolett.0c00305},
  number = {4},
  journal = {Nano Letters},
  publisher = {American Chemical Society (ACS)},
  author = {Fan,  Xiaopeng and Ji,  Zhurun and Fei,  Ruixiang and Zheng,  Weihao and Liu,  Wenjing and Zhu,  Xiaoli and Chen,  Shula and Yang,  Li and Liu,  Hongjun and Pan,  Anlian and Agarwal,  Ritesh},
  year = {2020},
  month = mar,
  pages = {2667–2673}
}

@article{ci_2022,
author = {Ci, Penghong and Zhao, Yuzhou and Sun, Muhua and Rho, Yoonsoo and Chen, Yabin and Grigoropoulos, Costas P. and Jin, Song and Li, Xiaoguang and Wu, Junqiao},
title = {Breaking Rotational Symmetry in Supertwisted WS2 Spirals via Moiré Magnification of Intrinsic Heterostrain},
journal = {Nano Letters},
volume = {22},
number = {22},
pages = {9027-9035},
year = {2022},
doi = {10.1021/acs.nanolett.2c03347},
    note ={PMID: 36346996},

URL = { 
    
        https://doi.org/10.1021/acs.nanolett.2c03347
    
    

},
eprint = { 
    
        https://doi.org/10.1021/acs.nanolett.2c03347
    
    

}

}

@article{peng_2022,
author = {Peng, Jiangbo and Ren, Caixia and Zhang, Weili and Chen, Hu and Pan, Xiaoguang and Bai, Hangxin and Jing, Fangli and Qiu, Hailong and Liu, Hongjun and Hu, Zhanggui},
title = {Spatially Dependent Electronic Structures and Excitons in a Marginally Twisted Moiré Superlattice of Spiral WS2},
journal = {ACS Nano},
volume = {16},
number = {12},
pages = {21600-21608},
year = {2022},
doi = {10.1021/acsnano.2c10562},
    note ={PMID: 36475630},

URL = { 
    
        https://doi.org/10.1021/acsnano.2c10562
    
    

},
eprint = { 
    
        https://doi.org/10.1021/acsnano.2c10562
    
    

}

}

@article{sb_spiral,
author = {Huang, Ding-Ming and Wu, Xu and Chang, Kai and Hu, Hao and Wang, Ye-Liang and Xu, H. Q. and Zhang, Jian-Jun},
title = {Strain Effects in Twisted Spiral Antimonene},
journal = {Advanced Science},
volume = {10},
number = {19},
pages = {2301326},
doi = {https://doi.org/10.1002/advs.202301326},
url = {https://advanced.onlinelibrary.wiley.com/doi/abs/10.1002/advs.202301326},
eprint = {https://advanced.onlinelibrary.wiley.com/doi/pdf/10.1002/advs.202301326},
year = {2023}
}

@article{zhang2023vapour,
  title={Vapour-phase deposition of two-dimensional layered chalcogenides},
  author={Zhang, Tianyi and Wang, Jiangtao and Wu, Peng and Lu, Ang-Yu and Kong, Jing},
  journal={Nature Reviews Materials},
  volume={8},
  number={12},
  pages={799--821},
  year={2023},
  publisher={Nature Publishing Group UK London}
}

@article{Liu2025, 
author = {Peng Liu and Xinchao Wang and Xiumeng Bao and Xiaoyong Fan and Junyuan Chen and Yang Bai and Yuying Hao and Hongli Yang and Weihua Yang and Xiaopeng Fan},
title = {Supertwisted WS2 spirals synthesized on step-edge non-Euclidean surfaces: Twist angle modulation and optical properties},
year = {2025},
journal = {Nano Research},
volume = {18},
number = {6},
pages = {94907451},
url = {https://www.sciopen.com/article/10.26599/NR.2025.94907451},
doi = {10.26599/NR.2025.94907451}
}

@article{mannix2022robotic,
  title={Robotic four-dimensional pixel assembly of van der Waals solids},
  author={Mannix, Andrew J and Ye, Andrew and Sung, Suk Hyun and Ray, Ariana and Mujid, Fauzia and Park, Chibeom and Lee, Myungjae and Kang, Jong-Hoon and Shreiner, Robert and High, Alexander A and others},
  journal={Nature nanotechnology},
  volume={17},
  number={4},
  pages={361--366},
  year={2022},
  publisher={Nature Publishing Group UK London}
}

@article{li2024towards,
  title={Towards the scalable synthesis of two-dimensional heterostructures and superlattices beyond exfoliation and restacking},
  author={Li, Jia and Yang, Xiangdong and Zhang, Zhengwei and Yang, Weiyou and Duan, Xidong and Duan, Xiangfeng},
  journal={Nature Materials},
  volume={23},
  number={10},
  pages={1326--1338},
  year={2024},
  publisher={Nature Publishing Group UK London}
}

@article{foutty2023,
  title={Tunable spin and valley excitations of correlated insulators in $\Gamma$-valley moir{\'e} bands},
  author={Foutty, Benjamin A and Yu, Jiachen and Devakul, Trithep and Kometter, Carlos R and Zhang, Yang and Watanabe, Kenji and Taniguchi, Takashi and Fu, Liang and Feldman, Benjamin E},
  journal={Nature Materials},
  volume={22},
  number={6},
  pages={731--736},
  year={2023},
  publisher={Nature Publishing Group UK London}
}

@article{ma2025relativistic,
  title={Relativistic Mott transition in twisted WSe2 tetralayers},
  author={Ma, Liguo and Chaturvedi, Raghav and Nguyen, Phuong X and Watanabe, Kenji and Taniguchi, Takashi and Mak, Kin Fai and Shan, Jie},
  journal={Nature Materials},
  volume={24},
  number={12},
  pages={1935--1941},
  year={2025},
  publisher={Nature Publishing Group UK London}
}

@article{campbell2024interplay,
  title={The interplay of field-tunable strongly correlated states in a multi-orbital moir{\'e} system},
  author={Campbell, Aidan J and Vitale, Valerio and Brotons-Gisbert, Mauro and Baek, Hyeonjun and Borel, Antoine and Ivanova, Tatyana V and Taniguchi, Takashi and Watanabe, Kenji and Lischner, Johannes and Gerardot, Brian D},
  journal={Nature Physics},
  volume={20},
  number={4},
  pages={589--596},
  year={2024},
  publisher={Nature Publishing Group UK London}
}

@article{nuckolls2026higher,
  title={Higher-dimensional Fermiology in bulk moir{\'e} metals},
  author={Nuckolls, Kevin P and Paul, Nisarga and Chen, Alan and Gaggioli, Filippo and Wakefield, Joshua P and Auslender, Avi and Gardener, Jules and Akey, Austin J and Graf, David and Suzuki, Takehito and others},
  journal={Nature},
  pages={1--8},
  year={2026},
  publisher={Nature Publishing Group UK London}
}

@article{angeli2021gamma,
  title={$\Gamma$ valley transition metal dichalcogenide moir{\'e} bands},
  author={Angeli, Mattia and MacDonald, Allan H},
  journal={Proceedings of the National Academy of Sciences},
  volume={118},
  number={10},
  pages={e2021826118},
  year={2021},
  publisher={National Academy of Sciences}
}

@article{zhang_fu_2021,
  title = {Electronic structures, charge transfer, and charge order in twisted transition metal dichalcogenide bilayers},
  author = {Zhang, Yang and Liu, Tongtong and Fu, Liang},
  journal = {Phys. Rev. B},
  volume = {103},
  issue = {15},
  pages = {155142},
  numpages = {6},
  year = {2021},
  month = {Apr},
  publisher = {American Physical Society},
  doi = {10.1103/PhysRevB.103.155142},
  url = {https://link.aps.org/doi/10.1103/PhysRevB.103.155142}
}

@article{hpan_2023,
  title = {Realizing a tunable honeycomb lattice in ABBA-stacked twisted double bilayer ${\mathrm{WSe}}_{2}$},
  author = {Pan, Haining and Kim, Eun-Ah and Jian, Chao-Ming},
  journal = {Phys. Rev. Res.},
  volume = {5},
  issue = {4},
  pages = {043173},
  numpages = {9},
  year = {2023},
  month = {Nov},
  publisher = {American Physical Society},
  doi = {10.1103/PhysRevResearch.5.043173},
  url = {https://link.aps.org/doi/10.1103/PhysRevResearch.5.043173}
}

@article{bistritzer2011moire,
  title={Moir{\'e} bands in twisted double-layer graphene},
  author={Bistritzer, Rafi and MacDonald, Allan H},
  journal={Proceedings of the National Academy of Sciences},
  volume={108},
  number={30},
  pages={12233--12237},
  year={2011},
  publisher={National Academy of Sciences}
}

@article{balents2020review,
  title={Superconductivity and strong correlations in moir{\'e} flat bands},
  author={Balents, Leon and Dean, Cory R and Efetov, Dmitri K and Young, Andrea F},
  journal={Nature Physics},
  volume={16},
  number={7},
  pages={725--733},
  year={2020},
  publisher={Nature Publishing Group UK London}
}

@article{bernevig2025fractional,
  title={Fractional quantization in insulators from Hall to Chern},
  author={Bernevig, BA and Fu, L and Ju, L and MacDonald, AH and Mak, KF and Shan, J},
  journal={Nature Physics},
  pages={1--12},
  year={2025},
  publisher={Nature Publishing Group UK London}
}

@article{Geim2013,
  title = {Van der Waals heterostructures},
  volume = {499},
  ISSN = {1476-4687},
  url = {http://dx.doi.org/10.1038/nature12385},
  DOI = {10.1038/nature12385},
  number = {7459},
  journal = {Nature},
  publisher = {Springer Science and Business Media LLC},
  author = {Geim,  A. K. and Grigorieva,  I. V.},
  year = {2013},
  month = jul,
  pages = {419–425}
}

@article{Novoselov2016,
  title = {2D materials and van der Waals heterostructures},
  volume = {353},
  ISSN = {1095-9203},
  url = {http://dx.doi.org/10.1126/science.aac9439},
  DOI = {10.1126/science.aac9439},
  number = {6298},
  journal = {Science},
  publisher = {American Association for the Advancement of Science (AAAS)},
  author = {Novoselov,  K. S. and Mishchenko,  A. and Carvalho,  A. and Castro Neto,  A. H.},
  year = {2016},
  month = jul 
}

@article{Andrei2021,
  title = {The marvels of moiré materials},
  volume = {6},
  ISSN = {2058-8437},
  url = {http://dx.doi.org/10.1038/s41578-021-00284-1},
  DOI = {10.1038/s41578-021-00284-1},
  number = {3},
  journal = {Nature Reviews Materials},
  publisher = {Springer Science and Business Media LLC},
  author = {Andrei,  Eva Y. and Efetov,  Dmitri K. and Jarillo-Herrero,  Pablo and MacDonald,  Allan H. and Mak,  Kin Fai and Senthil,  T. and Tutuc,  Emanuel and Yazdani,  Ali and Young,  Andrea F.},
  year = {2021},
  month = mar,
  pages = {201–206}
}

@article{lucht2025,
  title={2.5-dimensional topological superconductivity in twisted superconducting flakes},
  author={Lucht, Kevin P and Pixley, JH and Volkov, Pavel A},
  journal={npj Quantum Materials},
  volume={10},
  number={1},
  pages={10},
  year={2025},
  publisher={Nature Publishing Group UK London}
}

@article{pixley2025_rev,
   author = "Pixley, J. H. and Pavel A. Volkov",
   title = "Twisted Nodal Superconductors", 
   journal= "Annual Review of Condensed Matter Physics",
   year = "2026",
   volume = "17",
   number = "Volume 17, 2026",
   pages = "183-205",
   doi = "https://doi.org/10.1146/annurev-conmatphys-031524-063257",
   url = "https://www.annualreviews.org/content/journals/10.1146/annurev-conmatphys-031524-063257",
   publisher = "Annual Reviews",
   type = "Journal Article"
  }

@article{mak2022semiconductor,
  title={Semiconductor moir{\'e} materials},
  author={Mak, Kin Fai and Shan, Jie},
  journal={Nature Nanotechnology},
  volume={17},
  number={7},
  pages={686--695},
  year={2022},
  publisher={Nature Publishing Group UK London}
}

@article{chen2025spinless,
  title={Spinless topological chirality from Umklapp scattering in twisted 3D structures},
  author={Chen, Cong and Zeng, Xu-Tao and Yao, Wang},
  journal={Reports on Progress in Physics},
  volume={88},
  number={1},
  pages={018001},
  year={2025},
  publisher={IOP Publishing}
}

@article{malla2018spinful,
  title={Spinful Aubry-Andr{\'e} model in a magnetic field: Delocalization facilitated by a weak spin-orbit coupling},
  author={Malla, Rajesh K and Raikh, Mikhail E},
  journal={Physical Review B},
  volume={97},
  number={21},
  pages={214209},
  year={2018},
  publisher={APS}
}

@article{wilkinson1984,
    author = {Wilkinson, M.},
    title = {Critical properties of electron eigenstates in incommensurate systems},
    journal = {Proceedings of the Royal Society of London. A. Mathematical and Physical Sciences},
    volume = {391},
    number = {1801},
    pages = {305-350},
    year = {1984},
    month = {02},
    issn = {0080-4630},
    doi = {10.1098/rspa.1984.0016},
    url = {https://doi.org/10.1098/rspa.1984.0016},
    eprint = {https://royalsocietypublishing.org/rspa/article-pdf/391/1801/305/65470/rspa.1984.0016.pdf},
}

@article{Varsano2020,
  title = {A monolayer transition-metal dichalcogenide as a topological excitonic insulator},
  volume = {15},
  ISSN = {1748-3395},
  url = {http://dx.doi.org/10.1038/s41565-020-0650-4},
  DOI = {10.1038/s41565-020-0650-4},
  number = {5},
  journal = {Nature Nanotechnology},
  publisher = {Springer Science and Business Media LLC},
  author = {Varsano,  Daniele and Palummo,  Maurizia and Molinari,  Elisa and Rontani,  Massimo},
  year = {2020},
  month = mar,
  pages = {367–372}
}

@article{Fei2017,
  title = {Edge conduction in monolayer WTe2},
  volume = {13},
  ISSN = {1745-2481},
  url = {http://dx.doi.org/10.1038/nphys4091},
  DOI = {10.1038/nphys4091},
  number = {7},
  journal = {Nature Physics},
  publisher = {Springer Science and Business Media LLC},
  author = {Fei,  Zaiyao and Palomaki,  Tauno and Wu,  Sanfeng and Zhao,  Wenjin and Cai,  Xinghan and Sun,  Bosong and Nguyen,  Paul and Finney,  Joseph and Xu,  Xiaodong and Cobden,  David H.},
  year = {2017},
  month = apr,
  pages = {677–682}
}

@article{Jia2021,
  title = {Evidence for a monolayer excitonic insulator},
  volume = {18},
  ISSN = {1745-2481},
  url = {http://dx.doi.org/10.1038/s41567-021-01422-w},
  DOI = {10.1038/s41567-021-01422-w},
  number = {1},
  journal = {Nature Physics},
  publisher = {Springer Science and Business Media LLC},
  author = {Jia,  Yanyu and Wang,  Pengjie and Chiu,  Cheng-Li and Song,  Zhida and Yu,  Guo and J\"{a}ck,  Berthold and Lei,  Shiming and Klemenz,  Sebastian and Cevallos,  F. Alexandre and Onyszczak,  Michael and Fishchenko,  Nadezhda and Liu,  Xiaomeng and Farahi,  Gelareh and Xie,  Fang and Xu,  Yuanfeng and Watanabe,  Kenji and Taniguchi,  Takashi and Bernevig,  B. Andrei and Cava,  Robert J. and Schoop,  Leslie M. and Yazdani,  Ali and Wu,  Sanfeng},
  year = {2021},
  month = dec,
  pages = {87–93}
}

@article{pesin2022,
	title={{Low-energy effective theory and anomalous Hall effect in monolayer $\mathrm{WTe}_2$}},
	author={S. Nandy and D. A. Pesin},
	journal={SciPost Phys.},
	volume={12},
	pages={120},
	year={2022},
	publisher={SciPost},
	doi={10.21468/SciPostPhys.12.4.120},
	url={https://scipost.org/10.21468/SciPostPhys.12.4.120},
}

@misc{jiang20242dtheoreticallytwistablematerial,
      title={2D Theoretically Twistable Material Database}, 
      author={Yi Jiang and Urko Petralanda and Grigorii Skorupskii and Qiaoling Xu and Hanqi Pi and Dumitru Călugăru and Haoyu Hu and Jiaze Xie and Rose Albu Mustaf and Peter Höhn and Vicky Haase and Maia G. Vergniory and Martin Claassen and Luis Elcoro and Nicolas Regnault and Jie Shan and Kin Fai Mak and Dmitri K. Efetov and Emilia Morosan and Dante M. Kennes and Angel Rubio and Lede Xian and Claudia Felser and Leslie M. Schoop and B. Andrei Bernevig},
      year={2024},
      eprint={2411.09741},
      archivePrefix={arXiv},
      primaryClass={cond-mat.mtrl-sci},
      url={https://arxiv.org/abs/2411.09741}, 
}

@article{kraus2012topological,
  title={Topological states and adiabatic pumping in quasicrystals},
  author={Kraus, Yaacov E and Lahini, Yoav and Ringel, Zohar and Verbin, Mor and Zilberberg, Oded},
  journal={Physical review letters},
  volume={109},
  number={10},
  pages={106402},
  year={2012},
  publisher={APS}
}

@article{Xi2015ising,
  title = {Ising pairing in superconducting NbSe2 atomic layers},
  volume = {12},
  ISSN = {1745-2481},
  url = {http://dx.doi.org/10.1038/nphys3538},
  DOI = {10.1038/nphys3538},
  number = {2},
  journal = {Nature Physics},
  publisher = {Springer Science and Business Media LLC},
  author = {Xi,  Xiaoxiang and Wang,  Zefang and Zhao,  Weiwei and Park,  Ju-Hyun and Law,  Kam Tuen and Berger,  Helmuth and Forró,  László and Shan,  Jie and Mak,  Kin Fai},
  year = {2015},
  month = nov,
  pages = {139–143}
}

@article{NavarroMoratalla2016,
  title = {Enhanced superconductivity in atomically thin TaS2},
  volume = {7},
  ISSN = {2041-1723},
  url = {http://dx.doi.org/10.1038/ncomms11043},
  DOI = {10.1038/ncomms11043},
  number = {1},
  journal = {Nature Communications},
  publisher = {Springer Science and Business Media LLC},
  author = {Navarro-Moratalla,  Efrén and Island,  Joshua O. and Mañas-Valero,  Samuel and Pinilla-Cienfuegos,  Elena and Castellanos-Gomez,  Andres and Quereda,  Jorge and Rubio-Bollinger,  Gabino and Chirolli,  Luca and Silva-Guillén,  Jose Angel and Agraït,  Nicolás and Steele,  Gary A. and Guinea,  Francisco and van der Zant,  Herre S. J. and Coronado,  Eugenio},
  year = {2016},
  month = mar 
}

@article{gottlob2025quasiperiodicity,
  title={Quasiperiodicity protects quantized transport in disordered systems without gaps},
  author={Gottlob, Emmanuel and Borgnia, Dan S and Slager, Robert-Jan and Schneider, Ulrich},
  journal={PRX Quantum},
  volume={6},
  number={2},
  pages={020359},
  year={2025},
  publisher={APS}
}

@article{lang_2012,
  title = {Edge States and Topological Phases in One-Dimensional Optical Superlattices},
  author = {Lang, Li-Jun and Cai, Xiaoming and Chen, Shu},
  journal = {Phys. Rev. Lett.},
  volume = {108},
  issue = {22},
  pages = {220401},
  numpages = {5},
  year = {2012},
  month = {May},
  publisher = {American Physical Society},
  doi = {10.1103/PhysRevLett.108.220401},
  url = {https://link.aps.org/doi/10.1103/PhysRevLett.108.220401}
}

@article{ganeshan_2013,
  title = {Topological Zero-Energy Modes in Gapless Commensurate Aubry-Andr\'e-Harper Models},
  author = {Ganeshan, Sriram and Sun, Kai and Das Sarma, S.},
  journal = {Phys. Rev. Lett.},
  volume = {110},
  issue = {18},
  pages = {180403},
  numpages = {5},
  year = {2013},
  month = {May},
  publisher = {American Physical Society},
  doi = {10.1103/PhysRevLett.110.180403},
  url = {https://link.aps.org/doi/10.1103/PhysRevLett.110.180403}
}

@article{hofstadter,
  title = {Energy levels and wave functions of Bloch electrons in rational and irrational magnetic fields},
  author = {Hofstadter, Douglas R.},
  journal = {Phys. Rev. B},
  volume = {14},
  issue = {6},
  pages = {2239--2249},
  numpages = {0},
  year = {1976},
  month = {Sep},
  publisher = {American Physical Society},
  doi = {10.1103/PhysRevB.14.2239},
  url = {https://link.aps.org/doi/10.1103/PhysRevB.14.2239}
}

@article{xu2015discovery,
  title={Discovery of a Weyl fermion semimetal and topological Fermi arcs},
  author={Xu, Su-Yang and Belopolski, Ilya and Alidoust, Nasser and Neupane, Madhab and Bian, Guang and Zhang, Chenglong and Sankar, Raman and Chang, Guoqing and Yuan, Zhujun and Lee, Chi-Cheng and others},
  journal={Science},
  volume={349},
  number={6248},
  pages={613--617},
  year={2015},
  publisher={American Association for the Advancement of Science}
}

@article{Yuan2019,
  title = {Quasiparticle interference evidence of the topological Fermi arc states in chiral fermionic semimetal CoSi},
  volume = {5},
  ISSN = {2375-2548},
  url = {http://dx.doi.org/10.1126/sciadv.aaw9485},
  DOI = {10.1126/sciadv.aaw9485},
  number = {12},
  journal = {Science Advances},
  publisher = {American Association for the Advancement of Science (AAAS)},
  author = {Yuan,  Qian-Qian and Zhou,  Liqin and Rao,  Zhi-Cheng and Tian,  Shangjie and Zhao,  Wei-Min and Xue,  Cheng-Long and Liu,  Yixuan and Zhang,  Tiantian and Tang,  Cen-Yao and Shi,  Zhi-Qiang and Jia,  Zhen-Yu and Weng,  Hongming and Ding,  Hong and Sun,  Yu-Jie and Lei,  Hechang and Li,  Shao-Chun},
  year = {2019},
  month = dec 
}

@article{gao2017zero,
  title={Zero-field magnetic response functions in Landau levels},
  author={Gao, Yang and Niu, Qian},
  journal={Proceedings of the National Academy of Sciences},
  volume={114},
  number={28},
  pages={7295--7300},
  year={2017},
  publisher={National Academy of Sciences}
}

@article{zhang2016quantum,
  title={Quantum oscillations from generic surface Fermi arcs and bulk chiral modes in Weyl semimetals},
  author={Zhang, Yi and Bulmash, Daniel and Hosur, Pavan and Potter, Andrew C and Vishwanath, Ashvin},
  journal={Scientific reports},
  volume={6},
  number={1},
  pages={23741},
  year={2016},
  publisher={Nature Publishing Group UK London}
}

@article{alexandradinata2018revealing,
  title={Revealing the topology of Fermi-surface wave functions from magnetic quantum oscillations},
  author={Alexandradinata, A and Wang, Chong and Duan, Wenhui and Glazman, Leonid},
  journal={Physical Review X},
  volume={8},
  number={1},
  pages={011027},
  year={2018},
  publisher={APS}
}

@book{vanderbilt2018berry,
  title={Berry phases in electronic structure theory: electric polarization, orbital magnetization and topological insulators},
  author={Vanderbilt, David},
  year={2018},
  publisher={Cambridge University Press}
}

@article{kaganov1983coherent,
  title={Coherent magnetic breakdown},
  author={Kaganov, Moiseui Isaakovich and Slutskin, AA},
  journal={Physics reports},
  volume={98},
  number={4},
  pages={189--271},
  year={1983},
  publisher={Elsevier}
}

@article{potter2014quantum,
  title={Quantum oscillations from surface Fermi arcs in Weyl and Dirac semimetals},
  author={Potter, Andrew C and Kimchi, Itamar and Vishwanath, Ashvin},
  journal={Nature communications},
  volume={5},
  number={1},
  pages={5161},
  year={2014},
  publisher={Nature Publishing Group UK London}
}

@article{alexandradinata2018semiclassical,
  title={Semiclassical theory of Landau levels and magnetic breakdown in topological metals},
  author={Alexandradinata, A and Glazman, Leonid},
  journal={Physical Review B},
  volume={97},
  number={14},
  pages={144422},
  year={2018},
  publisher={APS}
}

@article{glazman_rev,
   author = "Alexandradinata, A. and Glazman, Leonid",
   title = "Fermiology of Topological Metals", 
   journal= "Annual Review of Condensed Matter Physics",
   year = "2023",
   volume = "14",
   number = "Volume 14, 2023",
   pages = "261-309",
   doi = "https://doi.org/10.1146/annurev-conmatphys-040721-021331",
   url = "https://www.annualreviews.org/content/journals/10.1146/annurev-conmatphys-040721-021331",
   publisher = "Annual Reviews",
   issn = "1947-5462",
   type = "Journal Article",
  }

@article{dominguez2019aubry,
  title={The Aubry--Andr{\'e} model as a hobbyhorse for understanding the localization phenomenon},
  author={Dom{\'\i}nguez-Castro, GA and Paredes, R},
  journal={European Journal of Physics},
  volume={40},
  number={4},
  pages={045403},
  year={2019},
  publisher={IOP Publishing}
}

@article{waters2023mixed,
  title={Mixed-dimensional moir{\'e} systems of twisted graphitic thin films},
  author={Waters, Dacen and Thompson, Ellis and Arreguin-Martinez, Esmeralda and Fujimoto, Manato and Ren, Yafei and Watanabe, Kenji and Taniguchi, Takashi and Cao, Ting and Xiao, Di and Yankowitz, Matthew},
  journal={Nature},
  pages={1--6},
  year={2023},
  publisher={Nature Publishing Group UK London}
}

@article{mullan2023mixing,
  title = {Mixing of moiré-surface and bulk states in graphite},
  volume = {620},
  ISSN = {1476-4687},
  url = {http://dx.doi.org/10.1038/s41586-023-06264-5},
  DOI = {10.1038/s41586-023-06264-5},
  number = {7975},
  journal = {Nature},
  publisher = {Springer Science and Business Media LLC},
  author = {Mullan,  Ciaran and Slizovskiy,  Sergey and Yin,  Jun and Wang,  Ziwei and Yang,  Qian and Xu,  Shuigang and Yang,  Yaping and Piot,  Benjamin A. and Hu,  Sheng and Taniguchi,  Takashi and Watanabe,  Kenji and Novoselov,  Kostya S. and Geim,  A. K. and Fal’ko,  Vladimir I. and Mishchenko,  Artem},
  year = {2023},
  month = jul,
  pages = {756–761}
}

@article{Zhu_2026,
doi = {10.1088/1361-648X/ae8b67},
url = {https://doi.org/10.1088/1361-648X/ae8b67},
year = {2026},
month = {jul},
publisher = {IOP Publishing},
volume = {38},
number = {30},
pages = {303002},
author = {Zhu, Guang and Li, Renzhe and Xu, Nan},
title = {Moiré effects in low-dimensional heterostructures: from 2D materials to 2D–3D mixed-dimensional systems},
journal = {Journal of Physics: Condensed Matter}
}

@article{lifshitz1956,
  title={Theory of magnetic susceptibility in metals at low temperatures},
  author={Lifshitz, IM and Kosevich, Am M},
  journal={Sov. Phys. JETP},
  volume={2},
  number={4},
  pages={636--645},
  year={1956}
}

@book{shoenberg1984magnetic,
  title={Magnetic oscillations in metals},
  author={Shoenberg, David},
  year={1984},
  publisher={Cambridge university press}
}

@article{niu2015,
  title = {Geometrical effects in orbital magnetic susceptibility},
  author = {Gao, Yang and Yang, Shengyuan A. and Niu, Qian},
  journal = {Phys. Rev. B},
  volume = {91},
  issue = {21},
  pages = {214405},
  numpages = {12},
  year = {2015},
  month = {Jun},
  publisher = {American Physical Society},
  doi = {10.1103/PhysRevB.91.214405},
  url = {https://link.aps.org/doi/10.1103/PhysRevB.91.214405}
}

@article{Niu1996,
  title = {Berry phase, hyperorbits, and the Hofstadter spectrum: Semiclassical dynamics in magnetic Bloch bands},
  author = {Chang, Ming-Che and Niu, Qian},
  journal = {Phys. Rev. B},
  volume = {53},
  issue = {11},
  pages = {7010--7023},
  numpages = {0},
  year = {1996},
  month = {Mar},
  publisher = {American Physical Society},
  doi = {10.1103/PhysRevB.53.7010},
  url = {https://link.aps.org/doi/10.1103/PhysRevB.53.7010}
}

@article{tknn,
  title = {Quantized Hall Conductance in a Two-Dimensional Periodic Potential},
  author = {Thouless, D. J. and Kohmoto, M. and Nightingale, M. P. and den Nijs, M.},
  journal = {Phys. Rev. Lett.},
  volume = {49},
  issue = {6},
  pages = {405--408},
  numpages = {0},
  year = {1982},
  month = {Aug},
  publisher = {American Physical Society},
  doi = {10.1103/PhysRevLett.49.405},
  url = {https://link.aps.org/doi/10.1103/PhysRevLett.49.405}
}

@article{kohmoto_1989,
  title = {Zero modes and the quantized Hall conductance of the two-dimensional lattice in a magnetic field},
  author = {Kohmoto, Mahito},
  journal = {Phys. Rev. B},
  volume = {39},
  issue = {16},
  pages = {11943--11949},
  numpages = {0},
  year = {1989},
  month = {Jun},
  publisher = {American Physical Society},
  doi = {10.1103/PhysRevB.39.11943},
  url = {https://link.aps.org/doi/10.1103/PhysRevB.39.11943}
}

@book{haidemenakis2014,
  title={Physics of Solids in Intense Magnetic Fields: Lectures Presented at the First Chania Conference Held at Chania, Crete, July 16--29, 1967},
  author={Haidemenakis, Epimenides D},
  year={2014},
  publisher={Springer}
}

@incollection{stark_falicov,
title = {Chapter VI Magnetic Breakdown in Metals},
editor = {C.J. Gorter},
series = {Progress in Low Temperature Physics},
publisher = {Elsevier},
volume = {5},
pages = {235-286},
year = {1967},
issn = {0079-6417},
doi = {https://doi.org/10.1016/S0079-6417(08)60124-9},
url = {https://www.sciencedirect.com/science/article/pii/S0079641708601249},
author = {R.W. Stark and L.M. Falicov}
}

@article{mirlin_rev,
  title = {Anderson transitions},
  author = {Evers, Ferdinand and Mirlin, Alexander D.},
  journal = {Rev. Mod. Phys.},
  volume = {80},
  issue = {4},
  pages = {1355--1417},
  numpages = {0},
  year = {2008},
  month = {Oct},
  publisher = {American Physical Society},
  doi = {10.1103/RevModPhys.80.1355},
  url = {https://link.aps.org/doi/10.1103/RevModPhys.80.1355}
}

@article{lee_review,
  title = {Disordered electronic systems},
  author = {Lee, Patrick A. and Ramakrishnan, T. V.},
  journal = {Rev. Mod. Phys.},
  volume = {57},
  issue = {2},
  pages = {287--337},
  numpages = {0},
  year = {1985},
  month = {Apr},
  publisher = {American Physical Society},
  doi = {10.1103/RevModPhys.57.287},
  url = {https://link.aps.org/doi/10.1103/RevModPhys.57.287}
}

@article{Devakul2023,
  title = {Magic-angle helical trilayer graphene},
  volume = {9},
  ISSN = {2375-2548},
  url = {http://dx.doi.org/10.1126/sciadv.adi6063},
  DOI = {10.1126/sciadv.adi6063},
  number = {36},
  journal = {Science Advances},
  publisher = {American Association for the Advancement of Science (AAAS)},
  author = {Devakul,  Trithep and Ledwith,  Patrick J. and Xia,  Li-Qiao and Uri,  Aviram and de la Barrera,  Sergio C. and Jarillo-Herrero,  Pablo and Fu,  Liang},
  year = {2023},
  month = Sept 
}

@article{mora2023,
  title = {Supermoir\'e low-energy effective theory of twisted trilayer graphene},
  author = {Mao, Yuncheng and Guerci, Daniele and Mora, Christophe},
  journal = {Phys. Rev. B},
  volume = {107},
  issue = {12},
  pages = {125423},
  numpages = {16},
  year = {2023},
  month = {Mar},
  publisher = {American Physical Society},
  doi = {10.1103/PhysRevB.107.125423},
  url = {https://link.aps.org/doi/10.1103/PhysRevB.107.125423}
}

@article{kaxiras2020,
  title = {Twisted Trilayer Graphene: A Precisely Tunable Platform for Correlated Electrons},
  author = {Zhu, Ziyan and Carr, Stephen and Massatt, Daniel and Luskin, Mitchell and Kaxiras, Efthimios},
  journal = {Phys. Rev. Lett.},
  volume = {125},
  issue = {11},
  pages = {116404},
  numpages = {6},
  year = {2020},
  month = {Sep},
  publisher = {American Physical Society},
  doi = {10.1103/PhysRevLett.125.116404},
  url = {https://link.aps.org/doi/10.1103/PhysRevLett.125.116404}
}

@article{bernevig2019,
  title = {Flatbands and Perfect Metal in Trilayer Moir\'e Graphene},
  author = {Mora, Christophe and Regnault, Nicolas and Bernevig, B. Andrei},
  journal = {Phys. Rev. Lett.},
  volume = {123},
  issue = {2},
  pages = {026402},
  numpages = {6},
  year = {2019},
  month = {Jul},
  publisher = {American Physical Society},
  doi = {10.1103/PhysRevLett.123.026402},
  url = {https://link.aps.org/doi/10.1103/PhysRevLett.123.026402}
}

@article{popov2023,
  title = {Magic angles in equal-twist trilayer graphene},
  author = {Popov, Fedor K. and Tarnopolsky, Grigory},
  journal = {Phys. Rev. B},
  volume = {108},
  issue = {8},
  pages = {L081124},
  numpages = {6},
  year = {2023},
  month = {Aug},
  publisher = {American Physical Society},
  doi = {10.1103/PhysRevB.108.L081124},
  url = {https://link.aps.org/doi/10.1103/PhysRevB.108.L081124}
}

@book{landau_mechanics,
  title={Mechanics: volume 1},
  author={Landau, Lev Davidovich and Lifshitz, Evgenii Mikhailovich},
  volume={1},
  year={1982},
  publisher={Elsevier}
}

@book{ryndyk2015theory,
  title={Theory of Quantum Transport at Nanoscale: An Introduction},
  author={Ryndyk, Dmitry},
  year={2015},
  publisher={Springer}
}

@article{citro2023thouless,
  title={Thouless pumping and topology},
  author={Citro, Roberta and Aidelsburger, Monika},
  journal={Nature Reviews Physics},
  volume={5},
  number={2},
  pages={87--101},
  year={2023},
  publisher={Nature Publishing Group UK, London}
}


\newpage
\clearpage
\onecolumngrid
\appendix

\counterwithout{equation}{section}  
\setcounter{equation}{0}
\setcounter{figure}{0}
\setcounter{table}{0}

\renewcommand{\thefigure}{S\arabic{figure}}
\renewcommand{\theequation}{S\arabic{equation}}
\renewcommand{\thetable}{S\arabic{table}}

\begin{center}
\textbf{\large Supplemental Materials: Dimensional localization and topology in 3D supertwisted spirals}
\end{center}

\tableofcontents
\newpage
\section{Dimensional localization and Aubry-Andr\'e Model}
In this section, we discuss the properties and the variation of the Aubry-Andr\'e (AA) model in the supertwisted spiral system. Section \ref{app:AA-psi} discusses the local and long-ranged properties of Aubry-Andr\'e model's eigenstates. Section \ref{app:almost-loc} described the ``almost'' localization of the AA model in the case where $\beta$ is rational, corresponding to the commensurate twist angle in twisted spiral. Section \ref{app:AA-SOC} generalizes the AA model into the system with spin-orbit coupling (SOC) along with the symmetry characterization and the effect of SOC on the localization.

\subsection{Wavefunction Localization Length in Aubry-Andr\'e model}\label{app:AA-psi}

Let us comment on the localization length $\xi$ of the wavefunctions for $k>k_c$. In the AA model $\xi$ is known to be exactly given by  $\xi = \log^{-1} \frac{\Delta}{2t} = \log^{-1} \frac{|\kk|}{k_c}/2$ \cite{aubry1980analyticity} (in units of interlayer spacing $a$), quickly approaching values below $1$ for $k>k_c$. We note that $\xi$ refers here to the decay exponent of the wavefunction for distances much larger then the period of the spiral and does not imply that the wavefunction is localized within one layer. For example, for $\beta\ll1$, the wavefunction can be shown to localize within one layer only at  $|\kk| \gtrsim k_{c, 2} \sim {k_c}/(\pi\beta) $ as discussed below.

We note also that for $k<k_c$ while all eigenstates are extended and the system does not have a rigorous mobility gap, some precursors of localization occur. Specifically, the wavefunctions with energies $|E|>2|t|-|\Delta(|\kxy|)|$ have been shown to be ``almost localized" in the $\beta \ll 1$ limit \cite{zhang2015almost,wilkinson1984,malla2018spinful}. They have large effective mass along the $z$ axis and would be prone to localization due to nonzero long-range hopping, higher-harmonic anisotropy, and disorder \cite{vu2023generic}. The case of a rational $\beta = p/\lambda$, where $p$ and $\lambda$ are co-prime integers is not very different. $\epsilon_l$ acts as a periodic potential with a period of $\lambda$. At finite $\beta$, the anisotropy $\Delta$ will open up the anti-crossing gap and suppress the bandwidth exponentially leading to the appearance of almost localized states \cite{zhang2015almost}; for $k>k_c$ all states will become almost localized in this case.

The extent of the wavefunction within a single period of the  spiral is determined by the ratio between $t$ and the variation of $\epsilon_l$  between consecutive layers. Using a harmonic expansion \cite{zhang2015almost} in $l$ near the minimum of $\epsilon_l$ in the small twist limit $\beta \ll 1$ one obtains $\psi_{{\kxy},l} \sim e^{- 
     {(l - l_0)^2}/{2l_{\rm HO}^2}}/l_{\rm HO}^{1/2}$
where $l_{\rm HO} = (\pi \beta |\kk|/k_c)^{-1/2}$ describes the characteristic length of the harmonic oscillator. From this expression, one notices that the wavefunctions localize in a single layer only for $|\kk| \gtrsim k_{c, 2} \sim {k_c}/(\pi\beta) $ (for $\beta\ll1$). As a result, in addition to the actual transition (in the $L\to \infty$ limit) at $k=k_c$, we expect a crossover to a single-layer localization at $k\approx k_{c,2}$. For $\beta \sim 1$, this crossover will occur close to the localization transition, but for $\beta \ll 1$ there will be an extended range of $k$, where the wavefunctions, while being localized, extend over multiple layers.

\begin{figure}[h]
\centering
\includegraphics[width=0.9\linewidth]{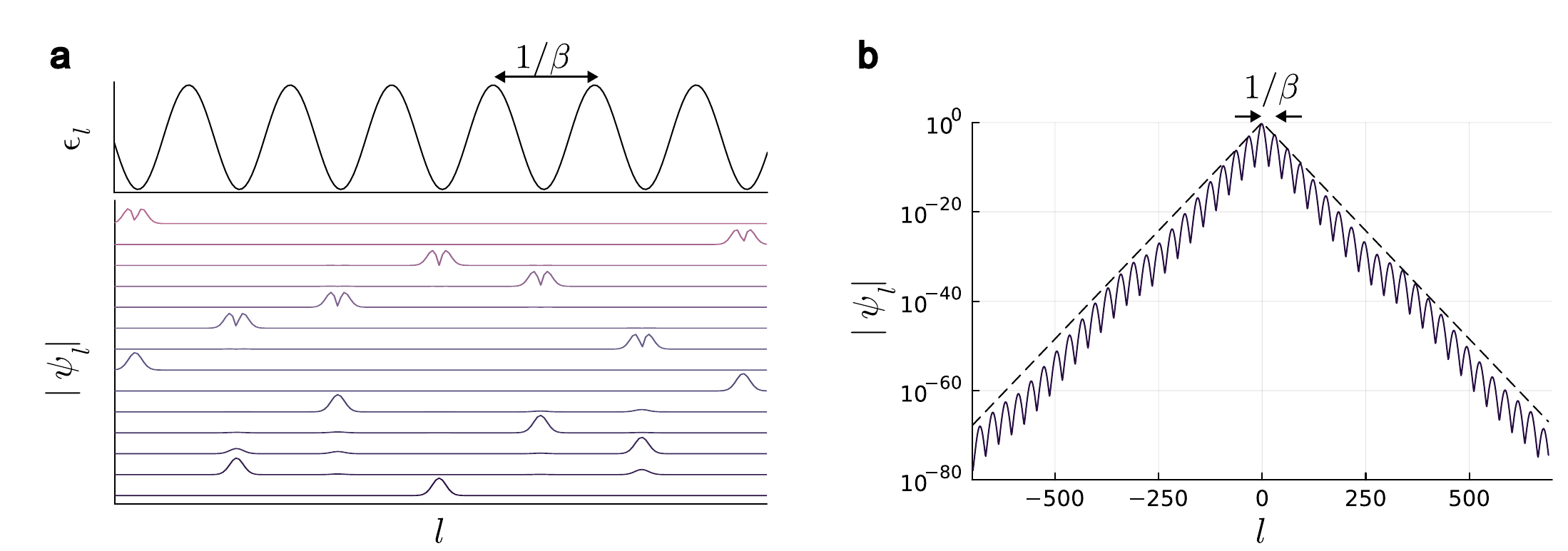}
    \caption{{\emph{Localization in Aubry-Andr\'e model at small $\beta$.}} (a) Schematics of the eigenstates of AA model in the localized phase $\Delta = 2.5 t$. (top) the layer-dependent potential $\epsilon_l$ as a function of layer $l$. (bottom) The wavefunction eigenstates $|\psi_l|$ as a function of layer number $l$ (shifted for clarity). We can see that the eigenstates forms bound states near the minima of $\epsilon_l$ which locally resembled a harmonic oscillator in a small $\beta$ limit. (b) The decay of the wavefunction $|\psi_l|$ at spectra minima when $\Delta = 2.5 t$ and $\beta = (1 + \sqrt{5})/100$. Here, the local form of the wavefunction shows $\beta, \Delta$-dependent feature, while the long-range features show exponential decays defines from $\langle \psi_{j} \rangle \sim \exp(- l / \xi)$ (dashed line) where $1/\xi = \log (2\Delta/t)$.}
    \label{fig:wave-fn-supp}
\end{figure}

To support this assertion, Fig. \ref{fig:wave-fn-supp} (a) shows the wavefunction amplitude $|\psi_l|$ as a function of layer $l$. We can see that the wavefunction forms bound states near each minima of $\epsilon_l$ and can be locally approximated by the harmonic oscillator solution $\psi_l \sim \exp(-l^2/2l_{\rm HO}^2)$. Fig. \ref{fig:wave-fn-supp} (b) shows the wavefunction in the length scale much larger than the oscillation period of the potential. We can see that average occupation for each period decays exponentially with an envelope function $ \langle \psi_{j} \rangle \sim \exp(-l/\xi)$, where $\langle ... \rangle$ defines averaging over the period $1/\beta$. 

\subsection{``Almost" localization in Aubry-Andr\'e model with rational $\beta$}\label{app:almost-loc}
\begin{figure}[h]
    \centering
    \includegraphics[width=0.6\linewidth]{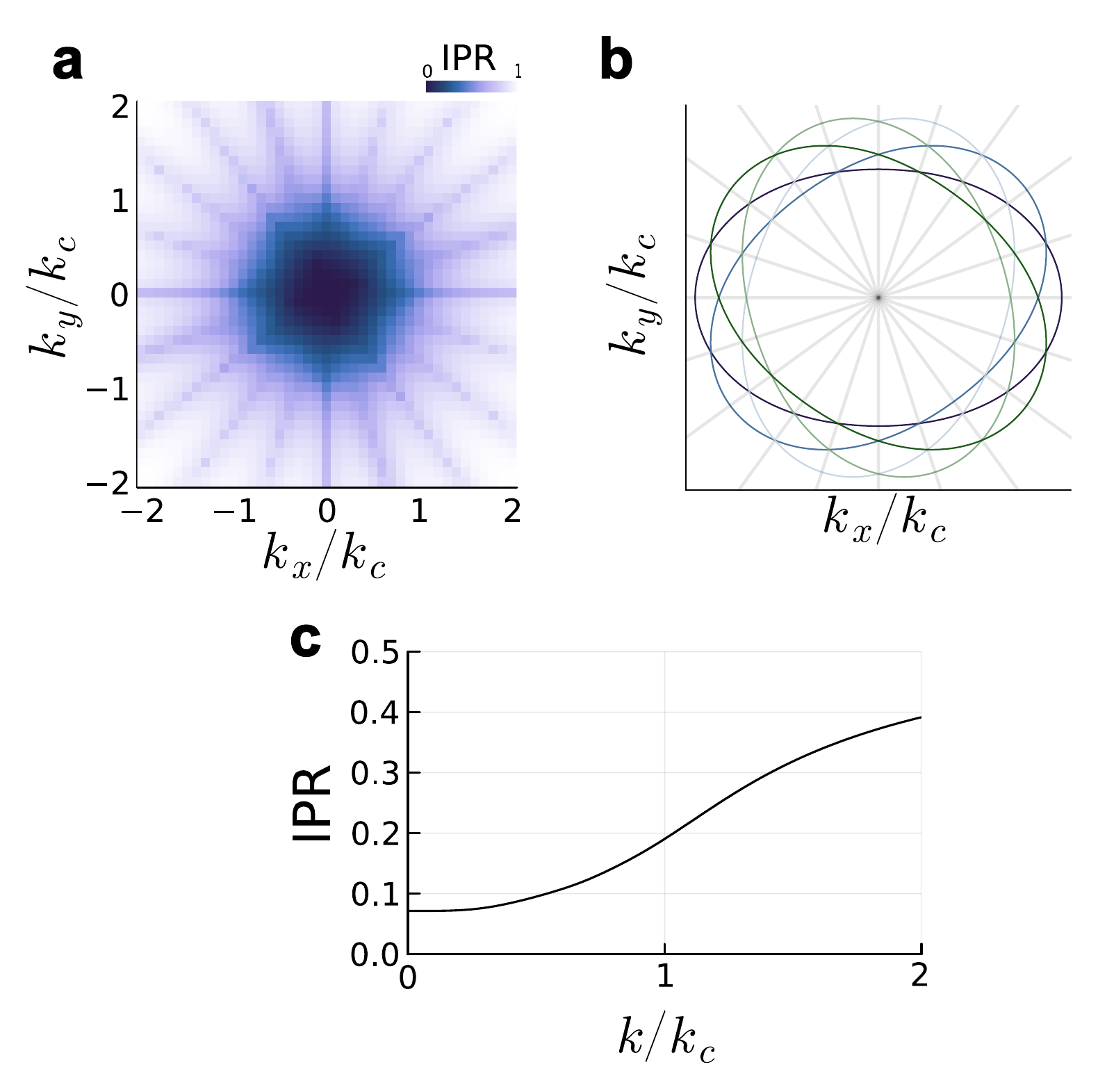}
    \caption{{\emph{``Almost'' Localization at rational $\beta$}.} (a) IPR as a function of $k_x$ and $k_y$ at $L = 20$, $\theta = \pi/5$, and $\varepsilon = 0.1$. The IPR shows weak crossover from extended state (IPR $\sim 1/L$) to localized state (IPR$\sim 1$) for most momenta except at the high-symmetry line originating from the overlap of the Fermi surface shown in (b). (c) average IPR as a function of $k$ at $L = 20$, $\theta = \pi/5$, and $\varepsilon = 0.1$. The average IPR shows a weaker crossover from localized to extended states than the irrational $\beta$. Here the average IPR saturate at IPR$\sim 0.4$ corresponding to electron localizing in $1/0.4 =2.5$ layers across $L = 20$ system.}
    \label{fig:IPR_commen_supp}
\end{figure}

Here, we consider the AA model localization in the rational $\beta$ limit. Fig. \ref{fig:IPR_commen_supp} (a) shows IPR averaging over all the eigenstates at $\kk$. While true localization is not expected in this case, we observe the bandwidth suppression as $|\kk|$ increases and the support of the wavefunction becomes ``almost" localized which can be seen by a crossover of IPR in Fig. \ref{fig:IPR_commen_supp} (c). In addition, we can see that the IPR is not rotationally symmetric but has maxima at high symmetry lines. These lines coincide with the ones where some of the twisted Fermi surfaces overlap, shown in Fig. \ref{fig:IPR_commen_supp} (b), resulting in stronger delocalization along the spiral axis.  

\clearpage
\subsection{Spin-Orbit effect on Aubry-Andr\'e mapping}\label{app:AA-SOC}
We consider the effect of spin-orbit coupling (SOC) and how it affects the localization transition. Under SOC, the modified eigen equation can be written as
\begin{align}
\label{eqn:SOC}
     E_z \psi_{\kk l} &= (\epsilon_l + \epsilon^{\rm SOC}_l) \psi_{\kk l} + (t + i (\vb{t}_1)_{l\theta} \cdot \bm{\sigma}) \psi_{\kk l+1} +  (t - i (\vb{t}_1)_{l\theta} \cdot \bm{\sigma})\psi_{\kk l-1}
    ,
    \\
    \epsilon^{\rm SOC}_l &= \sum_{a = x, y} \sum_{b = x, y, z}(\lambda_{ab})_{l\theta} k_{a} \sigma_b
    ,
\end{align}
where $\lambda$  and $\vb{t}_1$ are the in-plane and out-of-plane spin-orbit coupling (SOC) respectively. $(...)_{l\theta}$ represents the $l\theta$ rotation of the vector $\vb{t}_1$ and tensor $\lambda$. For system with rotational symmetry $C_n$ for $n > 2$, the layer dependence on the SOC terms vanish as we can simplify $\vb{t}_1 = (0, 0, t_1)$ and $\epsilon^{\rm SOC}_l = \alpha (k_x \sigma_x + k_y \sigma_y) + \beta (k_x \sigma_y - k_y \sigma_x)$. At $\kk = (k_x, k_y)$, we define $h_x = \alpha k_x - \beta k_y$ and $h_y = \alpha k_y + \beta k_x$ such that $\epsilon^{\rm SOC}_l = h_x \sigma_x + h_y \sigma_y$. We can define the basis 
\begin{align}
    \psi_{\kk l}^{\pm} &= \frac{1}{\sqrt{2}} (\ket{\uparrow} \pm e^{i\nu} \ket{\downarrow}),
\end{align}
where $e^{i\nu} = (h_x + ih_y)/\sqrt{h_x^2 + h_y^2}$, and rewrite the eigenvalue equation as
\begin{align}
    E_z \begin{pmatrix}
        \psi_{\kk l}^+\\
        \psi_{\kk l}^-
    \end{pmatrix}
    &=
    (\epsilon_l + \sqrt{h_x^2 + h_y^2} \sigma_z) 
    \begin{pmatrix}
        \psi_{\kk l}^+\\
        \psi_{\kk l}^-
    \end{pmatrix}
    + 
    \begin{pmatrix}
        t & i t_1 \\
        i t_1 & t
    \end{pmatrix}
    \begin{pmatrix}
        \psi_{\kk l+1}^+\\
        \psi_{\kk l+1}^-
    \end{pmatrix}
    +
     \begin{pmatrix}
        t & -i t_1 \\
        -i t_1 & t
    \end{pmatrix}
    \begin{pmatrix}
        \psi_{\kk l-1}^+\\
        \psi_{\kk l-1}^-
    \end{pmatrix}.
\end{align}
We can see that $\sqrt{h_x^2 + h_y^2}$ acts as a Zeeman splitting where $t_1$ acts as a SOC tunneling.  This model has been studied in the $\beta \ll 1$ limit \cite{malla2018spinful} and show localization suppression for states that are close in energy with the Zeeman splitting. For $|t_1| \ll \sqrt{h_x^2 + h_y^2} $ limit, the system decouple into two AA models with energy shift $E_z \pm \sqrt{h_x^2 + h_y^2}$ with the energy correction from $t_1$. For localized state, $t_1$ will introduce virtual hopping between layer with the energy correction of order  $|t_1|^2/(\delta\epsilon)$ where $\delta\epsilon$ is the energy difference between neighbor layers at $\kk$. For extended state, $t_1$ would couple the state of opposite spin in momentum space with the energy correction of order $|t_1|^2/\sqrt{h_x^2 + h_y^2}$.

\clearpage
\section{Details for conductance calculations at zero field }
This section outlines the calculation and additional results regarding the transport in AA model. We overview the transfer matrix method in \ref{supp:conductivity} which is used to calculate the conductance of the system. Additional transport results are also shown here. Section \ref{supp:param-transport}, we show the conductance dependence on $t$ while section \ref{supp:transport-rational-b} highlight the conductance near rational $\beta$.

\subsection{Transfer Matrix and Transmission Eigenvalue Calculation}\label{supp:conductivity}
Here, we show the detail calculation of the transmission eigenvalue. We start by defining the transfer matrix
\begin{align}
    \begin{pmatrix}
        \psi_{L+} \\
        \psi_{L-}
    \end{pmatrix}
     = 
    Q
      \begin{pmatrix}
        \psi_{0+} \\
        \psi_{0-}
    \end{pmatrix},
\end{align}
To calculate $Q$, we can rewrite Eq. \ref{eqn:single-channel-TB} as
\begin{align}
    \begin{pmatrix}
        \psi_{l+1}\\
        \psi_{l}
    \end{pmatrix}
    &=
   M_l
     \begin{pmatrix}
        \psi_{l}\\
        \psi_{l-1}
    \end{pmatrix}
    ,\\
    M_l &=
     \begin{pmatrix}
        (E_z - \epsilon_l)/t & -1\\
        1 & 0
    \end{pmatrix}.
\end{align}
Then, we can relate the wavefunction at $0$ and $L$ by 
\begin{align}
    \begin{pmatrix}
        \psi_{L+1} \\
        \psi_{L}
    \end{pmatrix}
    &= 
    M \begin{pmatrix}
        \psi_{1} \\
        \psi_{0}
    \end{pmatrix}
    ,\\
    M &= M_L M_{L-1} ... M_2 M_1
    .
\end{align}
We can write the right-moving and left-moving wavefunction in the form of the wavefunction at site $l$ as 
and $l-1$ using 
\begin{align}
    \begin{pmatrix}
        \psi_{L+1} \\ \psi_{L}
    \end{pmatrix}
    &= \begin{pmatrix}
        e^{i k_L (L+1)} & e^{-i k_L (L+1)}
        \\
        e^{i k_L L} & e^{-i k_L L}
    \end{pmatrix}
    \begin{pmatrix}
        \psi_{L+} \\ \psi_{L-}
    \end{pmatrix},
    \\
    \begin{pmatrix}
        \psi_{1} \\ \psi_{0}
    \end{pmatrix}
    &= \begin{pmatrix}
        e^{i k_0} & e^{-i k_0}
        \\
        1 & 1
    \end{pmatrix}
    \begin{pmatrix}
        \psi_{0+} \\ \psi_{0-}
    \end{pmatrix},
\end{align}
where $k_0$ is the wavevector of the incoming state at $l = 0$, and $k_L$ is the wavevector of the outgoing state at $l = L$. The transfer matrix $Q$, can then be calculated as 
\begin{align}
    Q = \begin{pmatrix}
        e^{i k_L (L+1)} & e^{-i k_L (L+1)}
        \\
        e^{i k_L L} & e^{-i k_L L}
    \end{pmatrix}^{-1} M \begin{pmatrix}
        e^{i k_0} & e^{-i k_0}
        \\
        1 & 1
    \end{pmatrix}
    .
\end{align}
Given the dispersion of the left and right lead $E_{\rm lead, left}, E_{\rm lead, right}$, the wavevector can be calculated from the relation
\begin{align}
    E_{\rm lead, left}(k_0, \kk) =  E_{\rm lead, right}(k_L, \kk) = E_F.
\end{align}
For normal metal leads with Fermi momentum $k_F$, the dispersion can be written as 
\begin{align}
    E_{\rm metal, left}(k_z, \kk) = E_{\rm metal, right}(k_z, \kk) = \frac{|\kxy|^2}{2M} + \frac{k_z^2}{2M} - \frac{k_F^2}{2M},
\end{align}
where $M$ is the electron effective mass in the metal lead. The incoming and the outgoing wavevectors are $k_L = k_0 = \sqrt{k_F^2 - |\kxy|^2 + 2ME_F}$. Here, we note that the majority of the conductivity contribution are from $|\kxy| \lesssim k_c$. Assuming that the fermi surface is large, i.e. $ |E_F|, k_c^2/2M \ll  k_F^2/2M$, we can approximate $k_0 = k_L \approx k_F$. Alternatively, we can consider the boundary condition where the leads are the untwisted version of the system and the dispersion can be written as 
\begin{align}
    E_{\rm untwist, left}(k_0, \kk) &= \frac{|\kxy|^2}{2\mu} (1 + \varepsilon \cos(2\phi)) + 2t\cos k_0
    \\
     E_{\rm untwist, right}(k_L, \kk) &= \frac{|\kxy|^2}{2\mu} (1 + \varepsilon \cos(2 L\theta + 2\phi)) + 2t\cos k_L
\end{align}
\begin{figure}\centering\includegraphics[width=0.8\linewidth]{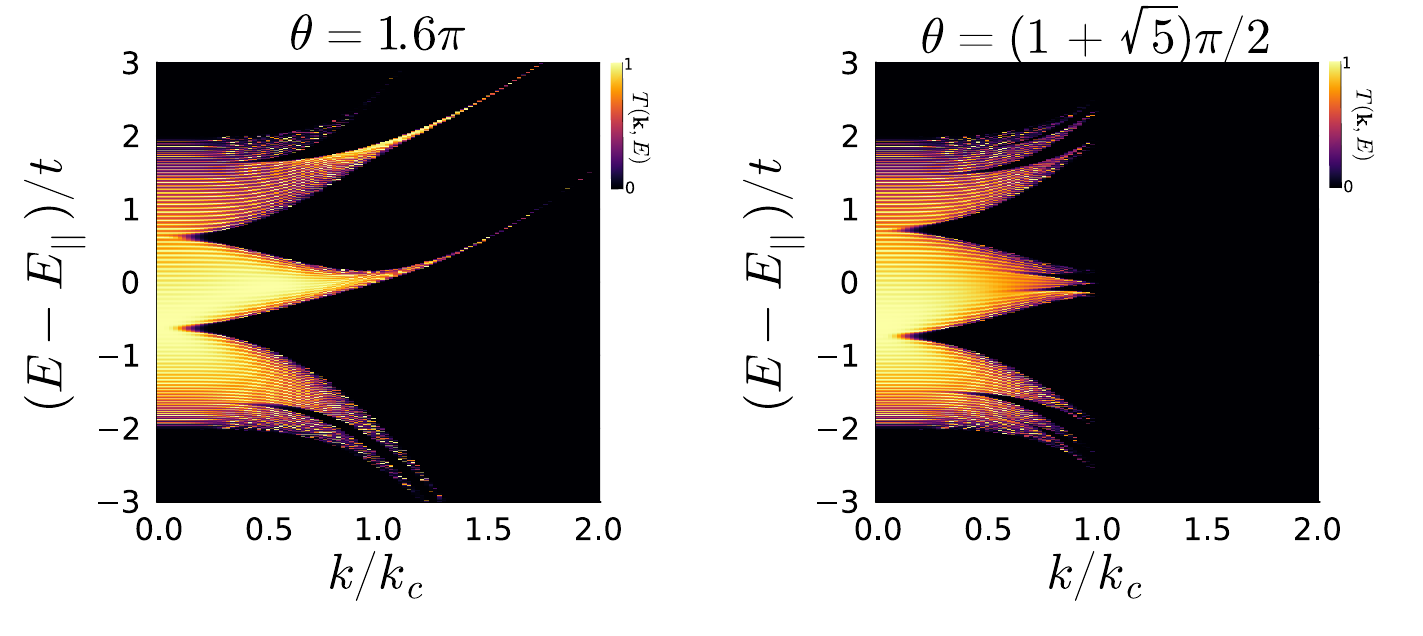}
    \caption{{\emph{Transmission map}.} $T(\kk, E)$ as a function of normalized doping $(E_F - E_{||})/t$ and normalized momentum $k/k_c$ at $t = 0.03$, $\varepsilon = 0.1$, $\phi = 0$, and $L = 100$. At rational $\beta$ (left), the transmission is finite within the band and exponentially decays outside. At irrational $\beta$ (right), the transmission shows similar behavior but is also highly suppressed in the localized regime $k > k_c$. Here, we note the fringes pattern which are finite-size effect associating with the energy resonating with the electronics state energy. At $\kk = \vb{0}$, the fringes corresponds to $E_z = 2t \cos (2\pi n /L)$ where $n = 0, 1, ...L-1$ and can be understood via Fabry–P\'erot interference.}
    \label{fig:transmission}
\end{figure}
To calculate the transmission eigenvalue from the transfer matrix, we consider the relation \cite{markos2008wave}
\begin{align}
    Q =  \begin{pmatrix}
        {1}/{S_{10}^*} & {S_{11}}/{S_{01}} \\
        -{S_{00}}/{S_{01}} & {1}/{S_{01}}
    \end{pmatrix},
\end{align}
where $S_{01} = S_{10}^*$ due to time-reversal symmetry. We can see that 
\begin{align}
    \Tr Q^\dagger Q = \frac{2}{|S_{10}|^2}
    + \frac{|S_{00}|^2}{|S_{10}|^2} + \frac{|S_{11}|^2}{|S_{10}|^2}.
\end{align}
Due to conservation of current, we can write $|S_{00}|^2 = |S_{11}|^2 = 1 - |S_{10}|^2$. Hence, the transmission coefficient $T$ can be written as 
\begin{align}
    T = \frac{4}{2 + \Tr Q^\dagger Q}.
\end{align}
The transmission in $L \gg 1$ limit is plotted in Fig. \ref{fig:transmission} as a function of energy $E_z$ and in-plane momentum $|\kk|$ where we can compare the transmission at rational and irrational $\beta$. For rational $\beta$, we can observe the bandwidth suppression i.e. ``almost localization" where as $|\kk|$ increase, the bandwidth with finite transmission exponentially shrunk in size. On the other hand, the irrational $\beta$ shows a hard localization cutoff at $|\kk| = k_c$ where the transmission drops to zero.

\clearpage

\subsection{Parameter Dependence in Universal Decays}\label{supp:param-transport}
\begin{figure}[h]
    \centering
\includegraphics[width=0.6\linewidth]{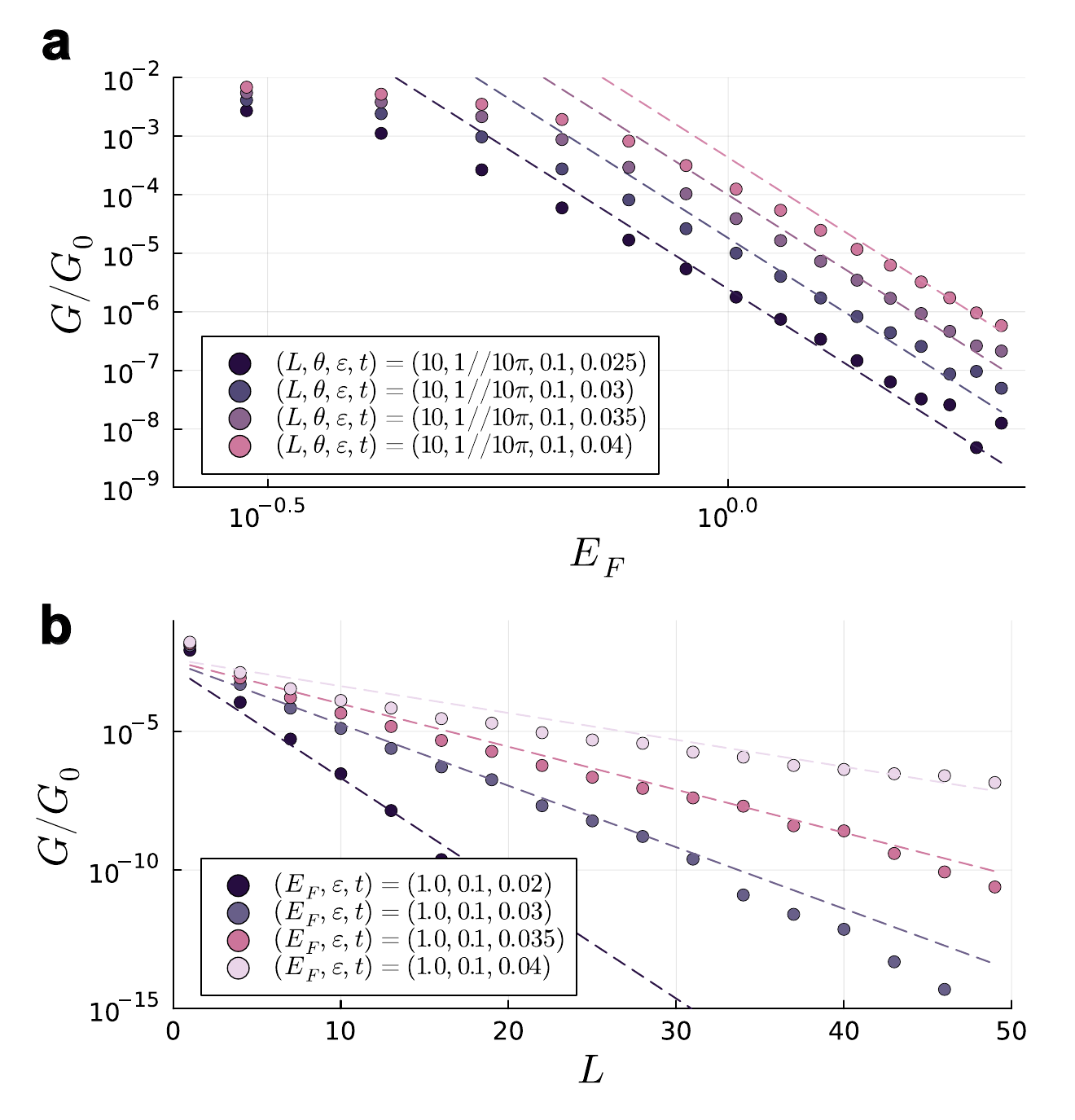}
    \caption{{\emph{Parameter dependence in universal conductance}.} (a) Normalized conductivity $G/G_0$ as a function of doping $E_F$ at different interlayer hopping. (b) Normalized conductivity $G/G_0$ as a function of system size $L$ at $\theta = (1 + \sqrt{5})\pi/2$ and different interlayer hopping. The dashed line indicate the corresponding empirical relation $G = 0.8 (e^2/h)\mathcal{N}t (E_c/E_F)^{L}$.}
\label{fig:universal-params}
\end{figure}

We show conductance suppression for rational $\beta$ at different $t$ in Fig.  \ref{fig:universal-params} (a); the plot  shows good fit to the universal conductance suppression \eqref{eq:univ_g} despite the small system size and periodicity. Fig. \ref{fig:universal-params} (b) shows conductance $G$ as a function of system size $L$ for incommensurate angle $\theta = (1 + \sqrt{5})/2\pi \approx 69 ^\circ$ at different $t$ which shows a good agreement to universal conductance suppression (\ref{eq:univ_g}).
\clearpage

\subsection{Conductance near rational $\beta$}
\label{supp:transport-rational-b}

Here, we consider the fragile behavior of conductivity near rational $\beta$. Consider an ``almost" rational angle $\theta = \theta_0 + \delta$, where $\theta_0$ corresponds to a rational $\beta_0 = 1/2, 1/3$ etc., and $\delta$ is small. For $L\delta \ll \pi$, the conductance feature resembles the feature at $\theta_0$ with plateaus at $L \gtrsim \lambda_0$. However, as $L$ increases, the twist mismatch also increases, leading to a suppression of conductance at finite $\delta$. The curve approaches Eq. \eqref{eq:univ_g} when $L\delta \gtrsim \pi$. We note that if $ \theta_0 + \delta$ is also commensurate, but with a longer period $\lambda$, we expect the conductance to saturate for $L\gtrsim \lambda$, albeit at a much smaller value.

\begin{figure}[h]
    \centering
\includegraphics[width=0.6\linewidth]{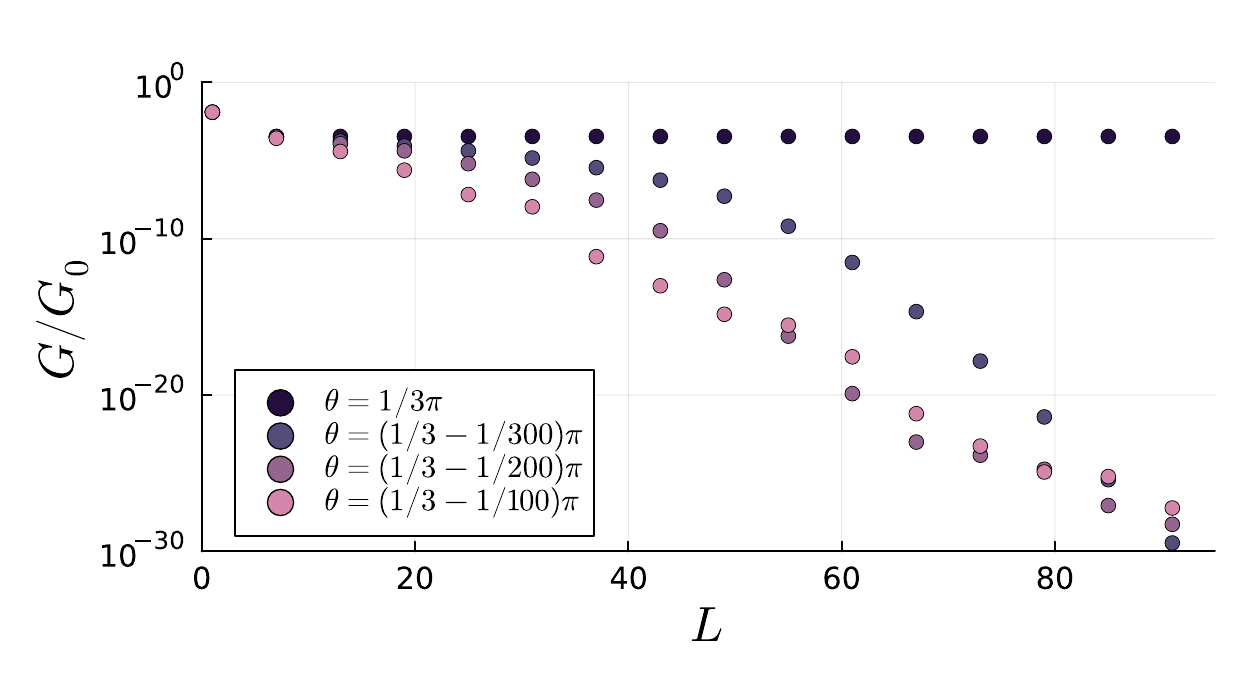}
    \caption{{\emph Conductance near the commensurate peak.} Normalized conductance $G/G_0$ as a function of $L$ at $\theta$ close to the prominent peaks at rational $\beta$. The figures are calculated at $\varepsilon = 0.1$, $t = 0.03$, and $E_F = 1.0$.}
    \label{fig:placeholder}
\end{figure}

\clearpage
\section{Surface states and their quantum oscillation signatures}
In this section, we discussed the surface state at a commensurate twist in \ref{supp:surface-commen-twist}. The effect of SOC on the surface states are shown in \ref{supp:SOC-surface}. We proposed the experimental signal from the surface states (\ref{sec:QO}) along with the limitation due to the bulk signal (\ref{app:bulk-oscillation}) and the breakdown of quasi-classical approximation (\ref{supp:classical-condition}).

\subsection{Surface state at commensurate twist}
\label{supp:surface-commen-twist}
\begin{figure}[h]
    \centering
    \includegraphics[width=0.8\linewidth]{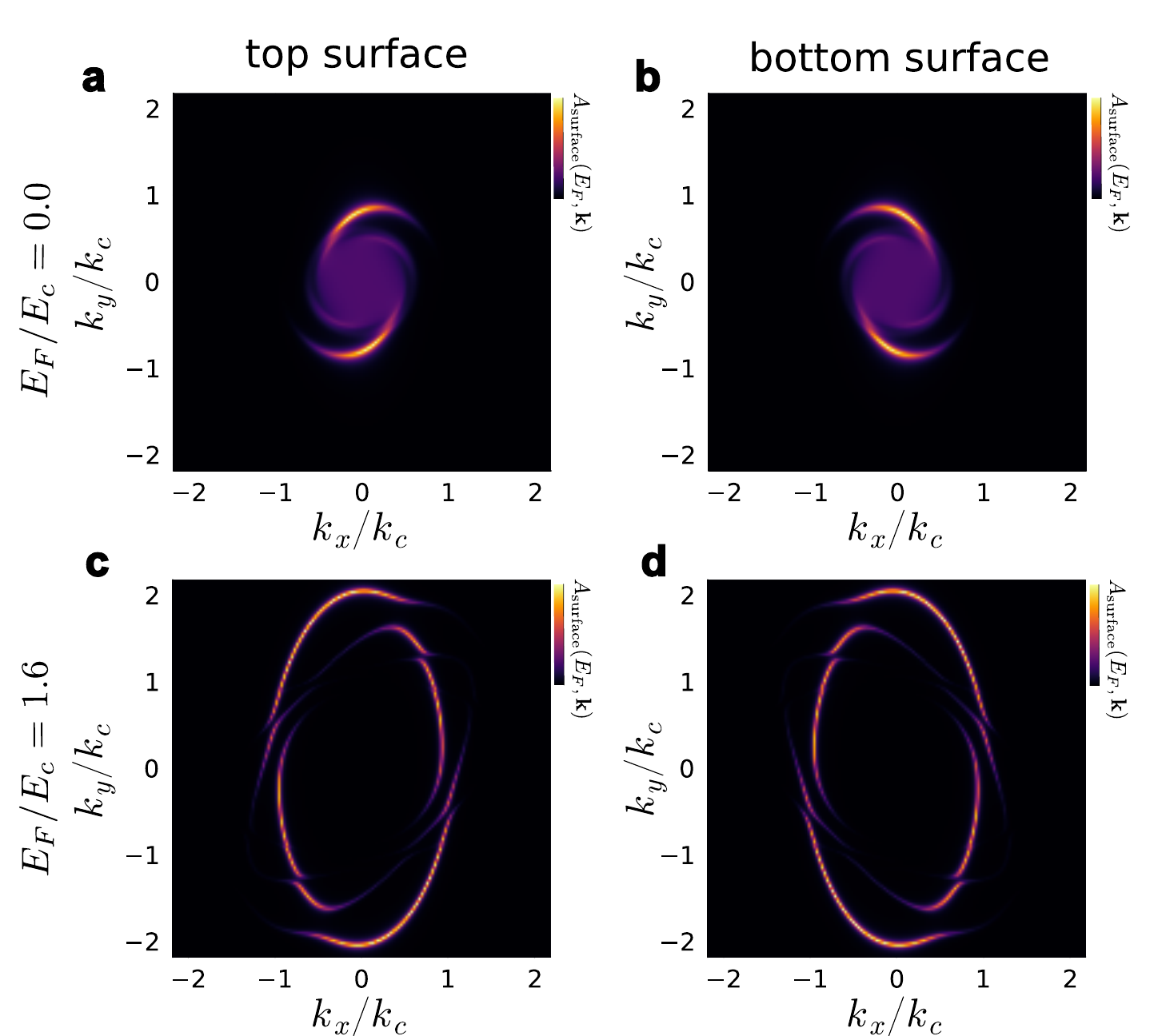}
      \caption{{\emph{Surface states in rational $\beta$.}} (a-b) Surface spectral function $A(\kk, E_F)$ as a function of $\kk$ for the localized phase $E_F/E_c = 0.0$ (a, b) and $E_F/E_c = 1.6$ (c, d) at top and bottom layer. The calculation is done at $\varepsilon = 0.6, \eta = 0.02, L = 103, \theta = 30^\circ$.}
    \label{fig:supp-surface-commen}
\end{figure}

We show the surface states of the system with commensurate twist $\theta = 30^\circ$ in Fig. \ref{fig:supp-surface-commen}. We observe qualitatively similar features to the one with incommensurate twist, indicating the robustness of the spectrum topology across the twist angle. 

\clearpage
\subsection{Spin-polarized surface state}
\label{supp:SOC-surface}
\begin{figure}[h]
    \centering
    \includegraphics[width=0.6\linewidth]{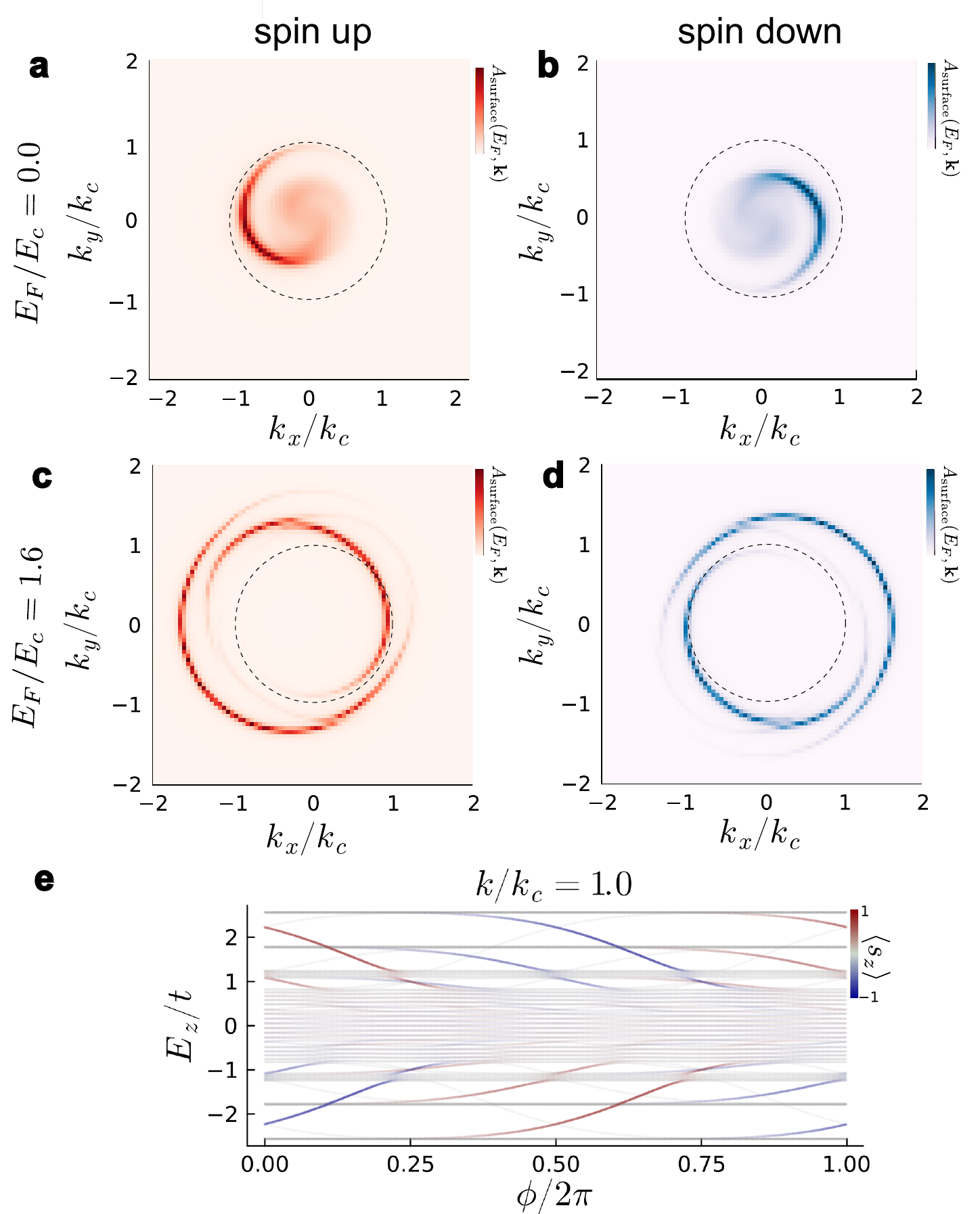}
    \caption{{\emph{Spin-polarized surface states.}} (a-b) The top layer surface spectral function $A(\kk, E_F)$ as a function of $\kk$ for (a, b) $E_F/E_c = 0.0$ (c, d) and $E_F/E_c = 1.6$ for (a, c) spin up and (b, d) spin down electrons. The calculation is done at $\varepsilon = 0.6, \eta = 0.02, L = 103, \theta = (1+\sqrt{5})/15 \pi \approx 0.21 \pi \approx 39^\circ$, and $\lambda = 1.0$. (e) $A(\kk, E_F)$ taking on a contour $k/k_c = 1.0$ (dashed line) as a function of $\phi = \tan^{-1} (k_y/k_x)$. The color depicted the spin of the surface states. The bulk states and bottom surface states are labeled by gray lines.}
    \label{fig:spin-surface}
\end{figure}

We consider the effect of spin-orbit coupling to the surface state by considering Hamiltonian from Eq. \ref{eqn:SOC} with $\epsilon_l = 0$, and $\lambda_{ab} = 0$ for all $(a, b) \neq (x, z)$:  
\begin{align}
    E_z \psi_{\kk l} = \lambda_{xz} (k_x)_{l\theta} \sigma_z \psi_{\kk l} + t \psi_{\kk l+1} + t \psi_{\kk l-1}. 
\end{align}
This Hamiltonian can be diagonalized into two independent AA model with $\beta = \theta/2\pi$, $\Delta = \pm \lambda_{xz} |\kk|$, and $\varphi = \phi$. For each spin, the model contains the localization transition at $
    k_c = \frac{2t}{\lambda_{xz}}$ and $ 
    E_c = \frac{2t^2}{m\lambda_{xz}^2}.$ Fig. \ref{fig:spin-surface} shows the surface state at the top layer for spin up and spin down component. We observe that instead of having $C_2$ symmetry like the symmetric counterparts, the surface state splits into spin up and spin down component related by time reversal symmetry instead. The spin splitting allows the surface state at low doping (a, b) to be completely separate and protected from back scattering by time-reversal symmetry. 

\newpage
\subsection{Experimental signatures of surface Fermi arcs in quantum oscillations} 
\label{sec:QO}

\begin{figure}[b]
    \centering \includegraphics[width=0.5\linewidth]{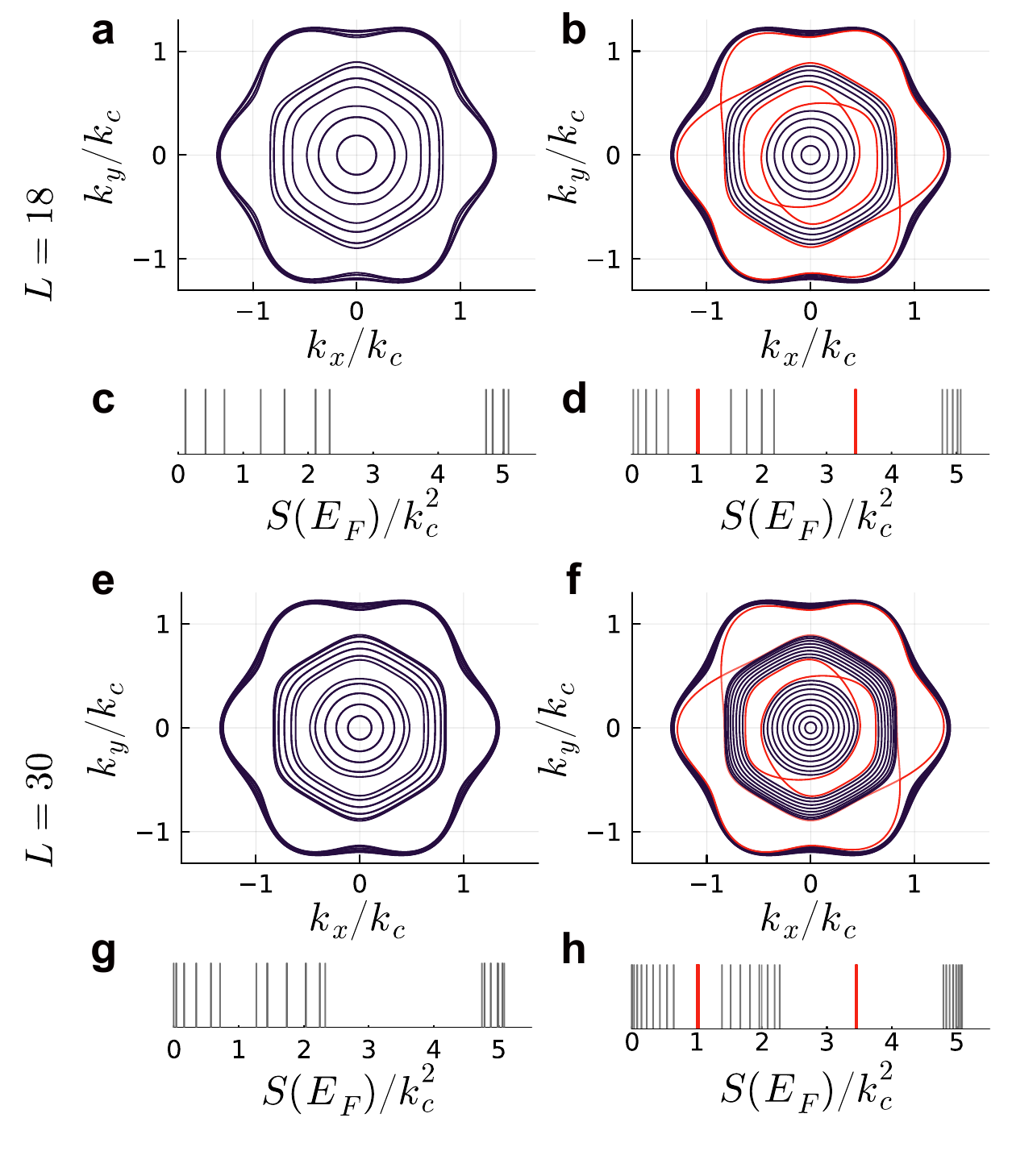}\caption{{\emph{Equal-energy orbits}} (a,b) The equi-energy surfaces of the twisted spiral at  commensurate twist $\theta = \pi/3 = 60^\circ$, $E_F/E_c = 0.6$, $L = 18$, $\varepsilon = 0.6$ with (a) open and (b) periodic boundary condition. (c, d) The area in momentum space $S(E_F)$ of the equal-energy orbits defined in (a, b). The red lines mark the surface state. (e-h) depicts the same plot with (a-d) but for a larger system $L = 30$.
    }
    \label{fig:equienergy}
\end{figure}

In the semiclassical picture \cite{lifshitz1956,shoenberg1984magnetic}, magnetic oscillations occur with minima $B_n$ related by \cite{gao2017zero}
\begin{align}
\frac{1}{B_{n}} - \frac{1}{B_{n+1}} = \frac{2\pi e}{\hbar c}\frac{1}{S(E)},
\label{eq:LK0}
\end{align}
where $S(E)$ is the area defined by the equal-energy orbits in ${\bf k}$-space. To address the influence of the Fermi arc states on the quantum oscillations, we present the equal-energy contours for finite-layer number spirals in Fig. \ref{fig:equienergy}, along with the corresponding $S(E)$ for a commensurate twist angle $\theta = \pi/3$.

The contours corresponding to the bulk and surface (Fermi arc) states are conveniently separated by comparing system with open  (Fig. \ref{fig:equienergy} (b,d,f,h)) and periodic (Fig. \ref{fig:equienergy} (a,c,e,g)) boundary conditions along the $z$ axis. The orbits corresponding to the cross-section of the bulk Fermi surfaces are shown in purple. For open boundary conditions, the contours arising from surface state (red), traverse the space between different bands. Moreover, different parts of this closed contour can be seen to be localized on different system surfaces. 
Such contour can be understood semi-classically as the electron orbiting at the top layer, traversing the bulk to the bottom layer, continuing the orbit at the bottom layer, and traveling back to the top layer via the bulk. Unlike Weyl semimetals, where bulk Fermi surface is reduced to a point, in spirals the bulk can have a finite Fermi surface and the Fermi arcs from top and bottom ``connect" via these bulk states. As a result, there emerges a quantum oscillation frequency from the surface state (d) not observed from the bulk (c). To confirm the robustness across system size, (e-h) show the surface state orbit at a larger system size persists while the density of the bulk orbit increases. We note that distinct Fermi-arc-induced orbits can be seen to cross; however, as at these momenta the wavefunctions are localized at opposite surfaces for the two orbits the hybridization is strongly (exponentially with thickness) suppressed and is not resolved in Fig. \ref{fig:equienergy}.

There are two limitations for observing the quantum oscillation signatures of the Fermi arcs. First, the oscillations from the bulk states (unlike in Weyl semimetals) may be expected to produce a stronger, extensive in $L$ signal at large $L$. We note however, that in the bulk $L\gg1$ limit, only the extremal orbits contribute for each bulk band, leading to a reduction of intensity from bulk oscillations $\sim\sqrt{B}$ at low fields (for a detailed estimate see Sec. \ref{app:bulk-oscillation}).

More importantly, the transition between localized and extended states along the red contour becomes very abrupt in the $L\gg 1$ limit. In this case, one expects the semiclassical picture to break down due to finite velocity of electrons, as they will not be able to traverse $L$ layers within the same time $\sim\omega_c^{-1}$, $\omega_c$ being the cyclotron frequency $\frac{e B\sqrt{1-\varepsilon^2} }{m c}$. More formally, the rapid change of the wavefunction along a semiclassical trajectory lead to the wavefunction-related corrections to quantum oscillation frequencies \cite{niu2015,gao2017zero}. Our estimates of these corrections (see \ref{app:bulk-oscillation}) indicate the validity of semiclassical approximation, Eq. \eqref{eq:LK0} to be limited to $\hbar \omega_c \ll t \frac{l_{loc}}{L}$. This introduces a velocity limit for electrons $v\sim L \omega_c \ll \frac{t}{\hbar} l_{loc}$, reducing to the interlayer hopping limit in the strongly localized case $l_{loc}\sim 1$.

\subsection{Estimates for bulk quantum oscillations in supertwisted spirals}\label{app:bulk-oscillation}

The lowest-harmonic oscillatory correction to the grand potential in a 3D system takes the form ((2.100) in  \cite{shoenberg1984magnetic}):
\begin{equation}
\delta \Omega \propto \int d k_z \cos\left(\frac{\hbar c S(k_z)}{e B} -1/2\right).    
\end{equation}
Therefore, the relative strength of the quantum oscillations produced by a particular band compared to the case of decoupled layers (where $S(k_z) = \text{const}$) can be estimated as:

\begin{equation}
    \frac{\delta \Omega}{\delta \Omega_{2D}}  = 
    \frac{\int d k_z \cos\left(\frac{\hbar c S(k_z)}{e B} -1/2\right) }{\cos\left(\frac{\hbar c S_0}{e B} -1/2\right)  \int d k_z }.
\end{equation}
$S(k_z)$ can be estimated approximating the dispersion with $E(k_\parallel, k_z) \approx \frac{\hbar^2 k_\parallel^2}{2m} + \frac{W}{2} \cos(k_z)$, where $W$ is the bandwidth. The spread of area as a function of $k_z$ can then be estimated as $\Delta S(k_z) = 2\pi k_F \Delta k_F \sim 2\pi k_F \frac{E_z(k_z)}{\hbar^2 k_F/m} = \frac{2\pi m E_z(k_z)}{\hbar^2}$.
As the integral is dominated by the saddle-point contributions, we get 
\begin{equation}
    \frac{\delta \Omega}{\delta \Omega_{2D}}  \approx
  \frac{\lambda}{2\pi}  
 \int d k_z \cos\left( \frac{\hbar c }{e B} \frac{2\pi m W k_z^2 \lambda^2}{4\hbar^2}\right)
  =
  \sqrt{\frac{B}{2 \pi \Phi_0 k_F^2} \frac{E_F}{W}},
\end{equation}
where we used $\int d k_z = \frac{2\pi}{\lambda}$ ($\lambda$ being the spiral period) and $\Phi_0$ is the (superconducting) flux quantum. Estimating $k_F^2 = 2\pi n $ with the density per layer $n\approx 10^{12}$ cm$^{-2}$, we get $\frac{\delta \Omega}{\delta \Omega_{2D}} \sim 0.036 \sqrt{\frac{E_F}{W}} \sqrt{B\text{[T]}}$.

\subsection{Quasiclassical conditions for Fermi arc quantum oscillations} \label{supp:classical-condition}
In Sec. \ref{sec:QO} we have demonstrated that in a finite supertwisted spiral slab, Fermi arcs form on opposite surfaces of the system and suggested that these may be observed in quantum oscillation experiments. This suggestion, however, relies on the applicability of the semiclassical approximation to these orbits, which we discuss here. Specifically, the wavefunction along this orbit evolves rapidly from localized at the surface layer to extended throughout all the layers and then localized again at the other surface.

Here we will estimate the corrections to the quantum oscillation frequencies due to wavefunction change and identify the regime when they are small. Such corrections arise systematically in the semiclassical expansion \cite{gao2017zero}:
\begin{equation}
    n B = \frac{\phi_0 S_0}{4 \pi^2}
    +
    [...]B
    + \frac{\phi_0 B^2}{2} \frac{\partial \chi}{\partial \mu}
+ O(B^3),
\label{sup:eq:qocorr}
\end{equation}
where both linear and quadratic in $B$ terms contain correction from wavefunction geometry. The former, however, only affects the phase of the oscillations, while the latter affects frequency. Moreover, the Berry curvature vanishes in the case of a finite slab, as the Bloch Hamiltonian (and thus the corresponding functions) are real for fixed ${\vb k}$.

In \cite{niu2015}, the geometric correction to susceptibility was derived for a general Hamiltonian that depends on the Hessian $\Gamma_{ij} = \partial_{k_i k_j}$, given in our case by a matrix in layer space:
\begin{equation}
    \Gamma_{ij}^l = 
    \frac{1}{m}
    \begin{pmatrix}
    1+ \varepsilon \cos(2\theta l) & - \varepsilon \sin(2 \theta l)\\
    - \varepsilon \sin(2 \theta l) & 1-\varepsilon \cos(2 \theta l)
    \end{pmatrix}.
\end{equation}
For the purpose of the estimate, we approximate $ \Gamma_{ij}^l  \approx \frac{1}{m} \delta_{ij}$ (which is valid for $\varepsilon \ll 1$), in which case the correction term becomes (cf. Eq. (7) in \cite{niu2015}):
\begin{equation}
    \frac{\partial \delta \chi_{geom}}{\partial \mu} =  \frac{\nu}{m} 
    \left\langle
    \frac{e^2}{c^2} [\langle \partial_{k_x} \psi_{AA}  
    |\partial_{k_x} \psi_{AA}\rangle
    +
    \langle \partial_{k_y} \psi_{AA}   |\partial_{k_y} \psi_{AA} \rangle]
    \right\rangle_{FS},
\end{equation}
where $\nu = \int \frac{d^2 k}{(2\pi)^2} ( -f')$ is the density of states at the Fermi energy (which can be approximated by $m/\hbar^2$ for $\varepsilon\ll1, t\ll E_F$ ), $f$ being the Fermi distribution.

The most pronounced change in the wavefunction occurs over the region where the wavefunction transforms from localized to extended. The extent of this region in momentum space can be estimated as $\Delta k \sim \frac{t}{E_F} k_F$ and 
$\langle \partial_k \Phi_{AA}  |\partial_k \Phi_{AA}\rangle$ as $\left(\frac{d l_{loc}/dk \sim L/\Delta k}{l_{loc}} \right)^2$. The condition that the wavefunction dependence correction ($\sim B^2$) in \eqref{sup:eq:qocorr} is small then becomes:

\begin{equation}
\frac{B^2}{2} \frac{\partial \chi}{\partial \mu}
\sim
\frac{e^2 B^2} {\hbar^2 c^2} \left(\frac{L}{l_{loc} (t/E_F) k_F}\right)^2 \ll \frac{S_0}{4\pi^2} \sim k_F^2,
\end{equation}
leading to 
\begin{equation}
    \frac{e B}{\hbar c} \frac{L}{l_{loc}} \frac{E_F}{t} \ll k_F^2,
\end{equation}
which can be rearranged as
\begin{equation}
    \hbar \omega_c \ll t \frac{l_{loc}}{L},
\end{equation}
using $\omega_c = eB/(m c)$ and $E_F = \hbar^2 k_F^2/(2m)$.
\newpage

\section{Details for the calculation of quantum oscillation phase offset}
This section describes the details regarding the phase offset. \ref{supp:phase-calculation} outlines the calculation for each terms in phase offset while \ref{supp:orb-vanish} shows the vanishing orbital contribution in the strongly-localized limit.
\subsection{Phase offset calculation}\label{supp:phase-calculation}
In this section, we outline the calculation for the phase offset in the commensurate limit. Given the Bloch wave function $\ket{\psi_{\kk k_z}^{ n_{b}}}$ where $n_{b}$ is the band index, we can write the Berry connection as 
\begin{align}
    \mathcal{A}^{n_{b}} (\kk, k_z) = \bra{\psi_{\kk k_z}^{ n_{b}}} i \nabla_{\kk} \ket{\psi_{\kk k_z}^{ n_{b}}}.
\end{align}
However, calculating Berry connection numerically can be difficult due to its quantity not being gauge invariant. To circumvent such issue, the Berry phase can be formulated in a gauge-invariant way by considering the contour $\mathcal{C}$ into series of $\kk_0, \kk_1, ..., \kk_N, \kk_{N+1}$ where $\kk_{N+1} = \kk_0$, the Berry contribution can, then, be written as \cite{vanderbilt2018berry}
\begin{align}
    \gamma_B^{n_{b}} = {\rm Im} \log \bra{\psi_{\kk_{N+1} k_z}^{n_{b}}}\ket{\psi_{\kk_{N} k_z}^{n_{b}}} ... \bra{\psi_{\kk_{1} k_z}^{n_{b}}}\ket{\psi_{\kk_{0} k_z}^{n_{b}}}.
\end{align}
On the other hand, the orbital magnetic operator can be written in a gauge invariant form as \cite{Niu1996,alexandradinata2018semiclassical} 
\begin{align}
\label{eqn:orbital-moment}
    m^{n_b}(\kk, k_z) = i \frac{\hat{z}} {2\phi_0} \cdot \sum_{n_b' \neq n_{b}} \frac{\vb{\Pi}_{n_b n_b'}(\kk, k_z) \times \vb{\Pi}_{n_b' n_b}(\kk, k_z)}{E_{n_b}(\kk, k_z) - E_{n_b'}(\kk, k_z)},
\end{align}
where $\vb{\Pi}_{n_b n_b'}(\kk)$ is the in-plane velocity operator which can be related to the Bloch Hamiltonian $H(\kk, k_z)$ via 
\begin{align}
    \label{eq:interband-velocity}
    \vb{\Pi}(\kk, k_z) =\nabla_{\kk} H (\kk, k_z).
\end{align}
We can, then, write an expression for the orbital contribution as
\begin{align}
\gamma_{\rm orb}^{n_b} &= - \phi_0\int d^2\kk \delta(E^{n_b}(\kk, k_z) - E) m^{n_b}(\kk)
\nonumber
\\
&= -\frac{i} {2}\int_{\mathcal{C}} \frac{d\kk}{|\Pi_{n_b n_b} (\kk, k_z)|} \cdot \sum_{n_b' \neq n_{b}} \frac{\hat{z} \cdot \left (\vb{\Pi}_{n_b n_b'}(\kk, k_z) \times \vb{\Pi}_{n_b' n_b}(\kk, k_z) \right )}{E_{n_b}(\kk, k_z) - E_{n_b'}(\kk, k_z)},
\end{align}
where $\vb{\Pi}_{n_b n_b} (\kk, k_z)$ is the diagonal part of the velocity operator i.e. the Fermi velocity of the band.
\clearpage
\subsection{Orbital phase in strongly localized limit}\label{supp:orb-vanish}
We consider the orbital phase contribution in the strongly localized limit. At a fixed-energy contour, the wavefunction can be well-approximated by the single-layer wavefunction up until the anti-crossing where the wave function of the adjacent layer would hybridized. Consider the wavefunction near the anti-crossing of layer 1 and 2, we can write the approximate Hamiltonian as 
\begin{align}
        H_{12}(\kk, k_z) =  E_{||} (\kk) + \begin{pmatrix}
            \Delta_1 (\kk) & t e^{-i k_z} \\
            t e^{ik_z}& \Delta_2 (\kk)
        \end{pmatrix},
\end{align}
 where $\Delta_1$($\Delta_2$) is the anisotropy term at layer 1(2). Suppose $\ket{\psi_{\pm}(\kk, k_z)}$ describes the eigenstate of such Hamiltonian,  the inter-band velocity operator \eqref{eq:interband-velocity} can be written in such basis as
 \begin{align}
     \Pi (\kk, k_z) = \nabla_{\kk} \left ( E_{||} (\kk) + \frac{\Delta_1(\kk) + \Delta_2(\kk)}{2}\right ) + \nabla_{\kk} \left (\frac{\Delta_1(\kk) - \Delta_2(\kk)}{2}\right ) 
     \bra{\psi_{\pm} (\kk, k_z)}
     \tau_z
     \ket{\psi_{\pm} (\kk, k_z)},
 \end{align}
 where $\tau_z = \begin{pmatrix}
     1 & 0 \\ 0 & -1
 \end{pmatrix}$. Since the orbital moment \eqref{eqn:orbital-moment} came from the off-diagonal matrix element in the velocity operator, only the second term can contribute to the orbital moment. However, since the second term in $\Pi^x$ and $\Pi^y$ only differs by the scalar multiplication, we can show the cross product between the velocity operator vanishes
 \begin{align}
     \Pi^x_{+-} \Pi^y_{-+} = \Pi^y_{+-} \Pi^x_{-+} = \partial_{k_x} \left (\frac{\Delta_1(\kk) - \Delta_2(\kk)}{2}\right )\partial_{k_y} \left (\frac{\Delta_1(\kk) + \Delta_2(\kk)}{2}\right ) |\bra{\psi_{+} (\kk, k_z)}
     \tau_z
     \ket{\psi_{-} (\kk, k_z)}|^2,
 \end{align}
 along with the orbital contribution.
 
\newpage

\section{General calculation of the Berry phase for the bulk orbits}
Here, we outline the systematic approach to calculate the Berry phase of the bulk orbit in the low-field (\ref{supp:low-field}) and intermediate-field (including magnetic breakdown effects, \ref{supp:intermediate-field}) limits. The implication of the semi-classical picture on the z-axis dispersion is also discussed in \ref{sup:sec:semiclVz}.

\subsection{Low fields} \label{supp:low-field}

\begin{figure}[h]
\centering
\includegraphics[width=0.4\linewidth]{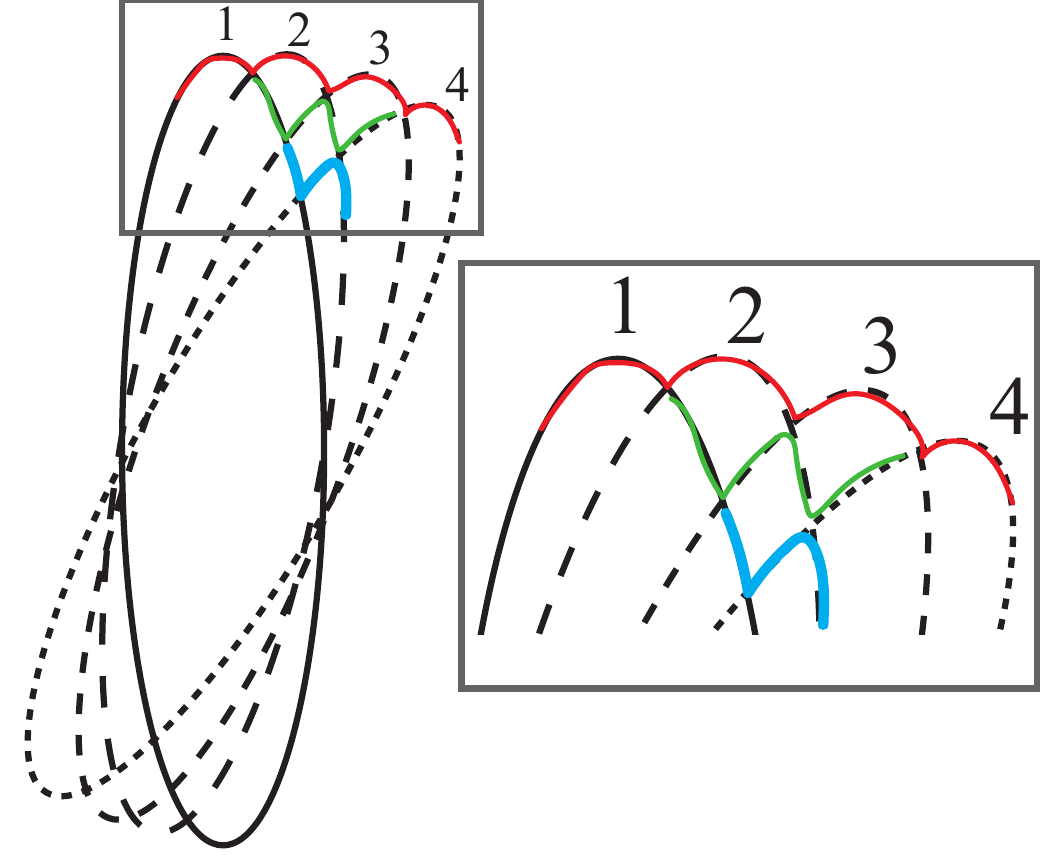}
    \caption{{\emph{Formation of closed momentum-space orbits in commensurate superspirals.}} Fermi surfaces of four consecutive layers of a superspiral are shown (ellipses with different dashing). Color shows the continuous orbits for semiclassical evolution of momentum in presence of magnetic field (magnified in the inset). The evolution is consistent with the cycle Eq. \eqref{sup:eq:lowfieldcycle} for $q=1$ (red) $q=2$ (green) and $q=3$ (blue).}
    \label{sup:fig:gensch-1}
\end{figure}

Here we consider the case of low fields where magnetic breakdown effects are negligible. We find that a systematic analysis is possible in the case of commensurate spiral $\lambda \in \mathbb{N}$ (corresponding to $\theta = \pi \frac{r}{\lambda}$, $r$ and $\lambda$ being co-prime). In such case, the electron traverses the layer with the following pattern (see Fig. \ref{sup:fig:gensch-1}):

\begin{equation}
    1 \to 1 + q \to 2 \to 2+q \to 3 \to ... \to N \to N+q \to 1
    \label{sup:eq:lowfieldcycle}
\end{equation}
for $q=1, 2, ..., \lambda-1$. Each step in the cycle leads to the electron moving along the $z$ direction. To calculate the Berry phase accumulated over a single closed orbit we note that advancing by one unit cell along the $z$ axis advances the phase by $k_z$ (due to the presence of $e^{i k_z}$ factor in the tunneling across the border between adjacent unit cells). Thus, we need to calculate the overall shift of the wavefunctions along the $z$ direction for each closed trajectory formed from a cycle of the form \eqref{sup:eq:lowfieldcycle}.

A critical consideration is that each tunneling process in \eqref{sup:eq:lowfieldcycle} can occur, in principle in two directions along the $z$ axis due to the periodicity of the system. In the $\Delta\gg t$ limit, however, one of the processes will typically be dominant --- the one where tunneling involves less intermediate layers. The exception occurs when this number is the same for both increasing and decreasing $z$; this occurs for even period $\lambda = 2k$ and is analyzed below.

\subsubsection{Odd period: $\lambda = 2k+1$.}

In this case, we can always identify the dominant tunneling direction for every overlap between Fermi surfaces of different layers. Specifically, for $ p \to p + q $ type of overlap the transition occurs in the positive $z$ direction (i.e., by $q$ layers forward) for $q\leq k$ and in the negative direction (i.e., by $\lambda-q$ layers back)  for $q>k$. For $p + q \to p+1$ overlap, a semiclassical wave packet would tunnel predominantly in the negative $z$ direction (by $q-1$ layers) for $q\leq k+1$ and forward (by $\lambda-q+1$ layers) for $q> k+1$.

We can now calculate the total layer displacement along $z$ axis that a semiclassical wave packet will accumulate when going along a closed orbit with arbitrary $q$:
\begin{equation}
    \Delta l = \begin{cases}
    2 \times \sum_{l=1}^{\lambda} +q + (1-q)  = 2 \lambda  &q\leq k\\
    2 \times \sum_{l=1}^{\lambda} (q-\lambda) + (1-q) = 2 (1-\lambda)\lambda  &q = k+1\\
      2 \times \sum_{l=1}^{\lambda} (q-\lambda) + (\lambda-q+1) = 2 \lambda   &q > k+1
    \end{cases},
\end{equation}
where the factor of $2$ is due to $C_2$ symmetry. In other words, as $\theta=\pi/N$, 1 full cycle \eqref{sup:eq:lowfieldcycle} would correspond to the wave packet being at the bottom arc of the Fermi surface of the layer number 1 in Fig. \ref{sup:fig:gensch-1}. The corresponding contributions to the Berry phase are then given by:
\begin{equation}
    \varphi_B = \frac{\Delta l \cdot k_z}{\lambda}
    =
        \begin{cases}
2 k_z  &q \leq k,\\
2(1-\lambda)k_z &q = k+1,\\
2 k_z   &q > k+1.
      \end{cases}
\end{equation}
leading to the results reported in the main text.

\subsubsection{Even period: $\lambda = 2k$.} In this case, more care needs to be taken as to tunneling direction; as we will see below there are cases here where the tunneling at a layer degeneracy point has equal amplitudes between the forward and backward directions. 

The overall scheme of the transitions between layers and formation of closed orbits is the same as for the odd-$\lambda$ case and represented by Fig. \ref{sup:fig:gensch-1} and Eq. \eqref{sup:eq:lowfieldcycle}. However, the directionality of transitions at individual crossing differs from that case. For $ p \to p + q $ type of overlap, the transition occurs in the positive $z$ direction (i.e., by $q$ layers forward) for $q< k$ and in the negative direction (i.e., by $\lambda-q$ layers back)  for $q>k$. For $q=k$, however, the transitions  $ p \to p + q $ and $ p \to (p + q -\lambda) $ involve the same number ($q=k$) of intermediate layers and would have the same (in magnitude) matrix elements. For $p + q \to p+1$ overlap, a semiclassical wave packet would tunnel predominantly in the negative $z$ direction (by $q-1$ layers) for $q< k+1$ and forward (by $\lambda-q+1$ layers) for $q> k+1$, while at $q=k+1$ the forward and backward tunneling would again be tied. As a result, for $q<k$ and $q>k+1$ we recover the previous results, but the cases $q=k,k+1$ warrant separate consideration. 

For $q=k$, the transitions $ p \to p + q $ would occur over $k$ intermediate layers in both positive and negative $z$ directions. Perturbatively, the effective hopping between layers $p$ and $p+q$ is given by $\tilde{t}_{p+q, p}+ \tilde{t}_{p-q, p}e^{-ik_z}$ for $p\leq k$ and $\tilde{t}_{p+q, p}e^{ik_z}+\tilde{t}_{p-q, p}$ for $p> k$, where $E_p$ are the energies of $p$th layer at the crossing point, and the phase factors $e^{i\pm k_z}$ arise when tunneling to the next unit cell occurs. $\tilde{t}_{p\pm q, p}$ is the effective hopping between layer $p$ and $p\pm q$ and is given by $\frac{t^k}{\prod_{i=1}^{q-1} (E_p - E_{p\pm i})}$ perturbatively. Without loss of generality (due to spiral screw symmetry), we will consider the effective hopping between the layer $0$ (zero twist) and layer $k$ ($\pi/2$ twist). Up to $\mathcal{O}(|\kk|^2)$, the single-layer energy can be described as 
\begin{align}
    E_p = E_{||} + \Delta \sin(r\frac{\pi}{k} p),
\end{align}
where $r \in \mathbb{Z}$, the period of the sinusoidal is fixed by the period of the spiral $\lambda = 2k$, and the phase of the sinusoidal is fixed by the resonance condition $E_0 = E_{\pm k}$. In that case, we get $(E_0 - E_i) = - (E_0 - E_{-i})$, such that of $t_{0, k}$ and $t_{0, -k}$ are equal of to a relative sign of $(-1)^{k-1}$. Using $ (1 \pm e^{i k_z}) = e^{i k_z/2} (e^{-i k_z/2} \pm e^{i k_z/2})$, we find that the tunneling in this case will produce a Berry phase of $k_z/2 + m \pi/2$. Note that the hybridization vanishes exactly for $k_z=0$ or $\pi$, so there will be a contribution from breakdown orbits around these $k_z$ values that will be small at low fields. The multiple of $\pi/2$, however, does not contribute to the final result as there are at least $4$ segments over which phase will accumulate for $k>1$. Similar analysis can be carried out for $q=k+1$, in which case the Berry phase of $+ k_z/2$ will accumulate over the $p+1\to p+1$ crossings for $p>k$ and $- k_z/2$ otherwise. Analogous analysis can also be made for $q = k+1$ case, resulting in the same Berry phase.

The final result for the Berry phase is then given by:
\begin{equation}
    \varphi_B = 
    \begin{cases}
2 k_z  &q < k,\\
    \frac{2k_z}{\lambda = 2 k} \times \left(
    \sum_{l=1}^{k} -\frac{2k-k}{2} +(1-k) +\sum_{l=k+1}^{2k}\frac{k}{2}+ (1-k) \right) = (2 - \lambda)k_z  &q = k,\\
    \frac{2k_z}{\lambda = 2 k} \times \left(
    \sum_{l=1}^{k} -(2k-k) +\frac{1-k}{2} +\sum_{l=k+1}^{2k} -(2k-k) +\frac{2k - k+1}{2} \right) = (2 - \lambda)k_z  &q = k+1,\\
2 k_z   &q > k+1,
      \end{cases}
\end{equation}
such that the total Berry phase for all closed orbits is zero.

\newpage
\subsection{Intermediate fields: nearest-layer tunneling approximation}\label{supp:intermediate-field}

Here we consider the case where all but nearest-layer tunneling can be neglected due to magnetic breakdown. This constitutes a wide range of fields for low interlayer tunneling.

\begin{figure}[h]
\centering
\includegraphics[width=0.4\linewidth]{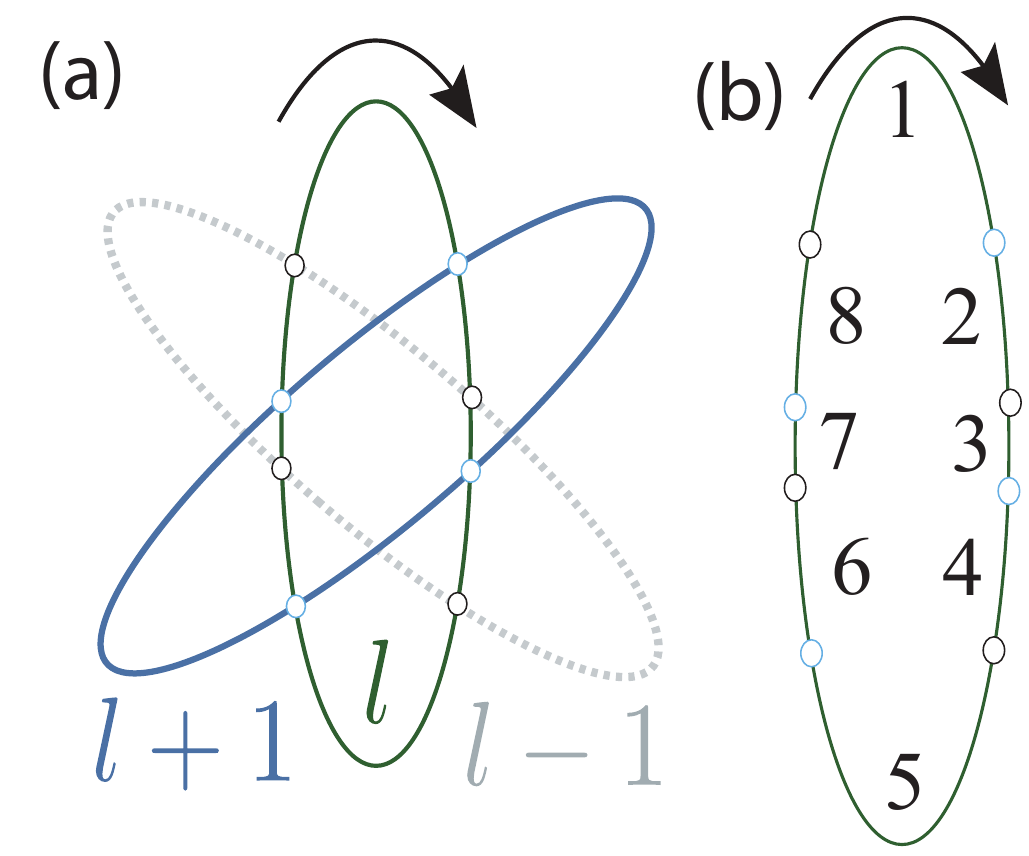}
    \caption{{\emph{General schematic for tunneling between adjacent layers in a superspiral in the low-tunneling limit.}} (a) Schematic of a Fermi surface for $n$-th layer and adjacent layers on the top (red) and bottom (blue) of it. Circles mark the intersection points of the respective Fermi surfaces. (b) Fermi surface of a layer in the superspiral with points of degeneracy with adjacent layers. At sufficiently weak fields, these points are where the tunneling between layers occurs. Arrow points in the direction of the momentum change in the presence of magnetic field. }
    \label{sup:fig:gensch}
\end{figure}

As a result, in each layer we can identify 8 different Fermi surface segments, Fig. \ref{sup:fig:gensch} (b). In the presence of magnetic field, the electron momentum evolves along the Fermi surface in a direction fixed by magnetic field (here, without loss of generality we assume this to be clockwise direction). Once the electron momentum reaches a crossing point, it tunnels to the adjacent layer, where the schematic Fig. \ref{sup:fig:gensch} can again be applied. As a result, we can identify the following cycles in the ``Fermi surface segment space", where each arrow corresponds to traversing one avoided crossing:
\begin{equation}
    \begin{gathered}
        1 \xrightarrow{+} 1; \; 3\xrightarrow{+} 3; \;5 \xrightarrow{+} 5; \;7 \xrightarrow{+} 7; \;\\
        2\xrightarrow{-} 4 \xrightarrow{-} 6 \xrightarrow{-} 8,
    \end{gathered}
    \label{sup:eq:highfieldcycles}
\end{equation}
where $\xrightarrow{+}$ means that traversing the anticrossing leads to the electron wavepacket moving from $n$-th layer to the $n+1$ layer (and conversely $\xrightarrow{-}$ indicates that transition occurs to $n-1$ layer).

\subsection{Semiclassical $v_z$ estimate}
\label{sup:sec:semiclVz}

Consider a semiclassical wavepacket moving along the Fermi surface in the presence of magnetic field. One can find its velocity along $z$ direction as $v_z = \frac{a_c}{T_i}$, where $T_i$ is the time that it takes for the wavepacket to traverse one of the Fermi surface segments ($i=1...8$), Fig. \ref{sup:fig:gensch}, before tunneling to a neighboring layer. Importantly, as the tunneling sequence \eqref{sup:eq:highfieldcycles} implies that after tunneling, the wavepacket propagates along a segment with exactly same length as the previous one, this definition is internally consistent and can be applied even in aperiodic systems.

Using semiclassical dependence $p_{x,y} = p_{x,y}^0 \sin(\omega_c t+ \psi_0)$, where $p_{x}^0 = \sqrt{\frac{2 m E}{1+\varepsilon}},\;p_{y}^0 = \sqrt{\frac{2 m E}{1-\varepsilon}}$ we can find the time a wavepacket will spend traversing each of the segments in Fig. \ref{sup:fig:gensch} (b):
\begin{equation}
\begin{gathered}
    T_{1,5} = \frac{2}{\omega_c} 
    \arcsin\left[ \sin \frac{\theta}{2} \sqrt{\frac{1+\varepsilon}{1-\varepsilon \cos \theta}} \right],\\
    T_{3,7} = \frac{2}{\omega_c} 
    \arcsin\left[ \sin \frac{\theta}{2} \sqrt{\frac{1-\varepsilon}{1+\varepsilon \cos \theta}} \right],\\
    T_{2,4,6,8} = \frac{1}{4} \left( \frac{2 \pi}{\omega_c} - 2 T_{1,5}-2T_{3,7} \right),\\
\end{gathered}
\label{sup:eq:semiclas}
\end{equation}
where the arguments of $\arcsin$ can be shown to be less than $1$ for the case $|\varepsilon|<1$ that we consider.

It can be further demonstrated that the fraction of the density of states corresponding to each of the segments is directly related to $T_i$, Eq. \eqref{sup:eq:semiclas}, via:
\begin{equation}
    \frac{\nu_i}{\nu_0} = \int_{p\in i} dp_x dp_y \delta(E({\bf k}_\parallel)-E).
\end{equation}
For segments where $p_y$ can be represented as single-valued function of $p_x$ (such as section $1$ and $5$) we can further simplify:
\begin{equation}
     \frac{\nu_i}{\nu_0} = \frac{2 \arcsin \frac{\Delta p_x}{2 p_x^0}}{2\pi},
\end{equation}
where $\Delta p_x$ is the length of the segments projection of the $p_x$ axis, which is equal to $\frac{\omega_c T_i}{2 \pi}$. Proceeding analogously for the other segments one finds that the formula applies there too. Importantly, this leads to the total field-induced current along $z$ vanishing in equilibrium: 

\begin{equation}
    \delta j_z (B) = \int d E \sum_{i=1}^{8} \nu_i v_i^z =  \int d E \left(4 \frac{\omega_c a_c}{2\pi}-4 \frac{\omega_c a_c}{2\pi}\right) =0,
\end{equation}
where we used that there are $4$ segments propagating in opposite directions each (1,3,5,7) and (2,4,6,8).

\subsubsection{Commensurate spirals}

In commensurate spirals, the cycles described above will lead to the emergence of closed orbits, commensurate with the spiral period. These orbits correspond to the case when magnetic breakdown dominates all Fermi surface crossings apart from nearest-neighbor ones.

Let us now calculate the Berry phases arising for the emergent orbits for $\theta = \pi \frac{p}{q}$, where $p<q$ are co-prime integers. We restrict our consideration to the case where $\theta< \pi/2$ as the cases with $\pi/2<\theta <\pi$ and $\theta>\pi$ can be mapped onto the aforementioned case using the $C_{2z}$ symmetry ($\theta\to \pi-\theta$ and $k_z\to -k_z$ for $\pi/2<\theta <\pi$, and $\theta \to \theta-\pi$ for $\theta>\pi$). 
Due to the $C_{2z}$ symmetry, under $\pi$ rotation, opposite segments in Fig. \ref{sup:fig:gensch} (b) transform into one another ($n\to n+4$, e. g. $1\to 5$). As a result, the first line in Eq. \eqref{sup:eq:highfieldcycles} leads to two closed orbits for odd $p$ (as after $q$ layers a $\pi$ rotation is accomplished), and 4 orbits for even $p$ (i.e. segments $1,3,5,7$ each lead to a separate trajectory). These are the ``innermost" and ``outermost" orbits. Since $p,q$ are co-primes, for any $n<q$, one has that $n \theta$ can not be equal to an integer multiple of $\pi$ and no transitions between unit cells occurs until $q$ transitions are accomplished. For even $p$, this implies that $q$ tunnelings constitute a full closed orbit, such that the Berry phase associated with it is $k_z$. For odd $p$, after $q$ interlayer transitions, the wavepacket will arrive at a $C_2$-related segment of the Fermi surface. Thus, a full closed orbit forms only after $2 q$ tunnelings among which transition between unit cells occurs twice, leading to a Berry phase of $2 k_z$ (but there are $2$ closed orbits as opposed to the even $p$ case, when there are $4$).





The orbits arising from the second line of \eqref{sup:eq:highfieldcycles} are constrained by the fact that the total cycle length $\delta l$ should be multiple of $4$. As a result, the cycle should satisfy two conditions: $\delta l \cdot \theta = 2\pi n_1$ and $\delta l = 4 n_2$, where $n_1,n_2\in \mathbb{N}$, where $\delta l$ is the total number of interlayer transitions during one cycle. For even $q =2 k$ (implying odd $p$) we find that $n_2=k,\; n_1=p$ satisfy the conditions above, resulting in cycles of $\delta l = 4k = 2 \lambda$ layers each and $2\pi \gamma_B= -2 k_z$. To cover all 4 segments $2,4,6,8$ of each layer's Fermi surface, two such orbits are required, consistent with vanishing total Berry phase of all orbits. For odd $q =2 k+1$, we find $n_1=2 p, n_2 = q$ to be the minimal integers satisfying the conditions above. As a result, we obtain a single cycle of $\Delta l = 4 q = 4 \lambda$ and total $2\pi\gamma_B = -4 k_z$. Remarkably, this trajectory covers {\it all} the $2,4,6,8$ segments for each Fermi surface of every layer, as can be shown for particular cases by directly drawing this trajectory. The total semiclassical area of this orbit is thus close to $\lambda$ times the Fermi surface of one layer.

We can now estimate the velocity using Eq. (16) of the main text, to demonstrate its consistency with the semiclassical estimate in \ref{sup:sec:semiclVz}. The correction to the velocity due to magnetic field is given by:
\begin{equation}
         \delta v_z
     =
     \frac{\lambda a}{\hbar}
     \left(\frac{\partial S}{\partial E} \right)^{-1}
     \frac{e B}{\hbar c} \frac{\partial (2\pi \gamma)}{\partial k_z}.
\end{equation}
$S$ represents here a total area of the closed trajectory. However, for trajectory consisting of identical segments $S=\delta l S_i$, where $S_i$ is the corresponding area of a segment in one layer. As a result, for an orbit $i$ characterized by the Chern number $C_i = \frac{\delta l}{\lambda}$, such that $\frac{\partial (2\pi \gamma_i)}{\partial k_z} = C_i$, we get:
\begin{equation}
\delta v_z^i
     =
     \frac{a}{\hbar}
     \left(\frac{\partial S_i}{\partial E} \right)^{-1}
     \frac{e B}{\hbar c}.
     \label{sup:eq:dvzchern}
\end{equation}

Classically, the time it takes the particle to traverse a segment of trajectory is given by $T_i = \int_{x_i^i}^{x_i^f} \frac{d x}{\sqrt{\frac{2}{m}(E-U(x))}}=\frac{\partial S_i}{\partial E}$
\cite{landau_mechanics}, $S^{cl}_i$ being the classical action on the segment. Correspondingly, the semiclassical time for motion in momentum space in presence of magnetic field is (using $\hat{y}_B \equiv \frac{c }{e B} \hat{p}_x$ playing the role of ``coordinate"  $[\hat{p}_x,\hat{y}_B]=-i\hbar$): $T_i = \frac{c}{eB} \hbar^2\frac{\partial S_i}{\partial E}$, where $\hbar^2 S_i = \int_i p_y dp_x$ along the $i$ segment, i.e. the area under the segment. The semiclassical velocity estimate can be then rewritten as:
\begin{equation}
    \delta v_z^{\rm semicl} = \frac{a}{T_i} = \frac{a e B}{\hbar^2 c}   \left(\frac{\partial S_i}{\partial E} \right)^{-1},
\end{equation}
which coincide with Eq. \eqref{sup:eq:dvzchern} identically.

\clearpage
\section{Perturbative analysis of the effects of the moir\'e potential}

We consider the eigenfunction 
$
    \psi_{\kk \alpha} (\rr)_l = u_{\kk\alpha l} e^{i \kk \cdot \rr},
$
where $\alpha$ is an index for the eigenfunction. We can project the moir\'e potential $V_{M, l}(\rr) = V (\cos g x_{l\theta}  + \cos g y_{l\theta})$ onto the eigenstate as 
\begin{align}
    \bra{\psi_{\kk\alpha}} &\hat{V}_M \ket{\psi_{\kk'\beta}} = \sum_{l} u^*_{\kk\alpha l} u_{\kk' \beta l} \int d\rr \, e^{-i (\kk - \kk') \cdot \rr}  V_{M, l}(\rr) 
    \nonumber
    \\
    &= V_M \pi \sum_{l, \g}  u^*_{\kk\alpha l} u_{\kk' \beta l}  \delta(\kk - \kk' - \g_{l\theta}),
\end{align}
where the moir\'e potential breaks the in-plane translational symmetry and couple states with different $\kk$ together. In the perturbative limit, this could open an energy gap with an order of magnitude $\bra{\psi_{\kk\alpha}} \hat{V}_M \ket{\psi_{\kk'\beta}}$ when the moir\'e potential connects two degenerate states, i.e., $\kk = \pm \g/2$. 

In the strongly localized limit, we can approximate the eigenfunction to be localized within one layer $u_{\kk \alpha l}^{\rm loc} = \delta_{l, z_\alpha}$. This limit requires the neighbor layer energy difference to dominate the hopping $|\kk| \gtrsim k_{c, 2} \sim {k_c}/(\pi\beta)$. Since the wavefunction is localized into a single layer, the moir\'e potential can open a gap and form moir\'e bands identically to the two-dimensional system with the gap of order $V_M$.  In the extended limit, the eigenfunction can be approximated by the plane wave. For the system with period $\lambda$, the wave function can be written as  $u_{\kk \alpha l}^{\rm ext} \sim e^{ik_{z\alpha} l} /\lambda^{1/2}$, resulting in the suppressed gap of order $V_M/\lambda$. In the case where each layer is not perfectly aligned, we can introduce a lattice mismatch $\vb{d}_l$ between each layer $l$. The shift would result in the moir\'e potential $V_{M, l}(\rr - \hat{z} \times \vb{\dd}_l/\theta)$ which corresponds to introducing a phase factor $e^{i \hat{z} \times \vb{d}_l/\theta}$. For $L \gg \lambda$ with random misalignment, the phase factor averages out along with any gap opening from moir\'e potential.

In an intermediate regime where $k_c \lesssim |\kk| \lesssim k_{c, 2}$, the wavefunction is spread across a finite region with an exponential tail. In the small twist limit $\beta \ll 1$, the spectra minima can be approximated locally with the Harmonic oscillator where $u \sim e^{- 
     {(l - l_0)^2}/{2l_{\rm HO}^2}}/l_{\rm HO}^{1/2}$
\end{document}